\documentclass[longauth]{aa}  

\usepackage{graphicx} 
\usepackage{txfonts}
\usepackage{natbib}
\bibpunct{(}{)}{;}{a}{}{,}
\usepackage{amsmath}
\usepackage[version=4]{mhchem}
\usepackage{longtable}
\usepackage{threeparttable}
\usepackage{booktabs}
\usepackage{subcaption}
\usepackage{placeins}
\usepackage{adjustbox}
\usepackage{ragged2e}
\usepackage{xcolor}
\usepackage{float}
\usepackage[colorlinks=true,linkcolor=blue,citecolor=blue,urlcolor=blue]{hyperref}  

\usepackage{amsmath}
\usepackage{graphicx}
\usepackage{setspace}
\usepackage{enumitem}
\usepackage{siunitx}
\usepackage{xcolor}
\usepackage{tikz}

\usepackage[version=4]{mhchem}

\begin{document}

   \title{MINDS: The mid-infrared spectrum of the disk, jets, and winds of DG Tau} 

      \author{Manuel G\"udel\inst{1, 2, 3} 
      \and P\'eter \'Abrah\'am\inst{1, 4} 
      \and Till Kaeufer\inst{5} 
      \and \'Agnes K\'osp\'al\inst{4} 
      \and J\'ozsef Varga\inst{4} 
      \and Nicole Pawellek\inst{1, 4} 
      \and Kamber Schwarz\inst{6}
      \and Valentin Christiaens\inst{7, 8, 9}
      \and Danny Gasman\inst{6}
      \and Matthias Samland\inst{6}
      \and Ewine F. van Dishoeck\inst{10, 11}
      \and Thomas Henning\inst{6}
      \and Aditya M. Arabhavi\inst{12}
      \and Alessio Caratti o Garatti\inst{13}
      \and Pac\^ome Est\`eve\inst{14}
      \and Adrian Glauser\inst{2}
      \and G\"oran Olofsson\inst{15}
      \and Beno\^{i}t Tabone\inst{14}
      \and G\"oran \"Ostlin\inst{15}
      }

   \institute{University of Vienna, Department of Astrophysics,
              T\"urkenschanz\-strasse 17, A-1180 Vienna
              \and ETH Z\"urich, Institute for Particle Physics and Astrophysics, Wolfgang-Pauli-Str. 27, 8093 Z\"urich, Switzerland
              \and ASTRON, Netherlands Institute for Radio Astronomy, Oude Hoogeveensedijk 4, 7991 PD Dwingeloo, The Netherlands
              \and Konkoly Observatory, HUN-REN Research Centre for Astronomy and Earth Sciences, MTA Centre of Excellence, Konkoly-Thege Mikl\'os \'ut 15-17, H-1121 Budapest, Hungary
              \and Department of Physics and Astronomy, University of Exeter, Exeter EX4 4QL, UK
              \and Max-Planck-Institut f\"{u}r Astronomie (MPIA), K\"{o}nigstuhl 17, 69117 Heidelberg, Germany
              \and Institute of Astronomy, KU Leuven, Celestijnenlaan 200D, 3001 Leuven, Belgium
              \and STAR Institute, Universit\'e de Li\`ege, All\'ee du Six Ao\^ut 19c, 4000 Li\`ege, Belgium
              \and Universit\'e Paris-Saclay, Universit\'e Paris Cit\'e, CEA, CNRS, AIM, F-91191 Gif-sur-Yvette, France 
              \and Leiden Observatory, Leiden University, PO Box 9513, 2300 RA Leiden, the Netherlands
              \and Max-Planck Institut f\"{u}r Extraterrestrische Physik (MPE), Giessenbachstr. 1, 85748, Garching, Germany
              \and Kapteyn Astronomical Institute, Rijksuniversiteit Groningen, Postbus 800, 9700AV Groningen, The Netherlands
              \and INAF – Osservatorio Astronomico di Capodimonte, Salita Moiariello 16, 80131 Napoli, Italy    
              \and Universit\'e Paris-Saclay, CNRS, Institut d'Astrophysique Spatiale, 91405 Orsay, France
              \and Department of Astronomy, Oscar Klein Centre, Stockholm University, AlbaNova University Center, 10691 Stockholm, Sweden
}

\date{}
\abstract 
{DG Tau is one of the best-studied young stellar objects in transition from a Class I protostar to a T Tauri star. It is no longer embedded in a dense molecular cloud, but still reveals features that are commonly seen in protostars at earlier evolutionary stages. Apart from a protoplanetary disk, DG~Tau hosts two energetic jets that can be studied down to the immediate environment of the star and disk at optical, infrared, and X-ray wavelengths.} 
{We use the Medium Resolution Spectrometer (MRS), an integral field unit of MIRI on the \textit{James Webb} Space Telescope (JWST), to study the gas and dust composition of the protoplanetary disk and investigate the dynamics and thermal evolution of the jet system.} 
{We studied the mineralogical composition of the disk with models including the hot inner rim of the disk, the disk midplane, and the dusty disk atmosphere. The gas disk composition is retrieved after spectral continuum subtraction, using slab models in local thermodynamic equilibrium. To study the ionization, thermal properties, and dynamics of the jets, we used spectral imaging across the entire MRS field of view and employed 1D shock models to study the emission and line ratios for ionized neon.}
{DG Tau reveals a mid-infrared spectrum consisting of a very bright continuum with various silicate features; the line spectrum is dominated by emission from CO, H$_2$O, OH, H\,{\sc i}, H$_2$, and lines from ions such as Ne$^{+}$, Ne$^{2+}$, Ar$^{+}$, Fe$^{+}$, and Ni$^{+}$; no lines from any other molecules are detected. Our dust analysis reveals substantial grain growth to at least 2~$\mu$m. 
Lines of H$_2$O and CO are formed at extremely high temperatures, in the range of 600-2000~K. The ratio of warm to hot water is the lowest among previous observations of disks with JWST. \\
The pair of jets reveals a rich spectrum of emission lines from ions, in particular producing forbidden lines of [Ne\,{\sc ii}], [Ne\,{\sc iii}], [Ar\,{\sc ii}], [Fe\,{\sc ii}], and [Ni\,{\sc ii}]. The forward jet shows a collimated structure as found previously, while the hitherto poorly described, receding  jet is more widely dispersed on scales
of several arcseconds, showing a layered structure with the fastest components being the most collimated. }
{Substantial grown dust in the inner disk of DG Tau at an early evolutionary stage points at rapid grain growth.  
Extreme water and CO temperatures in the inner disk and a very strong continuum point toward the relevance of the strong accretion of DG Tau and consequently strong stellar UV irradiation. 
In contrast, the ionization of the jet gas cannot be reconciled with high-energy radiation from the star itself. Instead, we propose a unified  nonequilibrium ionization shock model in which the previously studied X-ray jet plasma continuously cools, recombining ions until, at temperatures around $10^4$~K and below, metals at low ionization stages emit the observed ion lines.}


   \maketitle

\section{Introduction}
 
The \textit{James Webb} Space Telescope (JWST; \citealt{Gardner2023}) has fundamentally changed our view on protoplanetary disks (PPDs), particularly with respect to their structure, evolution, and composition, thanks to its order-of-magnitude increase in terms of the spatial resolution, spectral resolution, and sensitivity, especially in the mid-infrared (MIR) range \citep{Argyriou2023}. Currently, PPDs are revealing very diverse chemistry and compositions, ranging from carbon-rich compositions with highly abundant hydrocarbon molecules \citep{Arabhavi2025a} to disks where emission from water dominates \citep{Perotti2023, gasman2023, banzatti2025, temmink2025}. It is not yet clear whether such differences are entirely related to stellar mass (e.g., \citealt{grant2025, Arabhavi2025b}) or whether they might also reflect an evolutionary scenario (\citealt{mah2024, houge2025}; see  \citealt{grant2025} for a discussion of
the dichotomy seen among very low-mass stars and classical T Tauri stars [CTTSs]).

It has been empirically demonstrated that the mechanism of accretion through a disk is coupled to disk mass loss in the form of a molecular uncollimated wind or a collimated jet, often seen in forbidden lines of, for example, [O\ {\sc I}] and [S\,{\sc ii}] in the optical (e.g., \citealt{hartigan1995}), [Fe\,{\sc ii}] in the near infrared (NIR; e.g., \citealt{agraamboage2011, white2014a}), or [Ne\,{\sc ii}] in the MIR \citep{vanboekel2009, guedel2010}. 
Models for the jet and disk-wind mass loss include magnetohydrodynamic, and in particular, magnetocentrifugal winds that can radially extend from $\sim$0.1~au to tens of au in the youngest objects \citep{blandford1982, pelletier1992, ferreira1997}, including fast jets from the innermost disk region and slow winds in outer regions, revealing an ``onion-like'' velocity structure around the outflow axis (e.g., \citealt{bacciotti2000}, see \citealt{pascucci2023} and \citealt{Frank2014} for reviews). For more evolved CTTSs, thermally driven photoevaporative flows launched at typically $1-10$~au from the star result in slow winds. These show  velocities of a few km~s$^{-1}$ and may be important for the formation of disk gaps in the later evolution of a PPD \citep{ercolano2008, gorti2009a, gorti2009b, owen2010}.

There is convincing evidence that strong [Ne\,{\sc ii}]~12.81~$\mu$m forbidden line emission from CTTS originates in jets or winds as suggested by [Ne\,{\sc ii}]  line blueshift \citep{herczeg2007}, spatially extended [Ne\,{\sc ii}] flux from the jets and outflows of T Tau \citep{vanboekel2009}, and a \textit{Spitzer} Space Telescope (\textit{Spitzer}, henceforth) CTTS survey showing that the strongest [Ne\,{\sc ii}] emission originates from CTTS that drive jets and outflows \citep{guedel2010}. Further evidence can be found in \citet{sacco2012} and \citet{pascucci2020}.

The JWST guaranteed-time program MIRI Mid-Infrared Disk Survey (MINDS, PID: 1282, PI T. Henning, \citealt{henning2024}) is addressing structure, composition, and dynamics of a large sample of protoplanetary disks covering a wide range of stellar masses and evolutionary stages. 
Apart from disks around CTTS, the sample includes brown dwarf disks, disks around Herbig stars, and debris disks. Focusing on observations in the MIR utilizing the Mid InfraRed Instrument (MIRI; \citealt{Wright2015}), the program is specifically tuned to characterize inner disks; it also includes some objects that are known to drive disk winds, jets, and outflows. 

Among the latter, the MINDS program includes 
strongly accreting objects that are supposedly young but optically revealed CTTS. The Class~I-to-Class~II transition object (a so-called flat-spectrum source) DG Tau is an outstanding member of this category, because of its closeness, its brightness, modest optical extinction, and advantageous inclination ($i \approx 38^{\rm o}$, \citealt{eisloeffel1998}) that reveals a multitude of disk, wind, and outflow structures \citep{agraamboage2011, agraamboage2014, schneider2013a, guedel2018}, including a bipolar jet system seen at wavelengths from radio to X-rays and particularly in forbidden fine-structure lines \citep{eisloeffel1998, guedel2008, agraamboage2011, lynch2013}. 

In this paper, we present a new MIR characterization of the DG Tau star-disk-outflow system; we discuss disk mineralogy, disk chemistry, and the structure of jets based on observations taken by the Medium Resolution Spectrometer (MRS) of MIRI. We also address the origin of the strong forbidden-line emission itself.

\begin{figure}[h!]
    \centering
    \vskip 0.15cm
\includegraphics[width=1.0\linewidth]{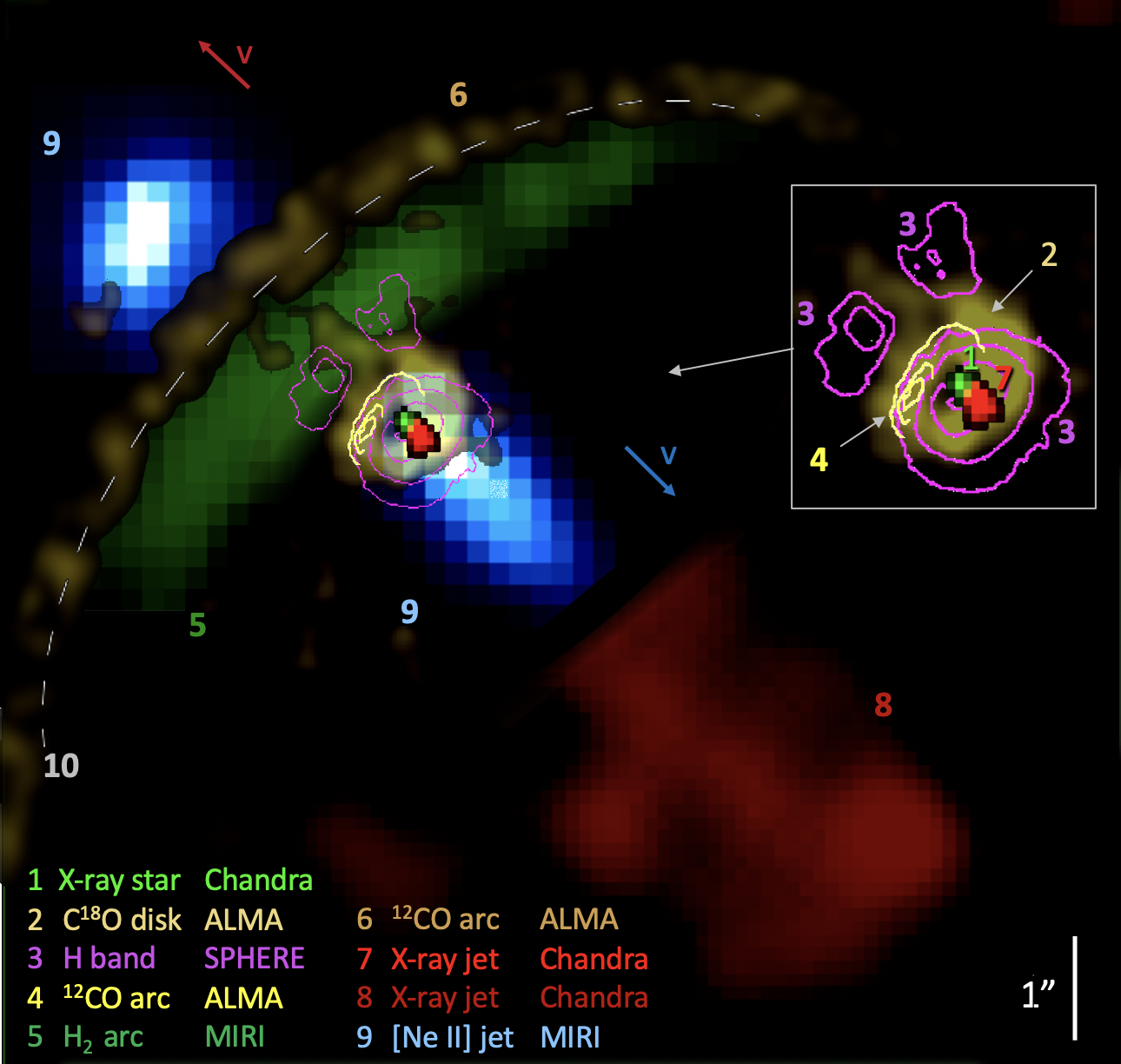}
    \caption{ 
    Overview of the DG~Tau system. The inset highlights the disk and features in its immediate vicinity. 
    \textbf{1. }Bright green dot in center: The central star DG Tau observed in 2--8~keV X-rays by Chandra  (X-ray data from ObsID~11009/11010/11011). 
     \textbf{2. }Faint yellow ellipsoidal structure around star: Protoplanetary disk in the C$^{18}$O line at 219.560~GHz observed by ALMA, smoothed \citep{guedel2018}. 
     \textbf{3. }Purple contours above disk: Scattered-light image in $H$-band from SPHERE, showing a steep cavity-like structure pointing along the forward jet axis, and filamentary extensions to the NE \citep{garufi2024}. A  similar structure is seen in H$_2$ \citep{agraamboage2014}.    
     \textbf{4. }Bright yellow contours at NE edge of disk: Wind in the $^{12}$CO line at 230.538~GHz observed by ALMA (from moment-1 map; \citealt{guedel2018}). 
     \textbf{5. }Broad green arc to the NE of disk: Arc in H$_2$ S(1)~17.03~$\mu$m observed by MIRI/MRS (smoothed), probably located in front of the disk, as discussed below.
     \textbf{6. }Brown arc to the NE of H$_2$ arc: Structure in the $^{12}$CO line at 230.538 GHz observed by ALMA \citep{guedel2018}. 
     \textbf{7. }Bright red elongated feature near star: Inner jet in 0.3--1.5~keV X-rays observed by Chandra \citep{guedel2011, Frank2014}; the X-ray stellar and jet sources have been deconvolved using subpixel event repositioning.
     \textbf{8. }Red elongated feature in the lower right: Outer blueshifted jet in 0.3--1.5~keV X-rays observed by Chandra \citep{guedel2011, Frank2014}; the image has been smoothed and cut off toward the star because of strong overlap with the PSF wings from the central X-ray source. The region between the inner and the outer jet cannot be reliably mapped. 
     \textbf{9. }Blue, elongated structure to the SW of the star and blob in the NE: The jets in the [Ne\,{\sc ii}]~12.81$~\mu$m line observed by MIRI/MRS, discussed further in this paper.  \textbf{10. }Dashed gray arc overlaid on $^{12}$CO arc: Part of a projected circle at an inclination of 38\degr\ to guide the eye. The two arrows along the jets indicate the flow direction. All observations have been corrected for proper motion of DG Tau to a common date. However, the SPHERE image required manual adjustment by a fraction of an arcsecond.
    For an alternative presentation of this image, see Fig.~\ref{fig:overview_alt}. 
    }
    \label{fig:overview}
\end{figure}

\section{The target: The DG Tau star-disk-jet system}
\subsection{The star} 
DG Tau\footnote{not to be confused with DG Tau B, a jet-driving protostar only about 50\arcsec\ south of DG Tau, see \citet{eisloeffel1998}} is a strongly accreting, optically revealed young stellar object of spectral type K6Ve \citep{hessman1997} at a distance of 125~pc \citep{GaiaCollaboration2020}. Mass estimates range from  $\sim$0.6~$M_{\odot}$ \citep{garufi2024} to $\sim$($0.75-0.84)~M_{\odot}$ \citep{herczeg2017}; the accretion rate is $\sim$(3$\pm 2)\times 10^{-7}~M_{\odot}$~yr$^{-1}$ \citep{agraamboage2011} or, from new H\,{\sc i}-based estimates, $(1.1-1.7)\times 10^{-7}~M_{\odot}$~yr$^{-1}$ \citep{shridharan2026}. Judging from its ``flat spectrum'' in the IR (defined by a $\lambda F_{\lambda}$ slope between $-0.3$ and $+0.3$ in the 2.2--10~$\mu$m range, \citealt{greene1994}), the star+disk are at a transition stage from red{C}lass~I to Class~II; DG Tau is indeed still surrounded by substantial extended structures that may constitute leftover material from the protostellar envelope \citep{kitamura1996, beck2008, varga2017, guedel2018, garufi2022}, although the optical extinction of $A_{\rm V} \approx 1.6$~mag to the star is modest \citep{gullbring2000}.

\subsection{Winds and jets}
The sketch in Fig.~\ref{fig:overview} gives an overview of features in the vicinity of DG Tau observed in the NIR and MIR at millimeter wavelengths, and in X-rays; most data were taken from respective archives as described in the figure caption and in previous publications. DG Tau drives a well-studied bipolar jet \citep{eisloeffel1998, takami2004, agraamboage2011, white2014a, white2014b} identified as Herbig-Haro object HH~158 (to the SE) with potentially related HH objects HH~702 and HH~830 at parsec-scale distances \citep{mcgroarty2007}.
The outflow system within a few arcseconds of the star is subject to an asymmetry in which the  forward jet with blueshifted line emission (blue jet, henceforth) is highly collimated and long, showing a number of bright knots that appear to be ejected semi-regularly with a period of around 2.5~yrs \citep{agraamboage2011} or 4.8~yrs \citep{rodriguez2012}; instead, the receding jet with redshifted line emission (red jet, henceforth)  has been described as short, better characterized as a broad, extended and expanding bubble limited to about 1\farcs 5
from the star in previous observations \citep{agraamboage2011, agraamboage2014, white2014a, white2014b}. The mass-loss rate in the blue jet alone amounts to, as given variously in the literature,  $\sim$($0.51, 0.8, 1.4, 3.3, 1-5, 6.7)\times 10^{-8}~M_{\odot}$~yr$^{-1}$, which is only a lower limit to the total mass-loss rate (values from \citealt{white2014b}, \citealt{maurri2014}, \citealt{lavalley2000}, \citealt{agraamboage2011}, \citealt{lynch2013}, and \citealt{coffey2008}, respectively), but the ratio betweeen mass-loss rate and mass accretion rate ratio overall agrees with values of order 10\% among large samples of T Tauri stars \citep{nisini2019}. The jet has also been identified as a strong, soft X-ray source extending from 0\farcs 2 to $\sim$5\arcsec\ \citep{guedel2005, guedel2008, schneider2013b, Frank2014}; unexpectedly, there is also a hot, receding, and most likely collimated X-ray jet toward the NE that is more or less symmetric to the SW jet \citep{guedel2008}; other evidence of this extended and collimated receding jet has been lacking, to the best of our knowledge, but we recover new evidence of a collimated, high-velocity jet to the NE from our MIRI observations below.

The jets are surrounded by a system of slow, wide-angle winds launched from the protoplanetary disk, best visible in line emission of H$_2$ \citep{agraamboage2011, agraamboage2014, white2014a} and $^{12}$CO \citep{guedel2018}, appearing in the form of a conical outflow also recognized in scattered-light images in the NIR \citep{garufi2024}.

In the 
\textit{Spitzer} survey of CTTS in the [Ne\,{\sc ii}]~12.81$\mu$m line by \citet{guedel2010} DG Tau stands out as the strongest well characterized single CTTS source; indeed, the survey sample suggests that jets are the dominant origin of strong forbidden-line emission in CTTS. The production mechanism is, however, unclear. [Ne\,{\sc ii}] could be related to internal jet shocks but also to irradiation by X-rays from either the jet itself or from the more distant star (e.g., \citealt{liu2016}, who presented observations of the jet in a [Ne\,{\sc iii}] line at 3869~\AA, together with an interpretation of the ionization of jet gas by high-energy radiation). We address this issue below.

\subsection{The disk}
DG Tau's MIR continuum and line spectrum was observed extensively with \textit{Spitzer}. Repeated observations revealed strong variability in the 6--12~$\mu$m silicate feature on time scales as short as a few weeks \citep{bary2009}. The disk's mineralogy was found to be dominated by crystalline forsterite. 

The innermost disk of DG Tau was resolved with IR interferometric instruments. From VLTI/MIDI observations, \cite{varga2017} measured a half-light radius of $\sim$$1$~au at $10\ \mu$m, corresponding to 0\farcs 008 at DG Tau's distance. They also found that the innermost zone of the disk ($r< 1-3$~au) shows $8-13\ \mu$m radiation with a silicate absorption feature caused by amorphous grains, while the disk region outside exhibits silicate emission by crystalline grains. Moreover, the time variability of the silicate feature mentioned above \citep{bary2009} is found to be associated with the outer zone. 

\textit{Spitzer} observations showed an abundance of MIR water lines, although they were described as faint at longer wavelengths while OH lines were prominently detected \citep{carr2011, carr2014}. Estimates indicated temperatures of $>$900~K and an emitting area with a radius of $<$0.5~au. OH would be produced by photodissociation of H$_2$O in a strong far-ultraviolet radiation field (FUV, mostly Ly$~\alpha$; \citealt{carr2014, neufeld2024, tabone2024}). Water was also detected in Herschel observations at 557~GHz and 1113~GHz, suggesting the presence of a strong UV field from the star that heats the disk upper layers up to 600~K at distances between 10~au and 90~au; if originating from the disk, the estimated total water mass in the disk is the equivalent of 100--1000 Earth water oceans in vapor and 100 times more in ice form \citep{podio2013}.  

Molecular lines of formaldehyde (H$_2$CO), CS, and CN have been detected in the disk at millimeter wavelengths, but the former two exclusively in a ring between $\sim$40$-70$~au \citep{podio2019, podio2020, garufi2021}. These molecules may preferentially be formed there, but some of their emission in the inner disk could also be suppressed by continuum optical depth.

\begin{figure*}
\centering
\includegraphics[width=0.96\textwidth,angle=0]{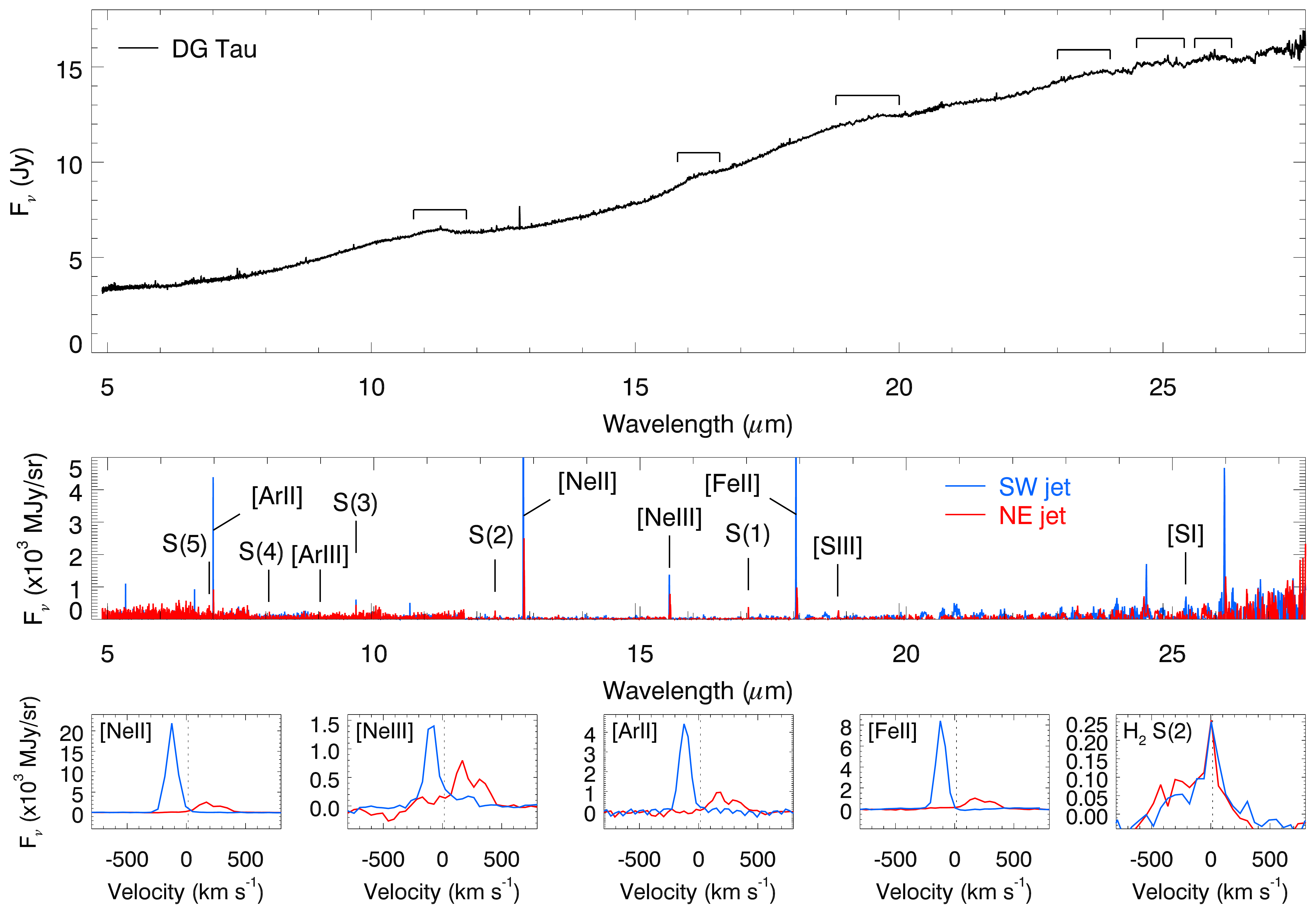}
\caption{MIRI/MRS spectra of DG Tau. \textit{Top panel:} Complete spectrum of the  PSF centered on DG Tau. Some atomic lines originate from the extended jet overlapping with the central PSF. Horizontal bars mark solid-state emission features, attributed to either amorphous or crystalline silicates, which are analysed in Sect.~\ref{sec:dust} to determine dust composition. \textit{Middle panel:} Spectra extracted from an aperture centered on the extended forward (in blue) and  receding jets (in red; the extraction locations are marked in Fig.~\ref{fig:RGBimages}a below). The [Ne\,{\sc ii}]~12.81~$\mu$m and
[Fe\,{\sc ii}]~17.94~$\mu$m lines are cut off to emphasize weaker lines. No significant continuum emission is recorded at those positions.  
\textit{Bottom panels:} Zoom-in on the jet and wind emission lines of [Ne\,{\sc ii}]~12.81~$\mu$m, [Ne\,{\sc iii}]~15.55~$\mu$m, [Ar\,{\sc ii}]~6.98~$\mu$m, [Fe\,{\sc ii}]~17.94~$\mu$m, and H$_2$~S(2)~12.28~$\mu$m. Velocities are in the barycentric system, and the systemic velocity of DG Tau at 15.4~km~s$^{-1}$  \citep{nguyen2012} is marked by the vertical dotted lines.
}\label{fig:fullspectrum}
\end{figure*}

\section{Observations and data reduction} \label{sec:obs}
\label{sec:observations}
We obtained a 5--28\,$\mu$m spectrum of DG~Tau on 2023 Feb. 20 with MIRI/MRS as part of the MINDS survey. 
Three grating settings were used to cover the whole wavelength interval. 
The recommended 4-point dither pattern optimized for point sources was used for our science target. Due to the brightness of DG Tau, the FASTR1 readout mode was utilized.

The MRS spectra were reduced utilizing a hybrid Python pipeline written within the MINDS collaboration (version 1.14, \citealt{mindspipeline}). 
The code combines routines from the standard JWST 
Science Data Processing Pipeline version 1.17.1 \citep{bushouse2025}
using Calibration Reference Data System (CRDS) context \emph{jwst$\_$1326.pmap}, and algorithms for background subtraction, bad pixel correction, and spectral extraction, taken from the Vortex Image Processing (VIP) package \citep{GomezGonzalez2017,Christiaens2023}. 
The results of this basic processing consist of 12 data cubes, one for each band. 
The field-of-view increases with wavelength from 
$\sim$3$\farcs5$ to $\sim$7$\farcs5$ across.
We experimented with subtracting the background emission in two different ways: taking a separate background measurement, or defining a sky annulus around the centroid position of the source. The two spectral extractions were very similar, deviations at the 1--2\% level occurring only longward of 20~$\mu$m. We adopted the second method, because it determines the background spectrum much closer to the source where diffuse emission is located \citep{kitamura1996, guedel2018}. 
The median background computed in the annulus was little affected by
the comparatively narrow bright jet parts that produce only ion lines (see Sect.~\ref{morphology}).
The background spectrum was subtracted from the data cubes at each wavelength. 

The 1D spectrum of the source was then extracted using aperture photometry with a wavelength-dependent aperture radius equal to twice the full width at half maximum (FWHM) of a point source. We corrected for aperture losses using the correction factors implemented in the JWST pipeline (Law et al. 2025), which increase with wavelength approximately from 1.11 to 1.16. In the next step, we applied an additional residual fringe correction on the spectrum. 
Then, as part of the pipeline processing, we stitched together the neighbouring bands using the overlapping parts of the band spectra, taking the shortest band as reference. Flux levels of different channels and sub-bands were consistent within 5\% in the overlapping wavelength ranges. The results were further checked via visual inspection, and the scaling factors were refined at the $\le$2\% level.
We fitted a continuum to the spectrum using the method of \citet{Temmink2024b}, and subtracted it to produce a line spectrum. The full point-source spectrum is shown in the top panel of Fig.~\ref{fig:fullspectrum}.

\section{Disk mineralogy}

\subsection{Analysis of dust in the disk}
\label{sec:dust}

The MIR spectral energy distribution (SED) of DG Tau (Fig.~\ref{fig:fullspectrum}, top panel) exhibits a general rising trend in $F_\nu$ characteristic of a flat spectrum source. A weak silicate emission feature at about 11~$\mu$m is clearly visible. Further solid state features at 16.3~$\mu$m and $\sim$19~$\mu$m, as well as several broad and shallow bumps in the 21--26~$\mu$m range seem also to be present. No PAH features are detected. 

We performed dust compositional fits to the line-subtracted and downsampled spectrum of DG Tau. The fits are done with the \texttt{DustComp}\footnote{Available at \url{https://github.com/jvarga42/DustComp}.} tool \citep{varga2026} based on the tools by \cite{Juhasz2009} and \cite{Kaeufer2024DuCKLinG}. Our simple model features components emitting pure blackbody radiation, and components accounting for the dust spectral features. The pure blackbody components are the central star, the hot inner rim, and the optically thick warm thermal emission from the disk midplane. The disk components are defined radially, with an inner radius ($r_\mathrm{in}$) representing the inner edge of the dusty disk. The midplane component has a power-law temperature dependence as a function of the radius, with a temperature of $T_\mathrm{midplane,in}$ at $r_\mathrm{in}$, and with a power-law index of $q_\mathrm{midplane}$. The rim component is defined as an annulus of constant surface brightness, with a radial width of $w_\mathrm{rim}$ and a temperature of $T_\mathrm{rim}$. The optically thin dust emission is modelled as a blackbody source function multiplied by the weighted sum of the dust opacity curves, where the weights can be expressed as mass fractions. The source function has a power-law radial temperature gradient, with a temperature of $T_\mathrm{dust,in}$ at $r_\mathrm{in}$, and with a power-law index of $q_\mathrm{dust}$. More details on the modelling approach can be found in \citet{varga2026}.
In such models the parameters determining the emitting area (the inner radius and temperature gradients) can be degenerate with those describing the surface density \citep[e.g.,][]{varga2026}.

For the set of dust opacities, Mg silicates and silica were considered based on earlier studies on dust mineralogy in protoplanetary disks \citep[e.g.,][]{Juhasz2010,jang2024}. 
Featureless species, such as Fe and C, were not part of this analysis.
We constructed two sets of opacity curves, which mainly differ in the applied grain shape model. With that, we aim to test the robustness of the compositional fits, as grain opacities strongly depend on the shape and size of the grains. Real circumstellar grains are likely aggregates and have highly varying shapes. From the various grain shape models developed over the years, we consider here the distribution of hollow spheres (DHS, \citealp{Min2005}), and the Gaussian random fields (GRF, \citealp{Min2007_GRF}), as both were found to provide a good match with observed dust spectral features \citep[e.g.,][]{jang2024}. \citet{varga2026} showed that different objects require different   opacity sets for optimum description.

\begin{figure*}[t!]
\centering
\includegraphics[width=0.49\hsize]{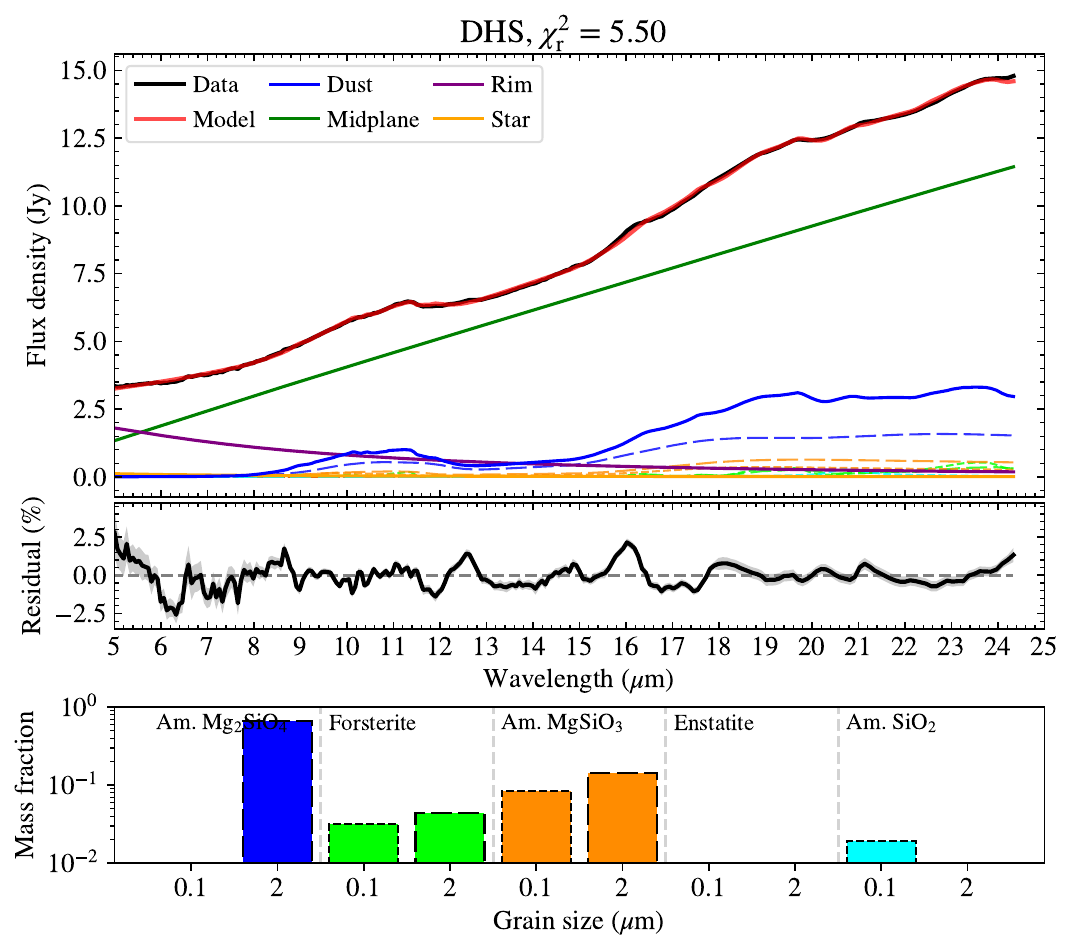} 
\includegraphics[width=0.49\hsize]{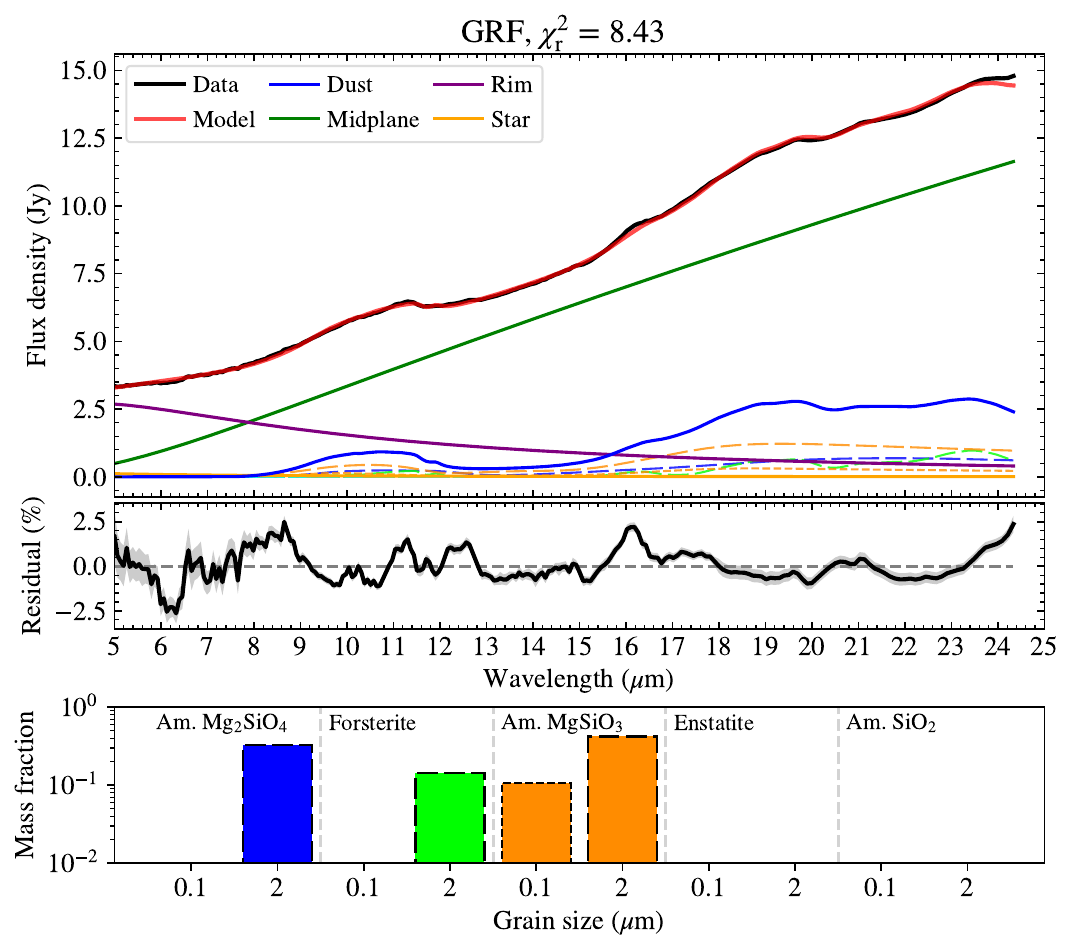} 
\caption{Silicate compositional fits to the line-subtracted JWST/MIRI spectrum of DG Tau, with different opacity sets. }\label{fig:sil_fits}
\end{figure*}
\begin{table}[]
    \centering
    \caption{Resulting dust mass fractions and disk parameters from the mineralogy fits.}
    \label{tab:sil_fits}
    \begin{tabular}{llrr}
    \hline
    \hline
         Dust & Grain & \multicolumn{2}{c}{mass fraction ($\%$)} \\
         species & size ($\mu$m)& \multicolumn{1}{c}{DHS} & \multicolumn{1}{c}{GRF}\\
         \hline
         am. Mg$_2$SiO$_4$ & 0.1 & 0.5 & $<1$\\ 
         am. Mg$_2$SiO$_4$ & 2 &  66.9 & 32.5\\
         cr. forsterite & 0.1 & 3.2 & $<1$\\
         cr. forsterite & 2 & 4.4 & 14.4\\
         am. MgSiO$_3$  & 0.1 & 8.5 & 10.8\\
         am. MgSiO$_3$  & 2 & 14.5 & 42.2 \\
         cr. enstatite & 0.1 & $<1$ & $<1$\\
         cr. enstatite & 2 & $<1$ & $<1$\\
         am. silica & 0.1 & 1.9 & $<1$\\
         am. silica & 2 & $<1$ & $<1$\\
         \hline
         $r_\mathrm{in}$ (au) & & $0.06_{-0.02}^{+0.01}$ & $0.23_{-0.01}^{+0.07}$\\
         $T_\mathrm{dust,in}$ (K)& &$709_{-204}^{+39}$ & $500_{-76}^{+29}$\\
         $q_\mathrm{dust}$ & &$-0.30_{-0.01}^{+0.01}$ & $-0.28_{-0.05}^{+0.04}$\\
         $T_\mathrm{midplane,in}$ (K) & & $986_{-32}^{+24}$ & $266_{-44}^{+24}$\\
         $q_\mathrm{midplane}$ & &$-0.49_{-0.02}^{+0.01}$ & $-0.49_{-0.03}^{+0.01}$\\
         $w_\mathrm{rim}$ (au) & &$0.04_{-0.01}^{+0.02}$ & $0.04_{-0.02}^{+0.01}$\\
         $T_\mathrm{rim}$ (K) & &$1530_{-202}^{+117}$& $1125_{-37}^{+233}$\\
         \hline
         $\chi^2_r$ & & 5.5 & 8.4\\
         \hline
    \end{tabular}
    \tablefoot{The estimated detection threshold is $\sim$$1\%$.}
\end{table}
The following dust species were included in the opacity sets: amorphous Mg$_2$SiO$_4$ (olivine-type, \citealp{Jager2003}), crystalline forsterite \citep{Suto2006}, amorphous MgSiO$_3$ (pyroxene-type, \citealp{Jager1998}), crystalline enstatite \citep{Dorschner1995}, and amorphous silica \citep{Kitamura2007}. In the GRF opacities, the choice of optical data was different for the following three minerals: amorphous Mg$_2$SiO$_4$ \citep{Henning1996}, crystalline forsterite \citep{Servoin1973}, and amorphous silica \citep{Henning1997}. 
For each species we chose two grain sizes: $0.1\ \mu$m, and $2.0\ \mu$m. These choices are motivated by earlier dust studies, such as \cite{vanBoekel2005survey} and \cite{Juhasz2010}.
The choice of discrete grain sizes is a simplification, as circumstellar grains have a continuous grain size distribution. With the selected grain sizes, we aim to represent both the submicron-sized and the few micron ($\mu$m) sized grain populations, which are the most significant contributors to the MIR emission. DHS opacities were calculated using \texttt{optool} \citep{Dominik2021}. 

The best-fit mass fractions of the dust species, along with the disk parameters, are presented in Table~\ref{tab:sil_fits}. The corresponding models are shown in Fig.~\ref{fig:sil_fits}. Both the DHS and the GRF opacities provide a good match with the data, as indicated by the small residuals, staying mostly within $2\%$. However, the reduced $\chi^2$ values are significantly larger than 1. We attribute this to the spectral complexity of the data. This complexity stems from more diverse dust mineralogy, and grain shape and size effects that our model cannot fully capture using a limited set of opacities and simplifying assumptions, such as a radially constant composition.
We tested whether having a third grain size in our grain size set would lead to a better fit. For that, we performed our dust analysis with the $\left[0.1, 1, 2 \right]$~$\mu$m and the $\left[0.1, 2, 5 \right]$~$\mu$m sets. With DHS opacities, we did not see any improvement over our adopted grain size set. With GRF opacities, we notice an improvement using the set which includes the $5\ \mu$m size ($\chi^2_\mathrm{r} = 5.8$). Still, our adopted grain size set, $\left[0.1, 2 \right]$~$\mu$m, with DHS opacities provides the overall best fit performance ($\chi^2_\mathrm{r} = 5.5$) with the smallest number of parameters.

We note that DG Tau is known for its time-variable silicate emission, where variable line-of-sight extinction can change the feature shape significantly \citep{bary2009,varga2017}. More details on that are provided in Sect.~\ref{sec:discussion_disk}. 

The resulting mass fractions from the DHS and GRF opacity models are broadly consistent. Our silicate compositional fits clearly detect forsterite grains (Table~\ref{tab:sil_fits} and Fig.~\ref{fig:sil_fits}). Crystalline enstatite is not detected at all, while silica may be present in a very small fraction.
Among the amorphous silicates, both Mg$_2$SiO$_4$ (olivine-type) and MgSiO$_3$ (pyroxene-type) are required by the model fits. However, the GRF and DHS models do not agree on the mass fractions of the $2\ \mu$m-sized amorphous Mg$_2$SiO$_4$ and MgSiO$_3$ components. We attribute this discrepancy to a degeneracy associated with grain size: as the grain size increases (particularly for sizes $\geq2\ \mu$m) the corresponding opacities become increasingly similar, making the individual components harder to distinguish. Nevertheless, the total mass fraction of the species in question can be robustly constrained by both models, with the combined mass fraction of the $2\ \mu$m-sized Mg-silicates being 81.4\% with the DHS and 74.7\% with the GRF opacities. 
The fraction of crystalline grains (forsterite) is 
$7.6\%$ in the DHS fit, and $14.4\%$ in the GRF fit. Further discussion of the grain properties and the fit residuals is provided in Sect.~\ref{sec:discussion_disk}.

Disk parameters resulting from 
the fits should be interpreted with caution. The relatively small wavelength range and the lack of spatially resolved data \citep[cf.][]{Varga2024} do not allow us to draw firm conclusions about the inner disk structure. We observed degeneracies between the temperature and geometric parameters, and uncertainties related to this issue are included in the formal error bars in Table~\ref{tab:sil_fits}.  Since these error bars are modest (typically within 30\%), the inner-disk temperatures and the power-law indices reflect reasonably well the overall ambient conditions in the warm inner part of the disk.

\subsection{Discussion: Disk mineralogy\label{sec:discussion_disk}}

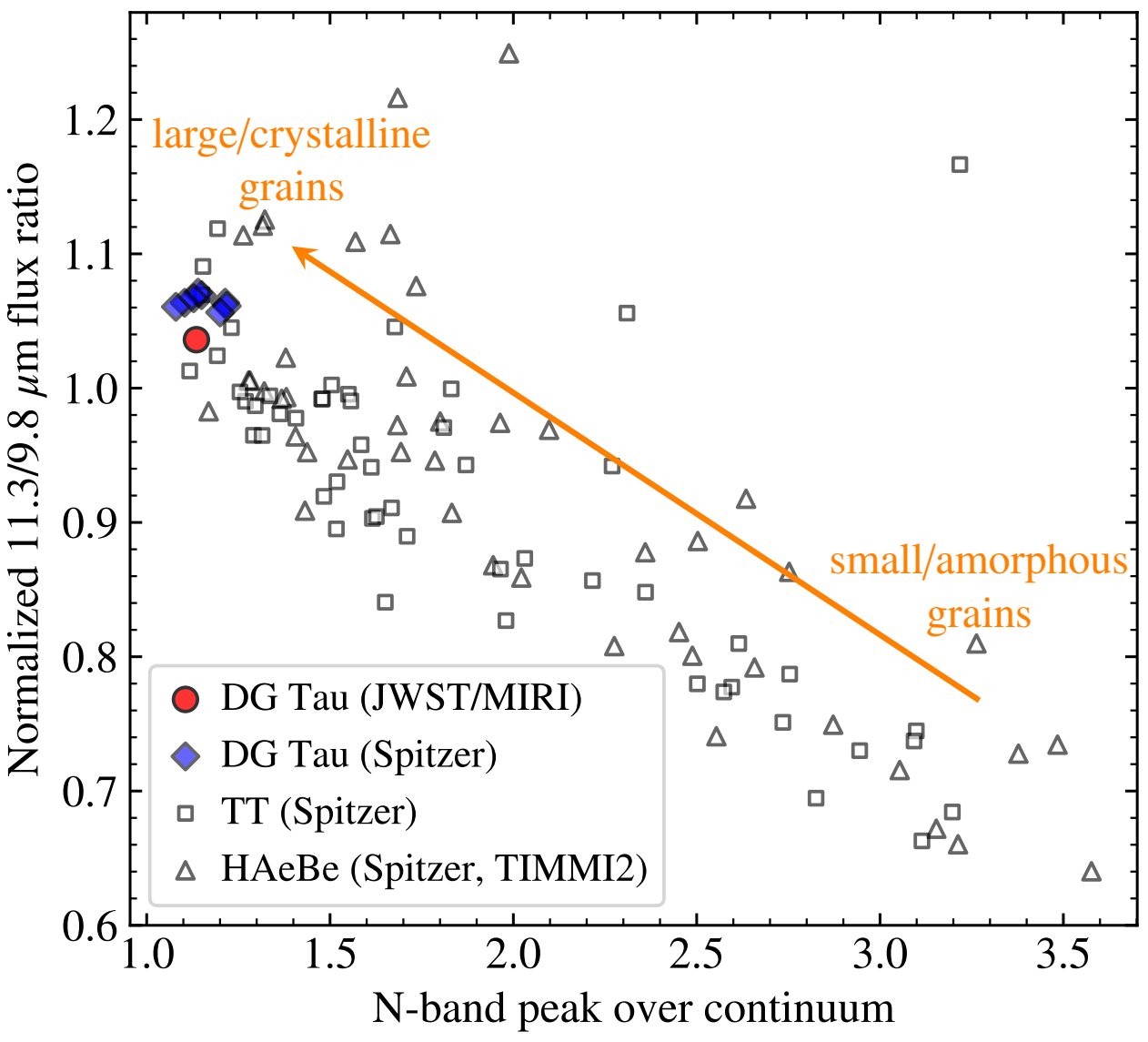
\begin{figure}
\centering
\vskip -0.1cm
\begin{tikzpicture}
\draw (0, 0)  node[inner sep=0] {\includegraphics[width=\columnwidth]{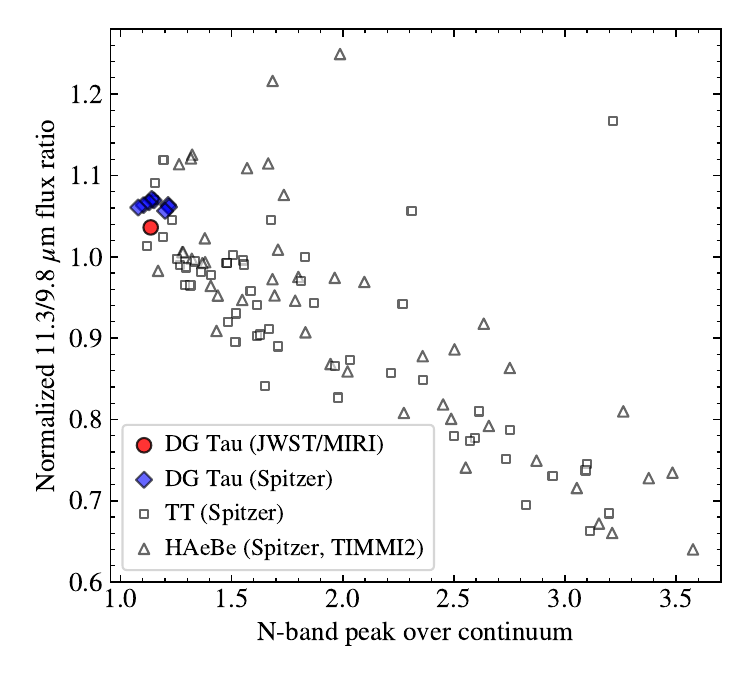}};
\draw [orange,very thick,{stealth}-](-2.0,2.0) -- (3.0,-1.3);
\node at (3.0,-0.5)  [align=center]{\footnotesize{\textcolor{orange}{small/amorphous}}\\ \footnotesize{\textcolor{orange}{grains}}};
\node at (-2.0,2.6)  [align=center]{\footnotesize{\textcolor{orange}{large/crystalline}}\\\footnotesize{\textcolor{orange}{grains}}}; 
\end{tikzpicture}
\caption{
Relation between the shape and the amplitude of the $N$-band silicate feature of DG Tau (red and blue points), compared to samples of CTTS and Herbig Ae/Be disks \citep{vanBoekel2005survey,Kessler-Silacci2006,Juhasz2010}. The orange arrow indicates the effect of grain growth or increasing crystallinity fraction \citep[see also][]{varga2026}.
}
\label{fig:shape_strength}
\end{figure}

To compare the silicate emission of DG Tau with that of other young disks, we derived two parameters from its $N$-band\footnote{The $N$-band is centered at $\sim$10.5~$\mu$m with a FWHM of $\sim$2.5~$\mu$m.} spectrum: one describing the feature strength and the other its shape \citep[][]{vanBoekel2003,przygodda2003}.\footnote{The feature strength and shape parameters have been calculated from the continuum-subtracted and normalized spectrum, according to the following formula: 
$F_{\nu,\mathrm{norm}} = 1 + (F_{\nu} -F_{\nu,\mathrm{cont}})/
\langle 
F_{\nu,\mathrm{cont}}\rangle$, 
where $F_{\nu,\mathrm{cont}}$ is a linear continuum under the 10~$\mu$m silicate feature, with anchor points at 7.85~$\mu$m and 13.15~$\mu$m. 
The feature strength is $\max\left(F_{\nu,\mathrm{norm}}\right)$, and the shape parameter is the 11.3~$\mu$m/9.8~$\mu$m flux ratio $F_{\nu,\mathrm{norm}}\left(11.3\,\mu\mathrm{m}\right) / F_{\nu,\mathrm{norm}}\left(9.8\,\mu\mathrm{m}\right) $. This is the same calculation as applied in \cite{varga2026}.}
The strength parameter, i.e., the $10~\mu$m flux peak over continuum, is defined such that a value of 1.0 corresponds to a featureless continuum, while values greater than 1 indicate silicate emission. 
We calculated these parameters not only for the new JWST spectrum of DG Tau, but also for earlier \textit{Spitzer}/IRS spectra of DG Tau, and plot the results in Fig.~\ref{fig:shape_strength}.
For comparison, we also included measurements for CTTS and Herbig Ae/Be disks from previous \textit{Spitzer} and TIMMI2 \citep{kaeufl2003} observations.

In Fig.~\ref{fig:shape_strength}, most points are distributed along a line from the lower right to the upper left, indicating an anti-correlation between the flatness of the silicate feature and its peak amplitude. Points in the lower right portion indicate a strong silicate feature with a peak at $9.8\ \mu$m, typical of ISM dust and characteristic of sub-$\mu$m-sized amorphous silicate grains. 
Increasing grain size suppresses the $9.8\ \mu$m peak and flattens the silicate feature, representing a shift to the upper left part of the diagram, as was demonstrated by, e.g., \citet{Kessler-Silacci2006}. A similar trend is produced by an increasing crystalline fraction. The broad distribution of points along this trend therefore indicates varying degrees of dust processing, resulting in larger grains, higher crystallinity, or a combination of both toward the upper left of the figure.
In the plot, DG Tau lies at the extreme left. 
This location suggests a high level of dust processing, in line with our mineralogy fits which predicted a mass fraction of $86$--$89\%$ for the larger ($2\ \mu$m-sized) grains. The presence of significantly processed dust in DG Tau, despite its young age, suggests that grain growth in the inner disk regions can proceed rapidly. We further note that the grain growth sequence does not correlate well with system age. For example, PDS 70, a transitional disk with an estimated age of $5\pm1$ Myr \citep{Keppler2018}, is located in the lower right region of the plot \citep{jang2024}, which is the locus of ISM-like, near-pristine grains. This implies a wide diversity of dust evolution time scales among planet forming disks, and suggests that secondary populations of small grains may be generated at later evolutionary stages (e.g., \citealt{oliveira2010, oliveira2011, Gaidos2024}).

Our dust mineralogy fits could reproduce the forsterite peaks both in the $N$-band and at longer wavelengths ($16.3\ \mu$m and $\sim$$20\ \mu$m). However, in the residuals, a peak at $16\ \mu$m still remains, indicating some additional emission from forsterite grains, not accounted for in our fits. There is also a residual peak at $12.5~\mu$m, which may correspond to annealed silica \citep{Fabian2000}. Our fits indicate that most of the forsterite crystals are in the form of larger grains, but their abundances could not be well constrained (we found $4\%$ in the fit with DHS opacities, and $14\%$ with the GRF opacities). In comparison, a systematic mineralogical study of a sample of CTTS disks by \cite{Olofsson2010} found an average crystalline fraction of $16-19\%$.  

\begin{figure}
\centering
\includegraphics[width=\columnwidth]{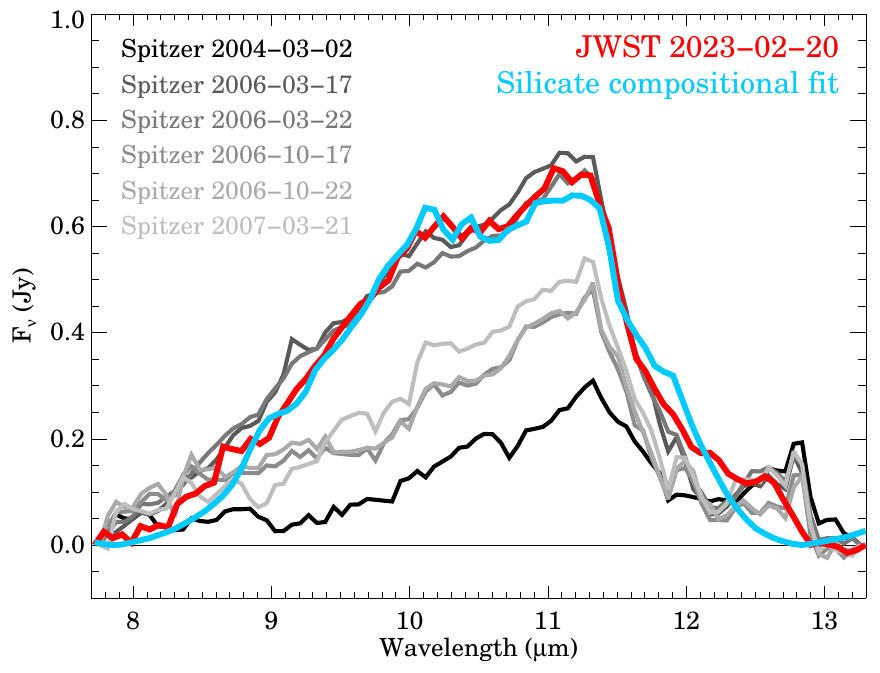}
\caption{$N$-band silicate feature of DG Tau observed with \textit{Spitzer} and JWST,  after subtracting a first-order polynomial continuum. The blue line shows the silicate compositional fit displayed in Fig.~\ref{fig:sil_fits} (left). Both the amplitude and the local peaks of the feature are well reproduced.
}\label{fig:silicate}
\end{figure}

In addition to the JWST/MIRI spectrum around the silicate feature at 8--13~$\mu$m, Fig.~\ref{fig:silicate} also displays several \textit{Spitzer}/IRS measurements obtained as part of a monitoring program \citep{bary2009}. The plot shows the well-known large-amplitude variability of the silicate feature in DG Tau that is independent of continuum flux \citep{bary2009}. With JWST, we observed the source at an epoch of one of the highest silicate features. Comparing the MIRI profile with the high amplitude \textit{Spitzer} spectra, MIRI outlines local excesses at 
10~$\mu$m and 12.3~$\mu$m.

The dust modelling implies a relatively shallow radial temperature gradient, with a power-law index of $-0.5$ for the midplane and $-0.3$ for the dust component. For comparison, in a standard flat accretion disk the power-law index is expected to be $-0.75$. The shallowness of the temperature profile in DG~Tau implies a relatively large area of warm MIR emission, which may suggest significant disk flaring. \citet{varga2017} presented a radiative transfer (RT) model for DG Tau fitted to $N$-band VLTI/MIDI and photometric data, where they assumed a flared disk geometry. We compared their model SED to our JWST/MIRI spectrum, and note that the spectral shape of the model is consistent with the new data. However, the flux levels are very different, as the MIRI spectrum has $\sim$$2\times$ as much flux as the RT model SED, and we observed a similar change in flux between the VLTI/MIDI single-dish spectra 
(i.e., total flux without interferometric combination, taken between 2011 and 2014) and the MIRI data. Based on these, the new MIRI data support the notion of the inner disk being flared, but it also highlights the spectacular MIR variability of DG Tau.

\citet{varga2026} presented a dust mineralogy survey of 26 T Tauri disks from the MINDS sample. DG Tau was not included in their sample. Comparing the dust spectral features of DG Tau with those of the other T Tauri disks in the MINDS sample, we find that DG Tau has one of the weakest normalized silicate emission features. Consequently, in the N-band feature shape versus strength plot (Fig.~\ref{fig:shape_strength}), DG Tau occupies the upper-left region, at the leftmost edge of the distribution of the MINDS T Tauri disks. Moreover, the 11.3~$\mu$m/9.8~$\mu$m flux ratio of its N-band feature is higher than that of any of the MINDS T Tauri disks. The closest object with a spectral shape very similar to that of DG Tau is DL Tau. However, our mineralogical fits suggest that its dust composition is not an outlier among the MINDS T Tauri disks, as its amorphous silicate and crystalline forsterite content place it not far from the sample average.

\section{Gas in the disk\label{sec:gas_in_disk}}

\subsection{Analysis of the line spectrum}\label{lineanalysis}

To examine the gas, we use the continuum-subtracted MIRI spectrum. The retrieval is performed with the Dust Continuum Kit with Line Emission from Gas \citep[\texttt{DuCKLinG},][]{Kaeufer2024DuCKLinG}. Even though \texttt{DuCKLinG} can produce the gas emission and dust continuum simultaneously, we use it only to fit the gas emission. This is because even though the continuum can be fitted with uncertainties of less than $\sim$2\% (similar to the results shown in Fig.~\ref{fig:sil_fits}), due to the brightness of DG\,Tau this translates to absolute residuals up to an order of $\sim$100\,mJy, which is similar  
to the typical molecular emission line strength ($\sim$200\,mJy).

The gas component of \texttt{DuCKLinG} is based on a grid of 0D local thermodynamic equilibrium (LTE) slab models \citep[introduced and explained by][]{Arabhavi2024}. This grid of models with different column densities and temperatures is used by \texttt{DuCKLinG} to interpolate and integrate, producing molecular fluxes from temperature and column density power laws. This is especially important for \ce{H2O} which emits under a wide range of conditions, justifying the need for more complex profiles than single temperature and column density \citep[e.g.,][]{Kaeufer2024DuCKLinG,Romero-Mirza2024,temmink2025}. The model fluxes are convolved to the JWST/MIRI spectral resolution and rebinned to the observed wavelengths following \citet{Kaeufer2024b}.

The retrievals are performed by nested sampling using \texttt{MultiNest} \citep{Feroz2008,Feroz2009,Feroz2019} via the Python interface \texttt{PyMultiNest} \citep{Buchner2014}. The details of this Bayesian sampling method are provided in the references above. For comparing the data to the observation a Gaussian likelihood function is defined \citep[equivalent to the one used by][]{Kaeufer2024DuCKLinG}. While most priors are dictated by the slab model grid's extent\footnote{We note that the upper temperature limit for \ce{CO} has been extended to $2500\,\rm K$ instead of $1500\,\rm K$ which 
is used for all other molecules.}, the slope of the temperature power law ($q_{\rm emis}$) ranges uniformly from $-1.5$ to $-0.1$ to ensure a radially decreasing temperature with slopes consistent with thermo-chemical models \citep[e.g.,][]{Fedele2016,Brittain2023}. The observational uncertainty ($\sigma$) is treated as a free parameter as well and has a uniform prior on a log scale ranging between $0.1\,\rm mJy$ and $100\,\rm mJy$. Fitting for the best value of $\sigma$ instead of using the observational uncertainty produced by the data reduction pipeline has been used in previous studies and leads to broader more realistic posteriors \cite[e.g.,][]{kaeufer2026}. With this setup, we analyze the gas composition and emitting conditions in the disk of DG\,Tau.

We analyse the gas species in the disk's JWST/MIRI spectrum in multiple steps. First, we investigate the presence of different species by fitting two wavelength windows ($4.9-8.5\,\rm \mu m$ and $13-18\,\rm \mu m$) where many molecular features have been detected in previous studies \cite[e.g.,][]{arulanantham2025,grant2025}. Thereafter, we extract representative emitting conditions for all relevant species.

\begin{figure*}[t!]
    \centering
    \vbox{
    \includegraphics[width=0.955\linewidth] 
    {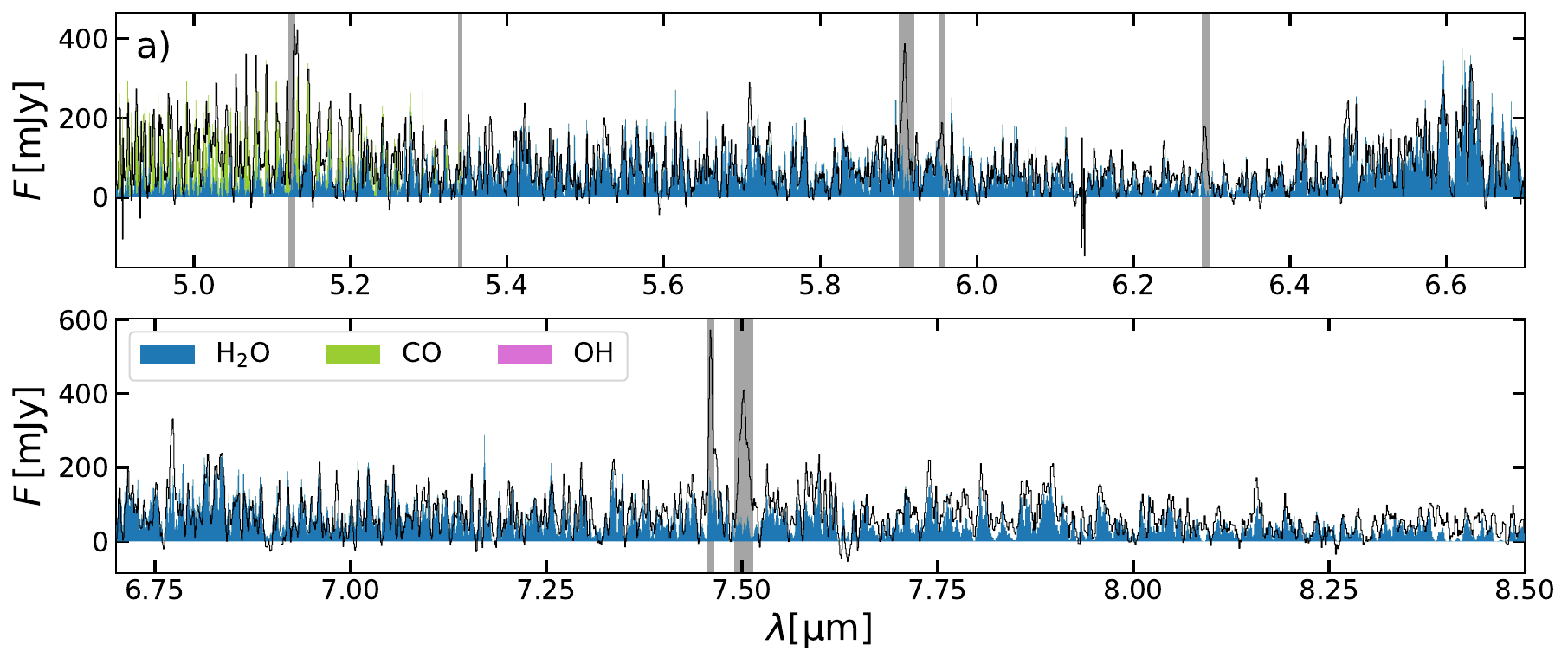}
   \includegraphics[width=0.955\linewidth]  
   {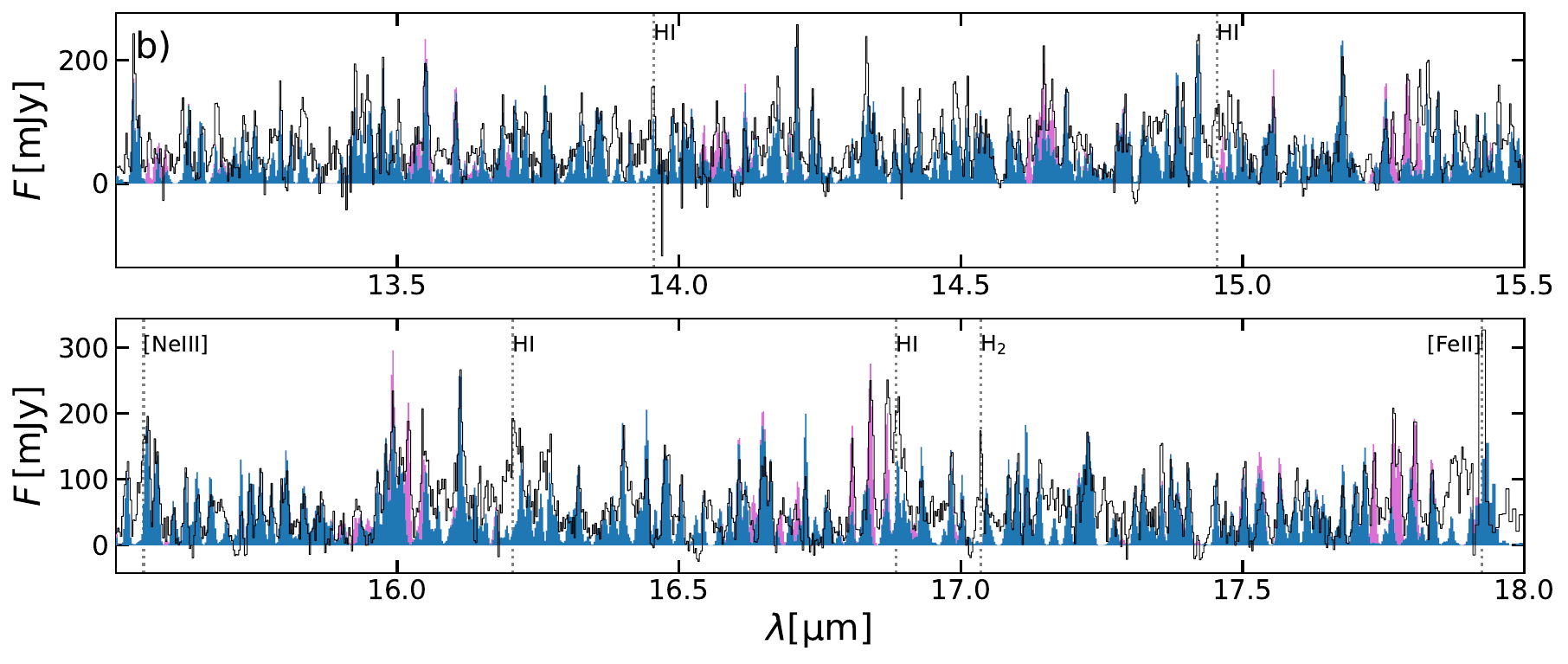}
   }
    \caption{Fit to the continuum-subtracted JWST/MIRI of the disk of DG\,Tau (black) from $4.9-8.5\,\rm \mu m$ (\textit{a}, top two panels) and $13-18\,\rm \mu m$ (\textit{b}, bottom two panels). The contributions of all used molecules to the median probability model are indicated by the colors explained in the legend. The regions excluded from the fit are indicated in grey. 
    }
    \label{fig:gas_short-middle_wave}
\end{figure*}

From a visual inspection, it became clear that the spectrum is dominated by \ce{H2O} lines, with additional emission from \ce{OH}, and the clearly visible 
lines of \ce{CO} at short wavelengths (Fig.~\ref{fig:gas_short-middle_wave}a).  
Therefore, we fitted the parts of the spectrum using these species before determining the evidence for the existence of more molecules. 

The first fit focused on the short wavelength range ($4.9-8.5\,\rm \mu m$) and can be seen in Fig.~\ref{fig:gas_short-middle_wave}a. This part of the spectrum exhibits a few strong atomic lines (e.g., H\,{\sc i}~(10--6) at $5.12\,\rm \mu m$, H\,{\sc i}~(9--6) at $5.9\,\rm \mu m$, or [Ar\,{\sc ii}] at $6.99\,\rm \mu m$), which were excluded from the fit to avoid biasing the retrieved molecular conditions. 
The fit quality is good, with a mean residual of $28.8\,\rm mJy$ and no clear molecular features left in the residual. Therefore, we are confident that this spectral region does not show traces of any other molecules above the residuals. 

After that, we focused on the line-rich spectral window between $13-18\,\rm \mu m$, where visually \ce{H2O} and \ce{OH} are identified, but also other molecules (e.g., \ce{HCN}, \ce{C2H2}, and \ce{CO2}) exhibit strong features for other objects. Testing if these latter molecules are detected in the spectrum of DG\,Tau as well, we set up multiple retrievals including and excluding them. To prohibit cases where one of the added molecules runs to unphysically large column densities or emitting areas, we limit the integrated fluxes in wavelength ranges that should not show any line emission to values smaller than $10^{-20}\,\rm W~m^{-2}$. These line-devoid wavelength regions are $13-14\,\rm \mu m$, $14.5-17\,\rm \mu m$, and $13-17\,\rm \mu m$, for \ce{CO2}, \ce{C2H2}, and \ce{HCN}, respectively.

The Bayes factor, which is a measure of the change of evidence, can tell if the differences in the residual justify the addition of extra molecules. 
The resulting Bayes factors are all very small ($\ln B(\mathrm{CO_2)}=0.75$, $\ln B(\mathrm{C_2H_2)}=0.91$, and $\ln B(\mathrm{HCN)}=0.46$), which means that the inclusion of neither of the three molecules significantly increases the fitting quality according to the definition by \cite{Trotta2008}. This conclusively shows that none of these commonly detected molecules (\ce{CO2}, \ce{C2H2}, and \ce{HCN}) can be seen in the JWST/MIRI spectrum of the disk of DG\,Tau. In particular, the non-detection of \ce{CO2} is surprising, given how commonly it is detected in other sources \citep[e.g. detected in $72\,\%$ of JWST/MIRI spectra of CTTS by][]{arulanantham2025}. A possible explanation might come from the temperature dependence of \ce{CO2} and \ce{H2O} formation. As shown by \cite{vanDishoeck2023}, both water and \ce{CO2} are formed using \ce{OH}. At high temperatures \ce{H2O} formation is favored, which, given the efficiency of water self-shielding at high column densities, can lead to a water-dominated state. 

The retrieval with \ce{H2O}- and \ce{OH}-only is displayed in Fig.~\ref{fig:gas_short-middle_wave}b. The mean residual ($29.7\,\rm mJy$) is comparable to the residual at short wavelengths. We did not mask any wavelength regions, to allow for the detection of molecular features overlapping with these lines. Therefore, the largest residuals are due to atomic lines, e.g. around $16.2\,\rm \mu m$ (H\,{\sc i}~(10-8)). However, we do not analyse the emission conditions from this fit specifically. To summarize, only molecules of \ce{CO}, \ce{H2O}, and \ce{OH} are detected in the disk around DG\,Tau; strong H$_2$ lines from partly extended regions are related to winds and outflows (see Sect.~\ref{h2extended}).

With the knowledge of the existence of different molecules, we can analyze their emission conditions. We refrain from analyzing the conditions of \ce{OH} in detail due to the lack of non-LTE effects in our models. 

\begin{table}
    \caption{Posterior parameter values and uncertainties for \ce{CO} and \ce{H2O}.}
    \centering
    \vspace{-2mm}
    \begin{tabular}{l|l}
\hline \hline & \\[-1.9ex] 
Parameter & Posterior \\ \hline  
 & \\[-1.9ex] 
 $\rm CO$:$r_{\rm eff}\,\rm [au]$ & $0.0619^{+0.0027}_{-0.0023}$\\ 
 & \\[-1.9ex] 
$\rm CO$:$T_{0.85}\,\rm [K]$ & $1270^{+80}_{-80}$\\ 
 & \\[-1.9ex] 
$\rm CO$:$T_{0.15}\,\rm [K]$ & $2040^{+120}_{-180}$\\ 
 & \\[-1.9ex] 
$\rm CO$:$N_{0.85}\,\rm [cm^{-2}]$ & $1.7^{+0.8}_{-0.5}(+20)$\\ 
 & \\[-1.9ex] 
$\rm CO$:$N_{0.15}\,\rm [cm^{-2}]$ & $1.1^{+0.6}_{-0.4}(+20)$\\ 
& \\[-1.9ex] 
  $\rm CO$:$q_{\rm emis}$ & $-1.30^{+0.27}_{-0.14}$\\ 
 & \\[-1.9ex] 
$\rm H_2O$:$r_{\rm eff}\,\rm [au]$ & $0.225^{+0.029}_{-0.023}$\\ 
 & \\[-1.9ex] 
$\rm H_2O$:$T_{0.85}\,\rm [K]$ & $670^{+50}_{-40}$\\ 
 & \\[-1.9ex] 
$\rm H_2O$:$T_{0.15}\,\rm [K]$ & $1240^{+80}_{-130}$\\ 
 & \\[-1.9ex] 
$\rm H_2O$:$N_{0.85}\,\rm [cm^{-2}]$ & $6.3^{+8}_{-3.1}(+18)$\\ 
 & \\[-1.9ex] 
$\rm H_2O$:$N_{0.15}\,\rm [cm^{-2}]$ & $7^{+6}_{-5}(+19)$\\  & \\[-1.9ex] 
$\rm H_2O$:$q_{\rm emis}$ & 
$-0.61^{+0.16}_{-0.13}$\\[+0.3ex]\hline
\end{tabular}
    \label{tab:gas_posterior}
\tablefoot{\ce{CO}: Extracted from the fit displayed in Fig.~\ref{fig:gas_short-middle_wave}a. \ce{H2O}: Retrieved from fitting the unblended water lines seen in Fig.~\ref{fig:gas_water-fit}. The notation $a(+b)$ means $a\times10^b$.}
\end{table}

\begin{figure}
    \centering
    \includegraphics[width=1.0\linewidth]{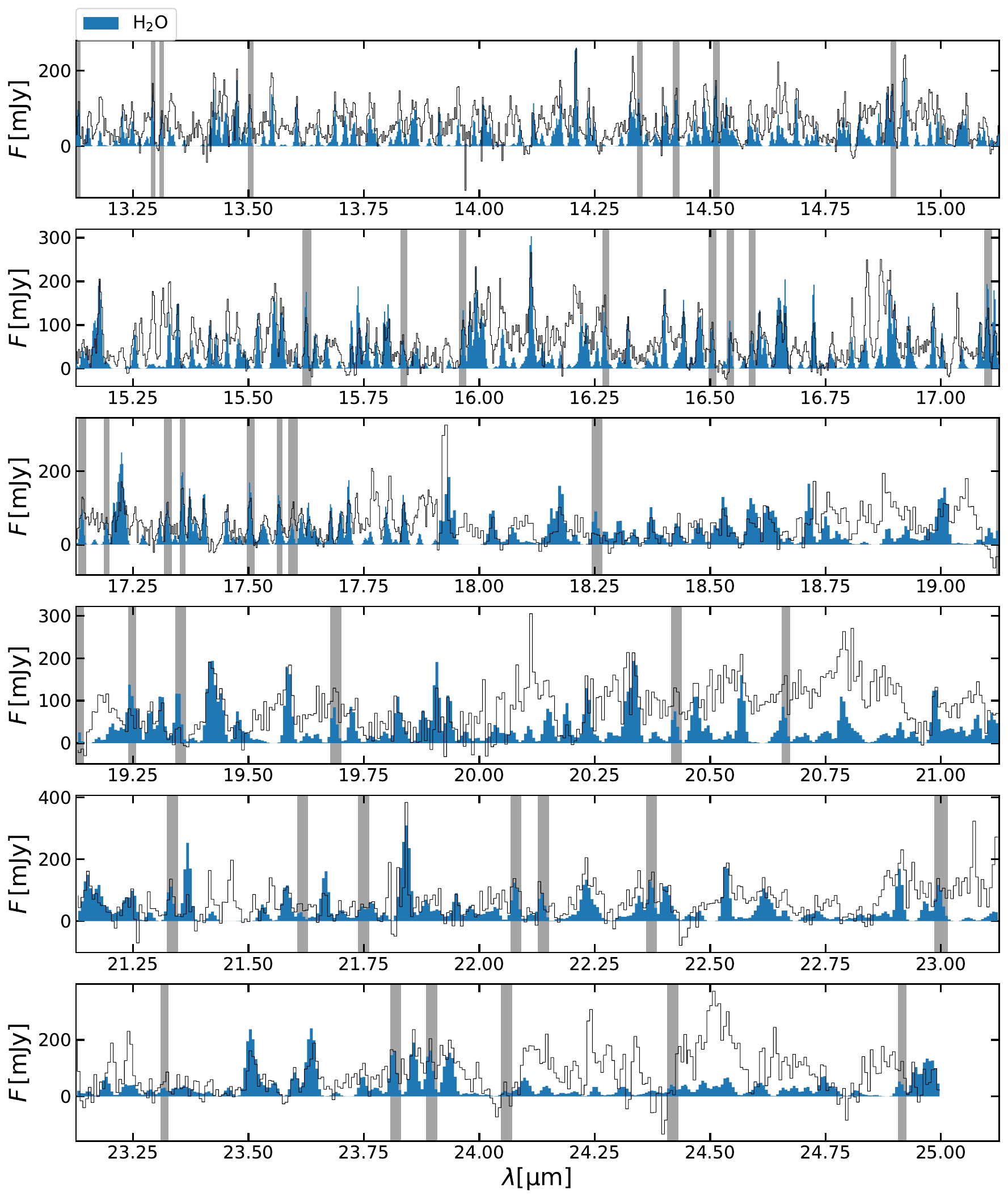}
    \caption{Fit to the unblended water lines in the continuum-subtracted MIRI of the disk of DG\,Tau (black) from $13-25\,\rm \mu m$. The contributions of \ce{H2O} to the median probability model are indicated in blue. The fitting regions are indicated in grey.}
    \label{fig:gas_water-fit}
\end{figure}

While we used the emission conditions extracted from the fit presented in Fig.~\ref{fig:gas_short-middle_wave}a for \ce{CO}, we ran a new, additional fit for water only using unblended rotational water lines. This was because the ro-vibrational water lines fitted in Fig.~\ref{fig:gas_short-middle_wave}a were not likely to be in LTE, which would complicate the interpretation with LTE models \citep{banzatti2023}. 
Additionally, fitting the rotational spectrum from $13-18\,\rm\mu m$ would not capture the coldest water lines and skew the extracted conditions due to the model's tendency to use the water component to minimize the residual at line-free noisy wavelength regions. All these different water fits can result in different extracted conditions, with a fit to the unblended lines resulting in the most representative picture for water \citep{kaeufer2026}. Therefore, we fit the unblended water lines between $13-25\,\rm\mu m$. The line windows were extracted from the \texttt{interactive Spectral-Line Analysis Tool} (\texttt{iSLAT}; \citealt{Jellison2024}), based on the method presented in \cite{banzatti2025}. In brief, iSLAT is an interactive interface used to explore and analyze molecular spectra. Due to the broad wavelength region that was fitted, we adjusted the likelihood function by assigning equal weight per wavelength, instead of per spectral point \citep[following][]{kaeufer2026}. The resulting fit is shown in Fig.~\ref{fig:gas_water-fit}. Although the overall spectrum shows many features that cannot be explained by water (e.g., as noise, possible artifacts of the continuum subtraction, and \ce{OH} lines), the unblended water lines are well fitted (mean residual of $29.0\,\rm mJy$).

\begin{figure}
    \centering
    \includegraphics[width=1.0\linewidth]{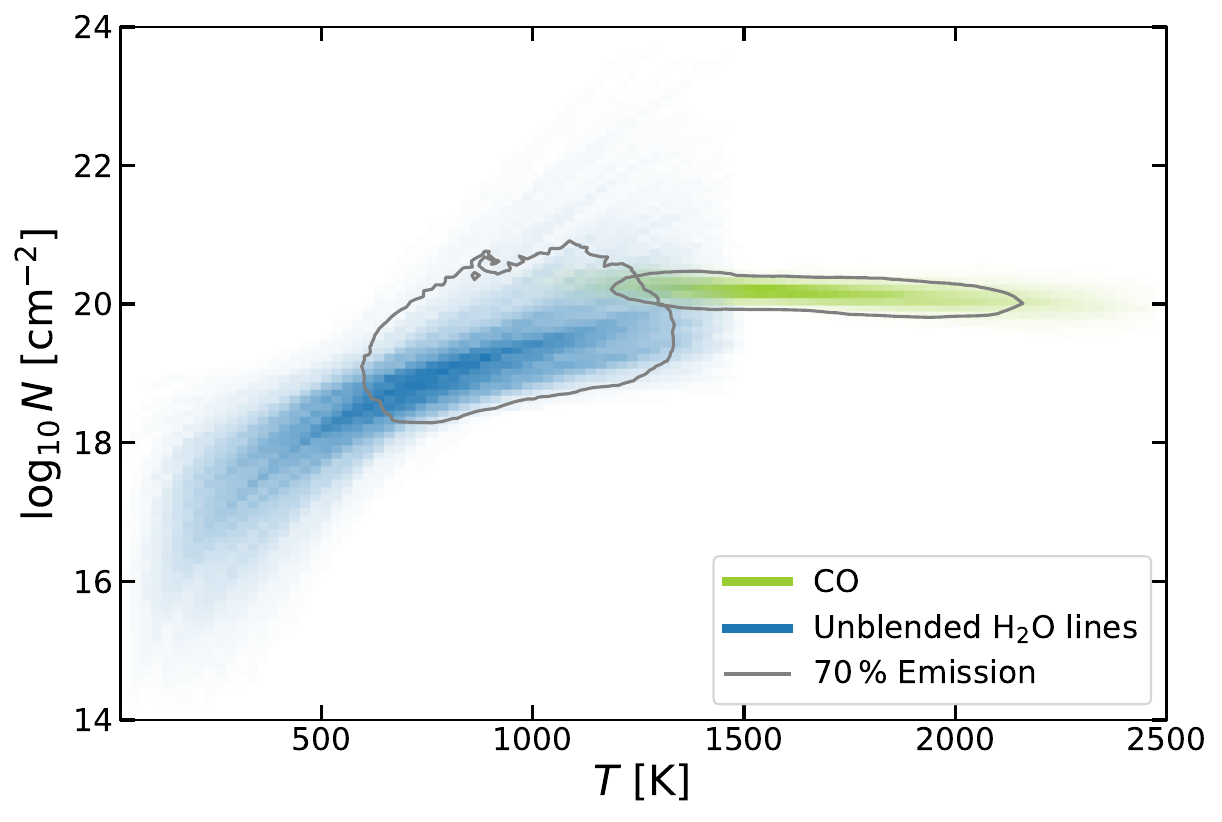}  
    \caption{Molecular column densities and temperatures derived from the continuum-subtracted MIRI spectrum. The colors show the emissions for the molecules listed in the legend, while the contours denote the region with the highest contribution to the respective molecular fluxes. The \ce{CO} conditions are derived from the fit presented in Fig.~\ref{fig:gas_short-middle_wave}a; the water conditions are extracted from a fit to the unblended water lines (Fig.~\ref{fig:gas_water-fit}).}
    \label{fig:gas_conditions}
\end{figure}
The extracted molecular conditions of \ce{H2O} and \ce{CO} are displayed in Fig.~\ref{fig:gas_conditions}. The colored areas depict the posterior distribution of column densities and temperatures for both molecules. The contours show the regions where most of the flux is originating \citep[details in, e.g.,][]{Kaeufer2024b}. The derived values for these quantities, relating to the values where $15\,\%$ and $85\,\%$ of the radial cumulative flux originate, are listed in Table~\ref{tab:gas_posterior}. $q_{\mathrm{emis}}$ is the radial temperature power law index, $r_{\mathrm{eff}}$ the radius of a disk with the same emitting area as the used radial profile, and $T$ and $N$, denoting the temperature and column density at the aforementioned positions, respectively.
The emission of \ce{CO} originates at near-constant column densities of $\sim10^{20}\,\rm cm^{-2}$ across a large temperature range (between $\sim 1000\,\rm K$ and $\sim2200\,\rm K$), with most emission arising between $1270^{+80}_{-80}\,\rm K$ and $2040^{+120}_{-180}\,\rm K$. The observable \ce{CO} lines in JWST/MIRI might bias the fit towards higher temperature \citep[e.g.,][]{Herczeg2011}. Therefore, we refrain from a detailed analysis and only suspect that \ce{CO} emits in the inner disk region, close to the star, where temperatures are the hottest.

Most of the \ce{H2O} emission arises from column densities of $\sim$10$^{19}\,\rm cm^{-2}$ and at temperatures above $670^{+50}_{-40}\,\rm K$. However, there is additional emission (with less resulting flux) from lower temperatures (i.e., few hundreds of Kelvins) and lower column densities {($\sim$10$^{16}-10^{19}\,\rm cm^{-2}$)}. Overall, both the temperature distribution and column densities are towards the upper end of typical values for Class II disks, which we further discuss in Sect.~\ref{gas}.

\subsection{Water line diagnostics}\label{gas}

As shown in Sect.~\ref{lineanalysis}
water seems to emit from high temperatures and column densities in the disk of DG\,Tau. We aim to test this model independently by calculating the line diagnostics introduced by \cite{banzatti2025}. For doing so, we extract the line strength of a few unblended water lines using \texttt{iSLAT}. 

\begin{figure}
    \centering
    \includegraphics[width=1.0\linewidth]{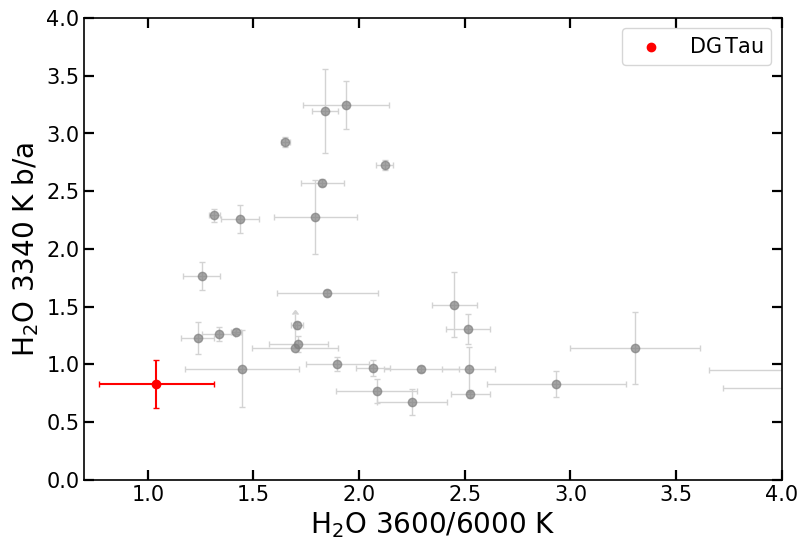} 
    
    \includegraphics[width=1.0\linewidth]{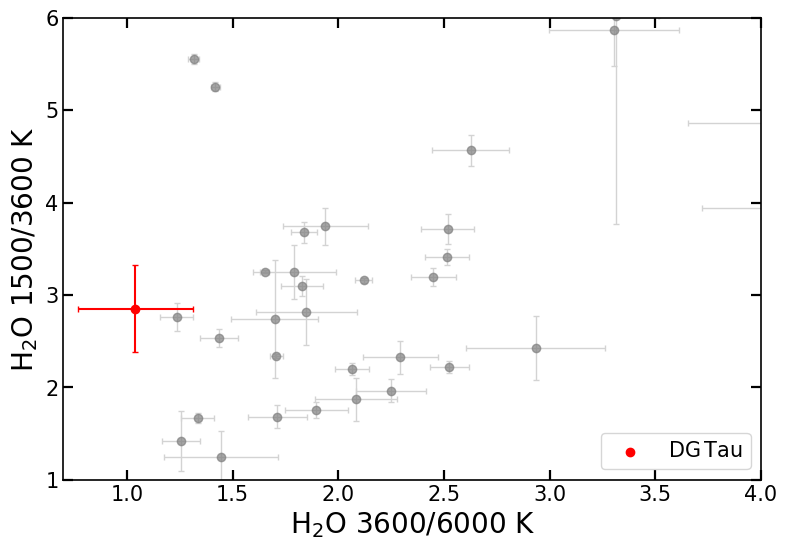}
    \caption{Water line diagnostics according to \citet{banzatti2025}. The temperatures denote the upper-level energies of the examined lines. The horizontal axes show the ratio of water lines tracing warm water emission ($\sim$400\,K) compared to hot water ($\sim$850\,K). The vertical axis of the upper panel displays the ratio of fluxes between two lines with the same upper-level energy (3340\,K) but very different Einstein-A coefficients ($25.29\,\rm s^{-1}$ for line b and $0.49\,\rm s^{-1}$ for line a) tracing the column densities of $\sim$400\,K water. The y-axis of the lower panel shows the ratio between water lines tracing cold emission ($\sim$200\,K) and warm emission. The grey dots are JWST/MIRI observations taken from \cite{temmink2025} and \cite{banzatti2025} with the latter values updated using the JDISCS reduction pipeline v9 as published by \cite{Mallaney2026}; the red dot with error bars indicates DG\,Tau.}
    \label{fig:water-line-diagnostics}
\end{figure}

The resulting line ratios are shown in Fig.~\ref{fig:water-line-diagnostics}. The vertical axis of the upper panel shows the ratio of two water lines with the same upper level energy but different Einstein A coefficients (b denoting the higher and a the lower Einstein A line). In the optically thin case, line b is stronger. With increasing column density and therefore opacity, this ratio gets lower. The value of $0.83$ for DG Tau indicates relatively high column densities, but nothing unseen in the sample of other JWST/MIRI observations. The horizontal axes measure the flux ratio of warm to hot water.  Here, DG\,Tau's ratio of $1.04$ is lower than any ratio seen in the previous observations with JWST. It indicates an overabundance of hot water or a relative lack of warm (and cold) water.
The ratio of warm to cold water (vertical axis of the lower panel) has a value of $2.85$, common for disks that do not show a cold water excess. To conclude, both the results from the Bayesian analysis and the model-independent line diagnostics indicate that DG\,Tau's disk exhibit high column densities and extremely high water temperatures (compared to the water emitting conditions in other sources). This might be an effect of the high accretion rate of DG\,Tau, which is shown to correlate with the water line strength, with a steeper relation for lines with higher upper level energy \citep{banzatti2025}. This is likely due to heating of the inner disk.

\subsection{Interpretation}\label{sec:gas-interpretation}

The DG Tau inner disk is, at the sensitivity level reached with MIRI/MRS, devoid of any molecular emission other than from  H$_2$, OH, 
H$_2$O, and CO. An abundance of water lines in the MIR is not unusual for CTTS disks (e.g., \citealt{romeromirza2024, temmink2024a, arulanantham2025, banzatti2025, gasman2025, temmink2025}) and seems to be related to the stellar mass or luminosity, and also to the evolutionary stage and the accretion process including dust and pebbles (e.g., \citealt{temmink2025, Arabhavi2025b, grant2025}); in contrast to a disk such as that of DG Tau, disks around very low mass stars are often carbon-rich with many lines from organic molecules (e.g., \citealt{tabone2023, Arabhavi2024}) although even those disks show water lines \citep{Arabhavi2025a}. Young disks indeed generally show very few molecular gas lines in the inner disks; see \citet{vanGelder2024} and \citet{vanDishoeck2025}.

We speculate that the absence of other detected molecules in MIRI/MRS is related to continuum optical depth but also to an underabundance of such molecules in the inner disk, compared to the outer disk; destruction of some molecules in the strong FUV and X-ray radiation fields in the inner part of the disk may also occur, while organic molecules have been detected in the outer disk \citep{guilloteau2013, podio2019, podio2020}.

\section{The winds and jets}

\subsection{Kinematic and spatial morphology of the jets}\label{morphology}

\begin{figure*}
\vskip -0.cm
\hbox{\hskip 0cm
\vbox{
\includegraphics[width=0.30\textwidth]{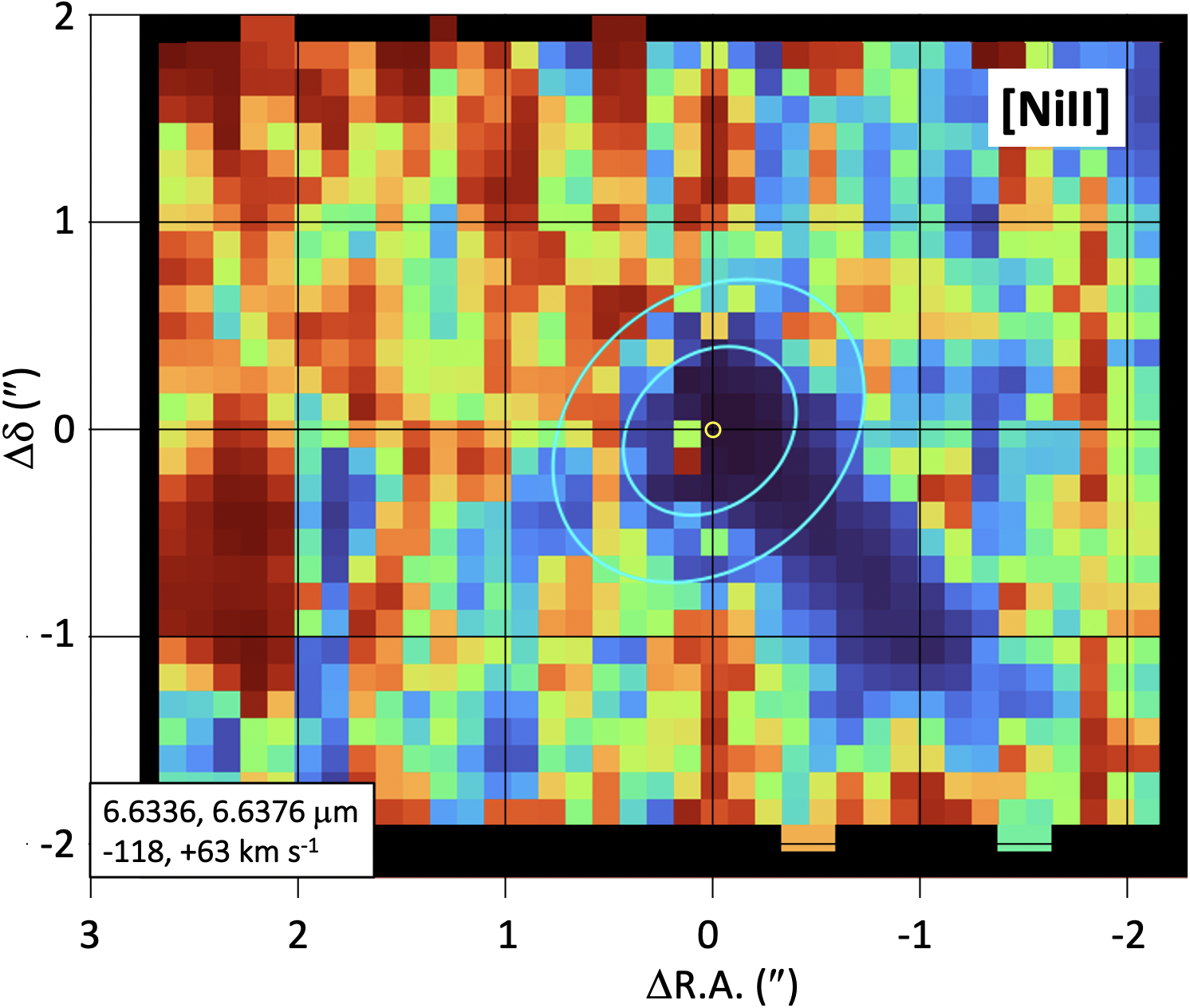}\\
\vskip 0.2cm   
\includegraphics[width=0.30\textwidth]{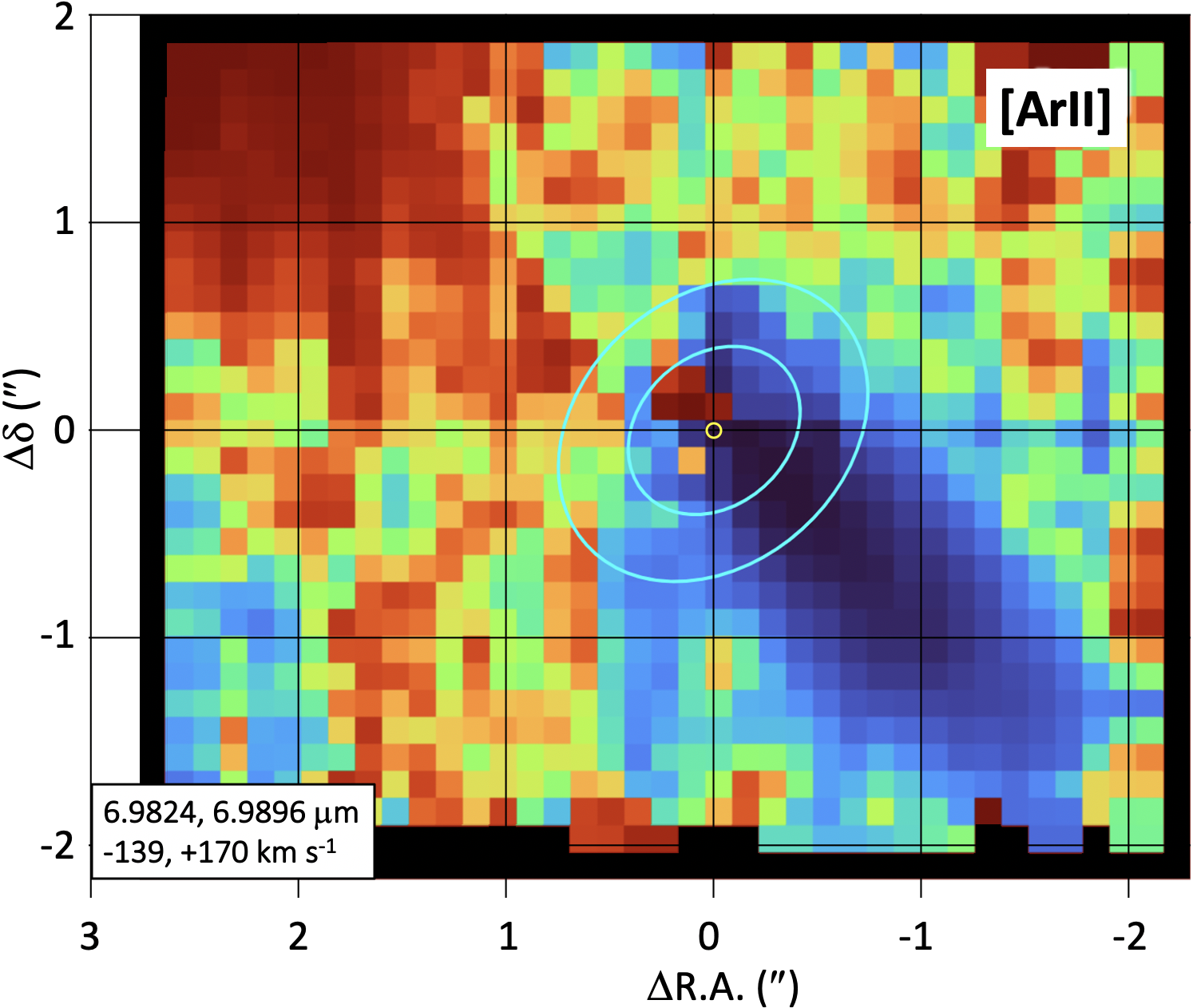}
}
\hskip -12.5cm
\vbox{\includegraphics[width=0.30\textwidth]{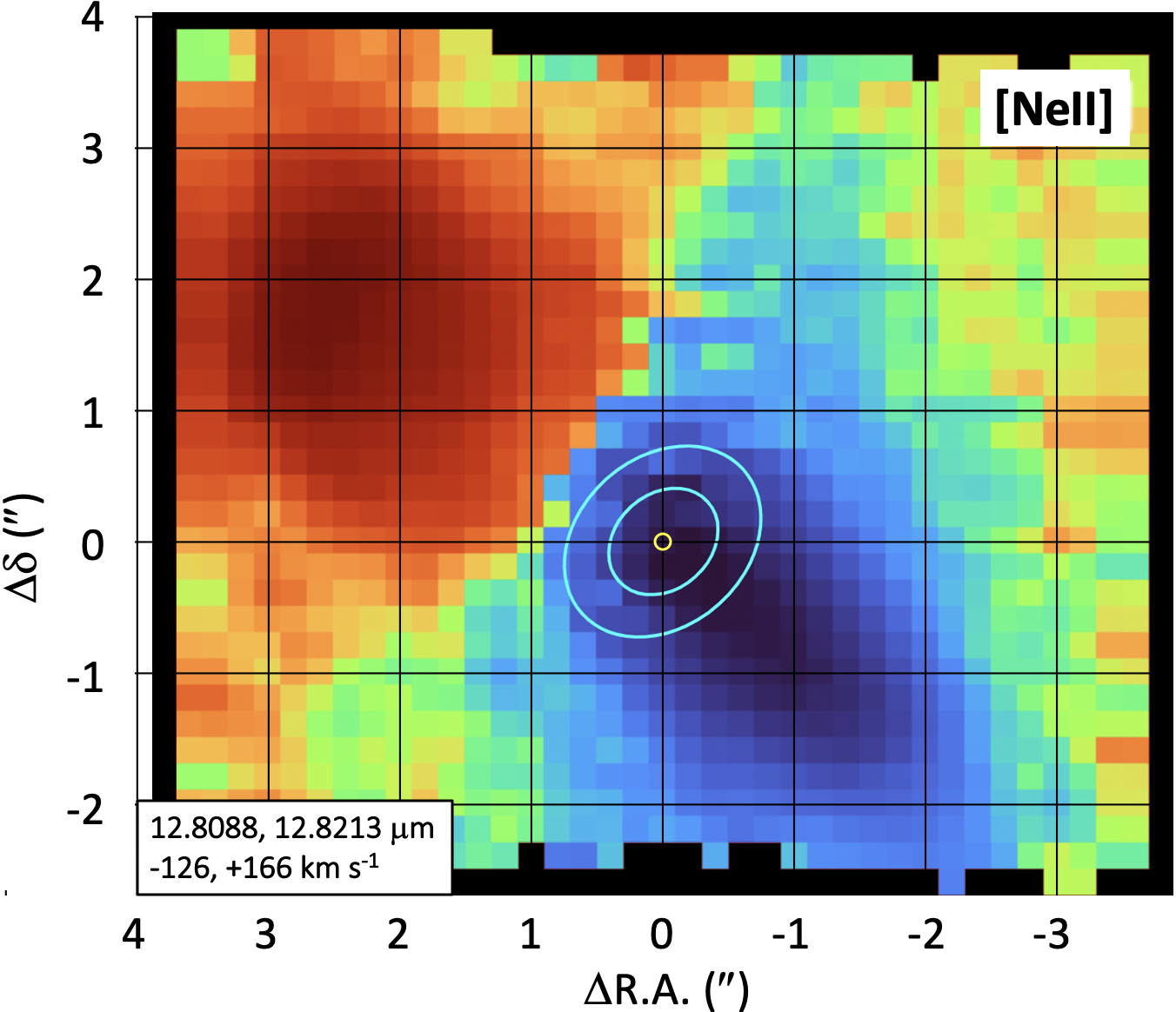}\\ 
\vskip -0.05cm   
\includegraphics[width=0.30\textwidth]{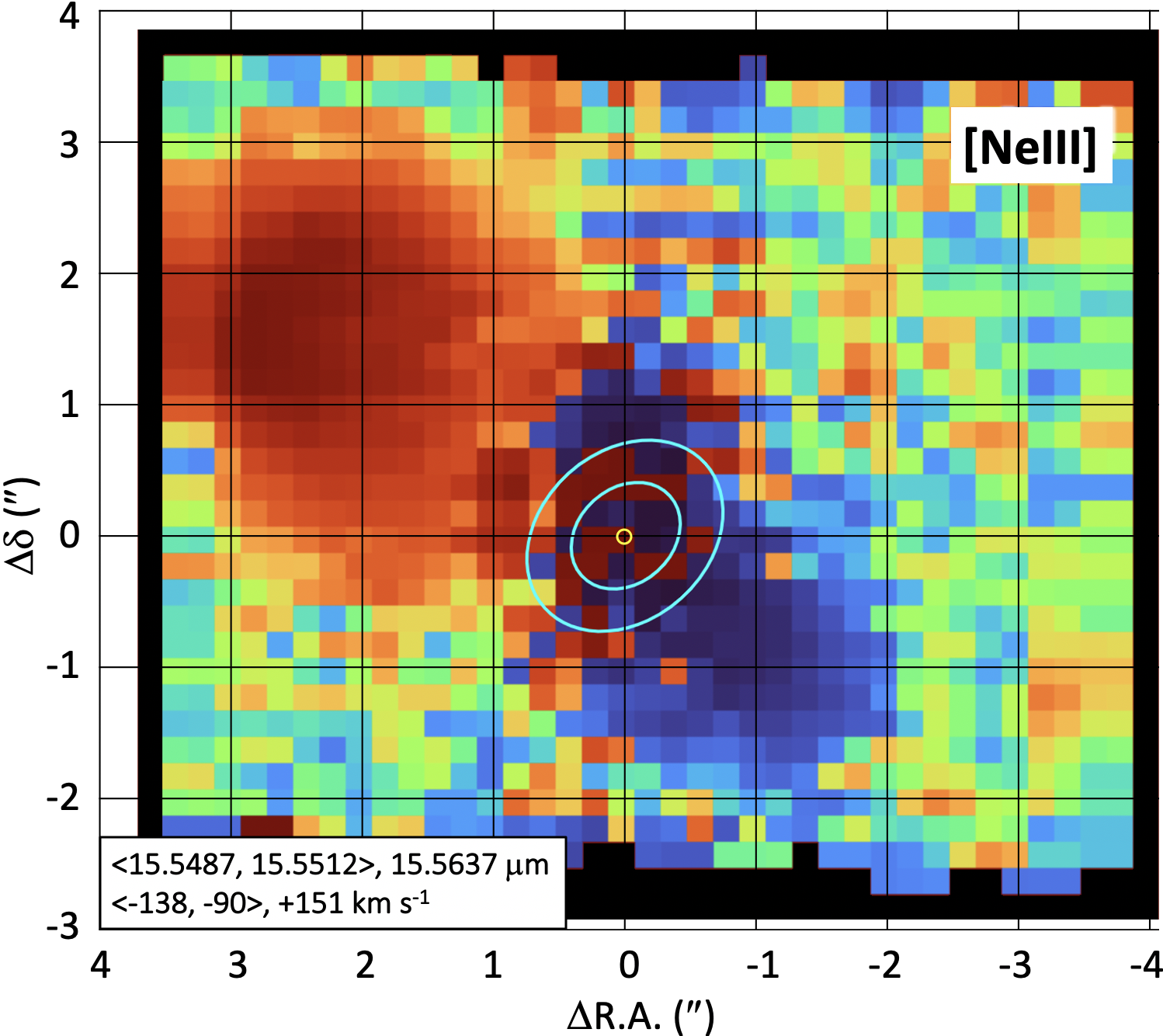}
}
\hskip -12.5cm
\vbox{
\includegraphics[width=0.30\textwidth]{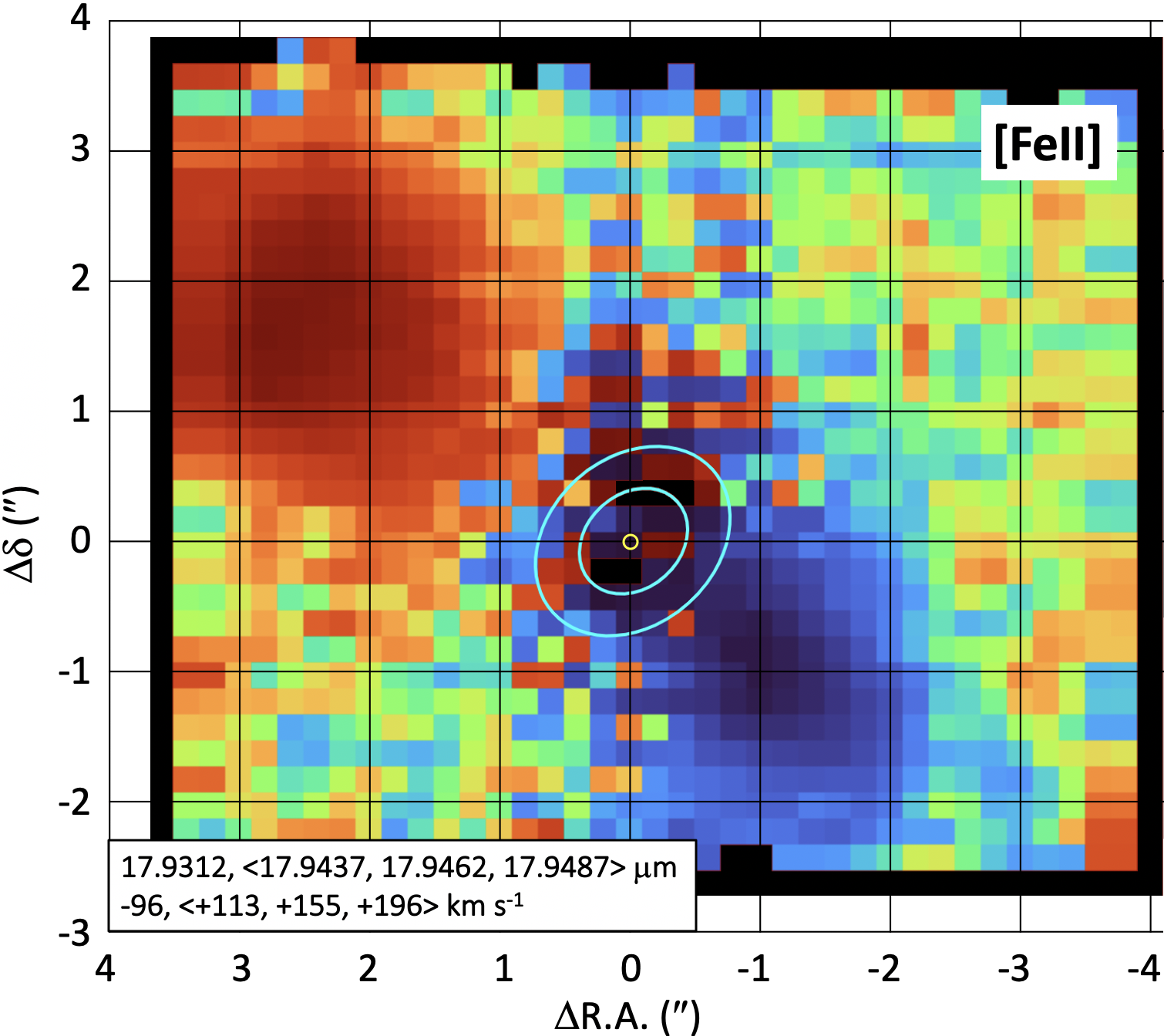}\\
\vskip 0.2cm  
\includegraphics[width=0.30\textwidth]{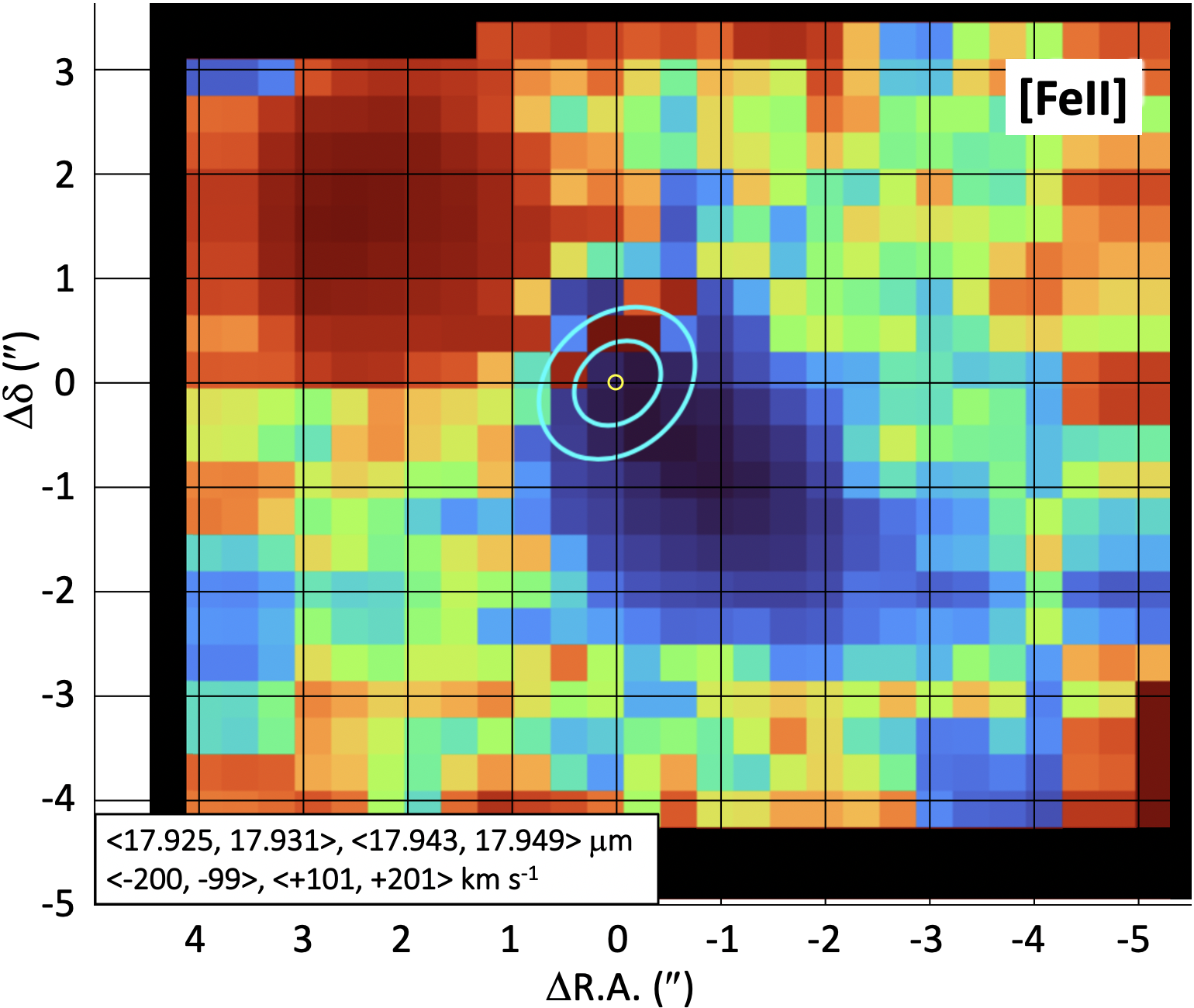}
}
\hskip -12.3cm
\vbox{
\includegraphics[angle=90, width=0.2cm, height=9.6cm]{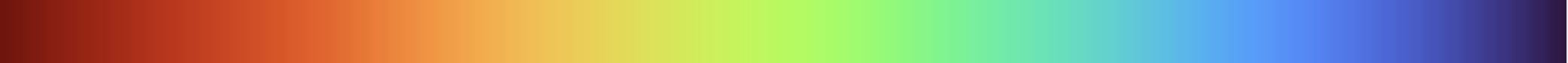}
\vskip 0.5cm  
}
}
\caption{Intensity images  
taken in fine-structure lines of [Ni\,{\sc ii}], [Ar\,{\sc ii}], [Ne\,{\sc ii}], [Ne\,{\sc iii}], and [Fe\,{\sc ii}].
The images were constructed by subtraction of the brightest spectral bins of the red jet from the brightest spectral bins of the blue jet, thus removing the local continuum emission. Schematic contours are overplotted for the gas CO disk (outer ellipse) and the continuum dust disk (inner ellipse), derived from ALMA observations \citep{guedel2018}. The star's and disk's center coordinates are, for 2023-02-20 (the star's position is marked by the small yellow circle at (0, 0)):
RA(2000.0) = 04$^{\rm h}$ 27$^{\rm m}$ 04\fs 701 although a slight pointing offset of 0\farcs 08 places the centroid at RA(2000.0) = 04$^{\rm h}$ 27$^{\rm m}$ 04\fs 695; and $\delta$(2000.0) = 
26\degr\ 06\arcmin\ 15\farcs 58; the pixel size is 0\farcs 20 except for the [Ar~{\sc ii}] and the [Ni~{\sc ii}] images (left column) that use  0\farcs 13, and the [Fe~{\sc ii}] image (bottom right) from channel 4-short that uses 
0\farcs 35. Red color indicates redshifted features, and blue color indicates blueshifted features. The insets in the lower left corner list the wavelengths of the blueshifted and the redshifted bins in the data cube in $\mu$m, and the corresponding radial velocities with respect to DG Tau in km~s$^{-1}$, positive values indicating redshift. If multiple spectral channels were averaged, their values are listed within brackets $<...>$, and the same holds for the corresponding radial velocities.
(Produced partly with JS9 software using color table \textsc{turbo} and scale = histeq; the color bar on the right defines the sequence of colors from red to blue; the scale uses the frequency of each data value in the image to distribute colors; it does not scale regularly and identically in each image but optimizes for visibility of low intensity levels). 
}
\label{fig:bluered}
\end{figure*}
\begin{figure*}[t!]
\hbox{\hskip 0cm
\vbox{
\includegraphics[width=0.36\textwidth]{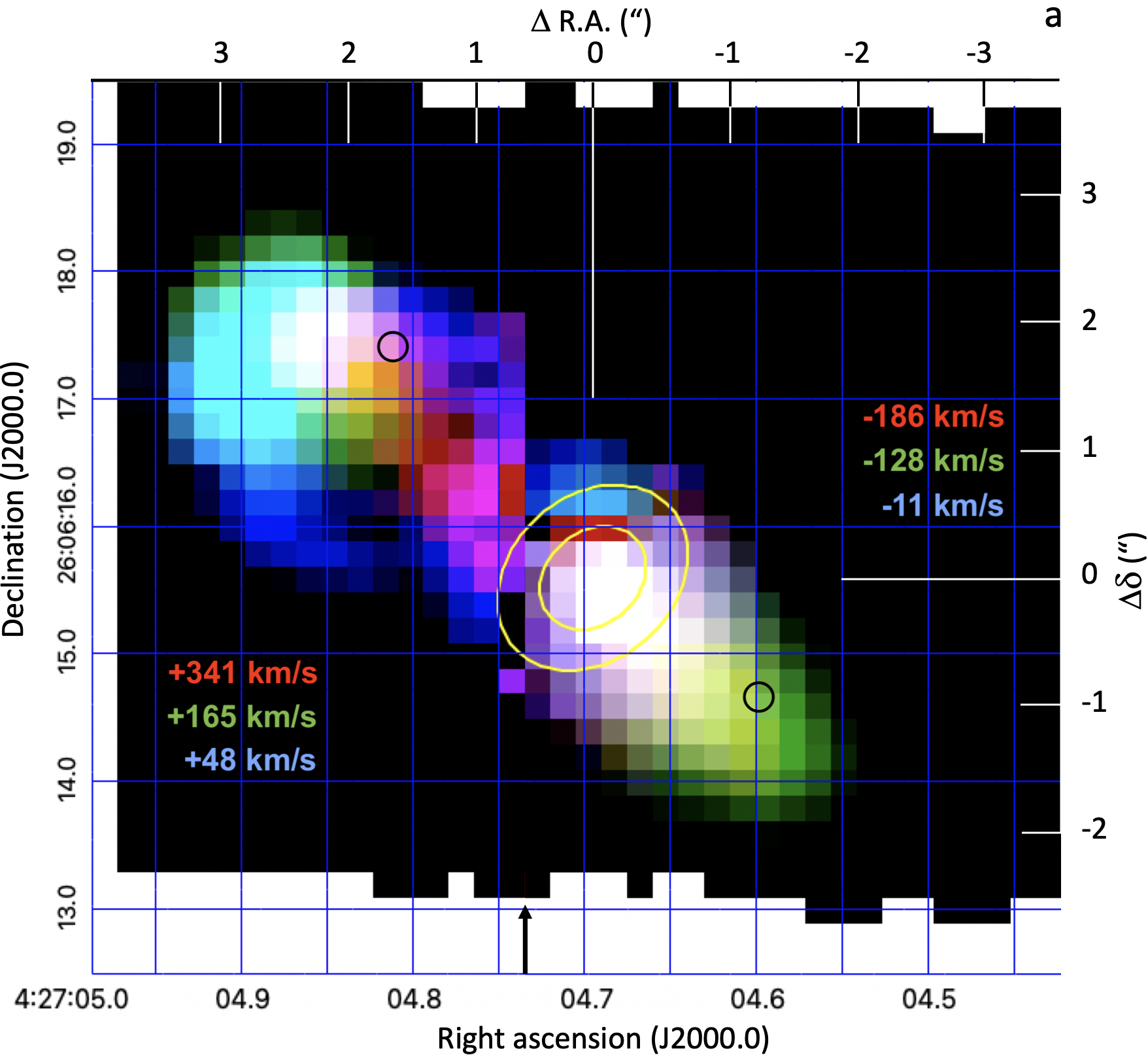}\\
\vskip -0.1cm
\includegraphics[width=0.36\textwidth]{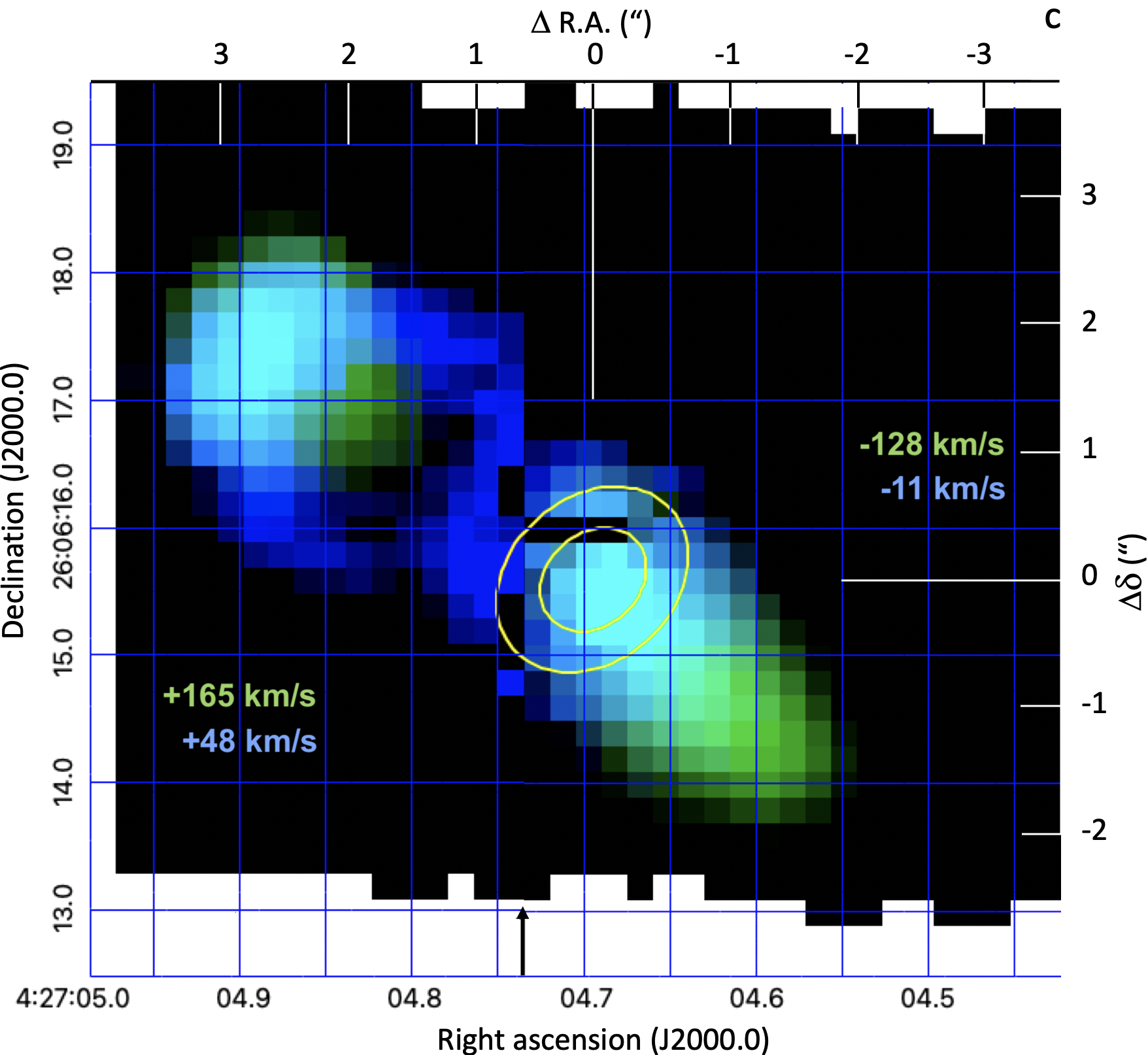}
}
\hskip -11.7cm\vbox{
\includegraphics[width=0.36\textwidth]{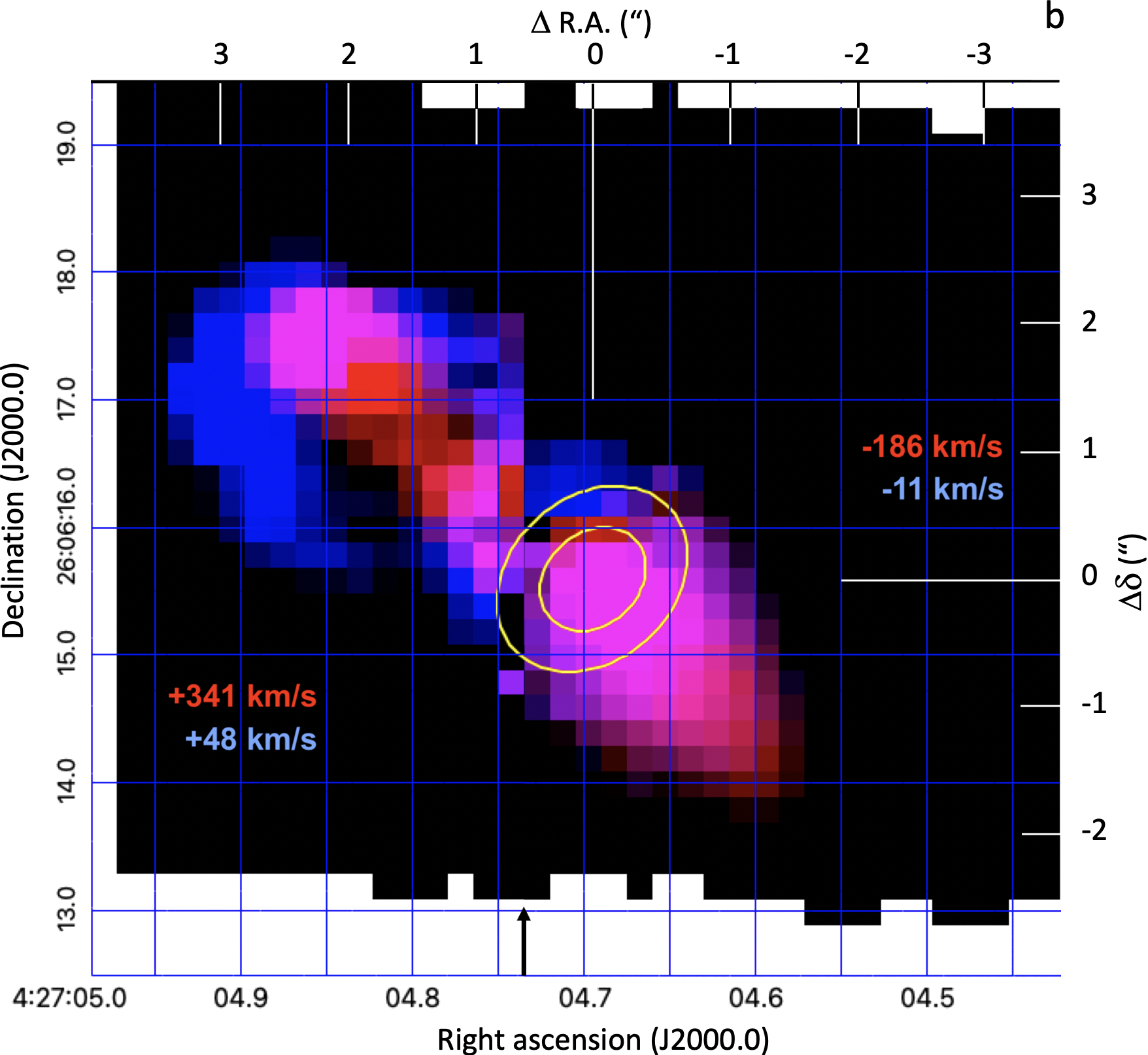}\\
\vskip -0.1cm
\includegraphics[width=0.36\textwidth]{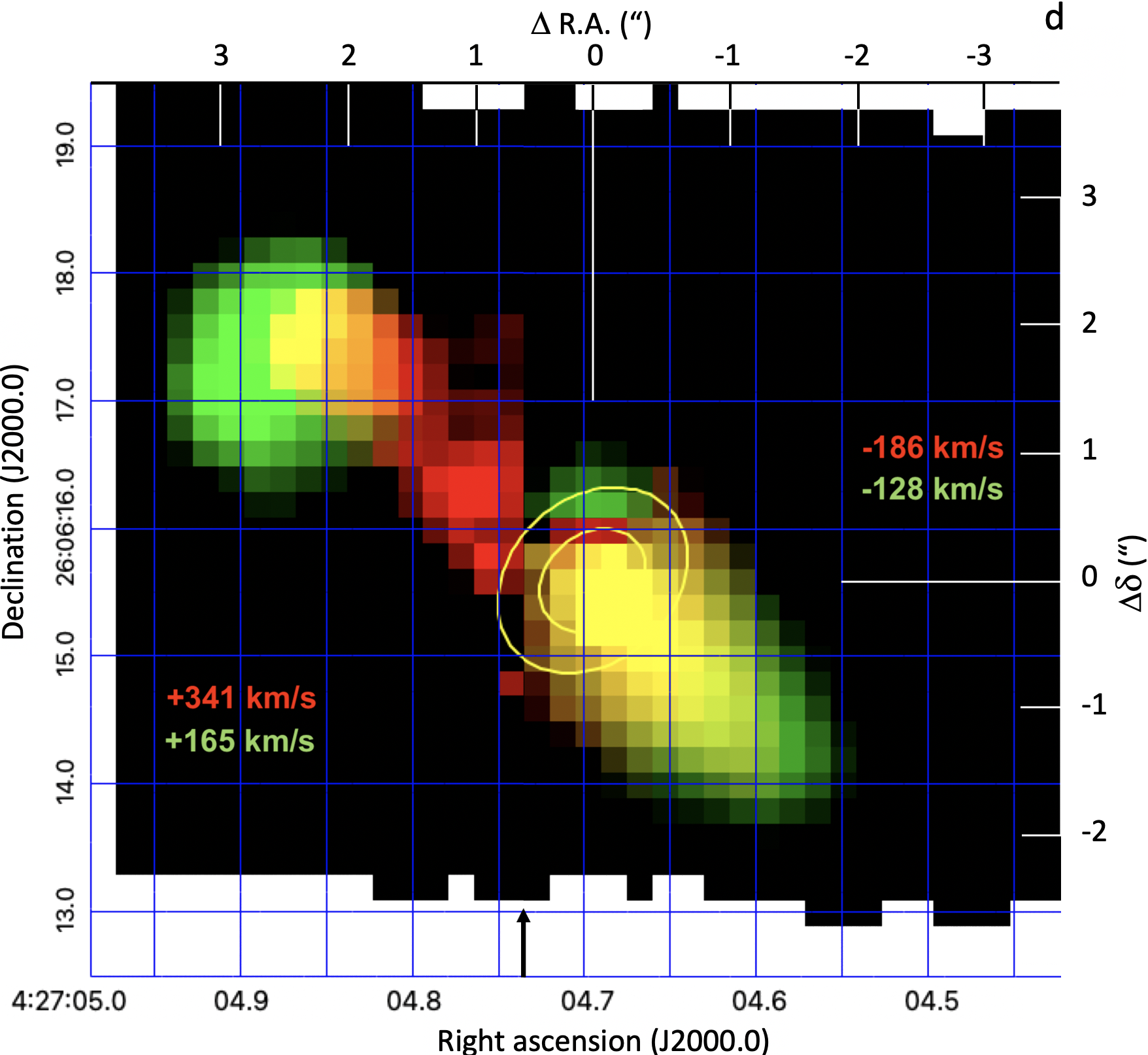}
}
\hskip -11.5cm
\vbox{  
\begin{minipage}{0.25\textwidth}
\vbox{\vskip -11.3cm
\caption{\label{fig:RGBimages} Different line-of-sight velocity components of the [Ne\,{\sc ii}] jet in RGB colors: R (red) is fastest, G (green) is intermediate, and B (blue) is slowest relative to the star; to the east of RA = 4$^{\rm h}$~27$^{\rm m}$~04\fs 73 (marked by an arrow) velocities are positive, to the west velocities are negative. \textit{a:} RGB. -- \textit{b:} RB. -- \textit{c:} GB. -- \textit{d:} RG. --
Radial velocities of the six components relative to the star, indicated in the plots, on the left side for the NE=receding jet, on the right side for the SW=approaching jet: SW jet  --186~km~s$^{-1}$ (R), --128~km~s$^{-1}$ (G), and --11~km~s$^{-1}$ (B); NE jet 
+341~km~s$^{-1}$ (R), +165~km~s$^{-1}$ (G), and +48~km~s$^{-1}$ (B).  The two black circles in panel (a) show the center positions of the spectral extractions for the jet spectra shown in the middle and bottom panels of Fig.~\ref{fig:fullspectrum}. The coordinate axes at the top and on the right side of the figures show offsets from the star in arcseconds; $(0, 0)$ marks the position of the star.  (Produced  with SAIOmageDS9 software.) 
}
}
\end{minipage}
}
}
\vskip -0.3cm
\end{figure*}

The spectral resolving power, $R$, of the MIRI MRS ranges between $\sim$3960 at 5~$\mu$m and $\sim$1400 at 25~$\mu$m, following the relation given in \citet{Argyriou2023}. Specifically, $R(12.81~\mu\mathrm{m}) \approx 2960$ and $R(15.55~\mu\mathrm{m}) \approx 2610$ for the [Ne~{\sc ii}] and [Ne~{\sc iii}] lines, respectively.
This allowed us to analyze the velocity structure of the jets in some detail across the entire field of view given a velocity broadening of the lines of several 100~km~s$^{-1}$ each in the blueshifted and the redshifted component. We can obtain a first impression from the middle and lower panels of Fig.~\ref{fig:fullspectrum}. These line spectra were extracted from positions on the blue and the red jets, respectively (colored accordingly; see  Fig.~\ref{fig:RGBimages} for the extraction locations), and are essentially continuum-free; the continuum is expected to originate only from the disk. The relative suppression of redshifted lines compared to blueshifted lines toward short wavelengths may be partly related to higher extinction toward the red jet. We find single-line flux peaks at similar radial velocities among the blueshifted components of the atomic lines, but the redshifted components show at least two peaks at different velocities, and their highest velocities of $\ga$500~km~s$^{-1}$ exceed those in the blue jet ($\sim$300~km~s$^{-1}$). The velocity peaks relate to different flow structures in the jet, as we discuss below.

To obtain a more detailed view of the jet extent and structure, we show in Fig.~\ref{fig:bluered} six intensity images taken in the fine structure lines of Ni$^+$, Ar$^+$, Ne$^+$, Ne$^{2+}$, and Fe$^+$ ions, i.e., [Ni~{\sc ii}]~6.63~$\mu$m (top-left), [Ar~{\sc ii}]~6.98~$\mu$m (bottom-left),
[Ne~{\sc ii}]~12.81~$\mu$m (top-center), [Ne~{\sc iii}]~15.55~$\mu$m from channel 3-long (bottom-center), and [Fe~{\sc ii}]~17.94~$\mu$m from channel 3-long (top-right) and channel 4-short (bottom-right).
We chose the one, two, or three spectral bins with the highest fluxes in the resolved blueshifted and redshifted parts of each line, took the average in case of two or three bins, and subtracted the MRS image in the redshifted line(s) from the image in the blueshifted line(s). 
Given the close proximity of the blue- and redshifted lines (peak-to-peak separation for the [Ne~{\sc ii}] line is 0.0125~$\mu$m, or a fraction of $\sim$10$^{-3}$ of the wavelength), this procedure optimally removes the local continuum and background flux at the wavelengths of the lines, leaving only line flux across the images and potentially some increased noise in the central PSF from the subtraction process. Figure~\ref{fig:bluered} shows that there are no systematic PSF-sized residuals at the star's location due to incomplete continuum subtraction.
The color scheme is such that red hues indicate (a dominance of) redshifted emission while blue hues represent blueshifted emission. The figures also show the approximate size and shape of the continuum dust (inner ellipse) and the gas (outer ellipse) disk derived from ALMA continuum and CO line observations at 220~GHz \citep{guedel2018}. 

All images show the known bipolar outflow in the NE-SW direction, the SW lobe approaching the observer. There are a number of noteworthy features: (i) The [Ne~{\sc ii}] image shows either dominant blueshifted or redshifted emission almost throughout the MIRI MRS field of view, with a clear demarcation line between the lobes. (ii) Redshifted flows are not recorded in the projected disk area because of the disk's optical depth (e.g., \citealt{white2014b}). (iii) The structure of the two lobes is  different. 
While the blueshifted jet is relatively narrow and collimated (see in particular the [Ni~{\sc ii}] and [Ar~{\sc ii}] images), 
the redshifted jet is more akin to a bubble structure, the brightest part being located about 3\arcsec\ away from the central star; this is reminiscent of the finding by  \citet{agraamboage2011} and \citet{white2014b} although those authors find this structural asymmetry at much closer distances from the star, within about one arcsecond. The higher-resolution [Ni~{\sc ii}] and [Ar~{\sc ii}] images show a more limited area but reveal more structure in the redshifted features. (iv) The [Fe~{\sc ii}] image from channel 4-short may reveal the blue jet out to $\sim$6\arcsec\ in the SW.
We also note that some of the panels (e.g., [Ne~{\sc ii}], [Ne~{\sc iii}], [Ar~{\sc ii}]) marginally show some blue features extending just north of the disk.  Blueshifted emission in this region was also identified by \citet{garufi2022} who interpreted the feature as an accretion streamer falling toward the disk from its
far side.

We go on to take a closer look at the spatial and velocity structure of the outflow lobes in [Ne\,{\sc ii}]. Figure~\ref{fig:RGBimages} shows composite red-green-blue (RGB) images in which the red (R), green (G), and blue (B) channels are used for three representative data cube slices at high, intermediate, and low velocities, respectively, relative to the systemic velocity. For the latter, we adopt a (barycentric) radial velocity of 15.4~km~s$^{-1}$ for DG Tau \citep{nguyen2012}.
For the blue jet, the three radial velocities are: 
$-186$~km~s$^{-1}$ (R), \nobreak{$-128$} km~s$^{-1}$ (G), and $-11$~km~s$^{-1}$ (B); for the red jet, +341~km~s$^{-1}$ (R), +165~km~s$^{-1}$ (G), and +48~km~s$^{-1}$ (B). The wavelength bins were chosen such that they optimally highlight three distinct morphological features (see below); the purpose of the figures is to spatially relate these features to each other. The four panels show different combinations of velocity channels: RGB (top-left), RB (top-right), BG (bottom-left), and RG (bottom-right); the east half of each image shows redshifted emission, the west half blueshifted emission; the separation is marked by an arrow. A more extensive collection of images in multiple individual velocity channels, including other ions, is given in Appendix~\ref{appendixfigures}. See Figs.~\ref{fig:ratio1}--
\ref{fig:ratio8}, which compare  [Ne\,{\sc ii}] and [Ne\,{\sc iii}] at similar velocities (and also ratio images, discussed in Sect.~\ref{cause}). See also Fig.~\ref{fig:arfe}, which shows different velocity channels for [Ar\,{\sc ii}]~6.98~$\mu$m, [Ni\,{\sc ii}]~6.63~$\mu$m, and [Fe\,{\sc ii}]~17.94~$\mu$m, and 
Fig.~\ref{fig:neIIvelocity}, which shows the [Ne\,{\sc ii}]~12.81~$\mu$m images used in Fig.~\ref{fig:RGBimages}.

The structure of the SW (approaching) jet is similar in all three velocity channels, with a hint at intermediate-velocity material (G) emitting the farthest from the center, while especially slow material (B) is seen close to the center only.
The NE (receding) jet is morphologically much more complex. We can distinguish three separate features related to the three velocities (see also Figs.~\ref{fig:ratio1}--\ref{fig:ratio8}). First, the slowest material (B) forms a wide, arc-shaped feature (arc henceforth) surrounding the faster material. The intermediate-velocity material (G) is located at the apex of the detected outflow structure, showing a featureless but extended patch of emission (blob). Instead, the fastest feature delineates what appears to be a well-collimated $\sim$3\arcsec\ long jet, perhaps the analog of the fast SW jet.

\subsection{Extended H$_2$ emission}\label{h2extended}

\begin{figure}[t!]
\centering
\vskip -0.1cm
\includegraphics[width=0.48\textwidth]{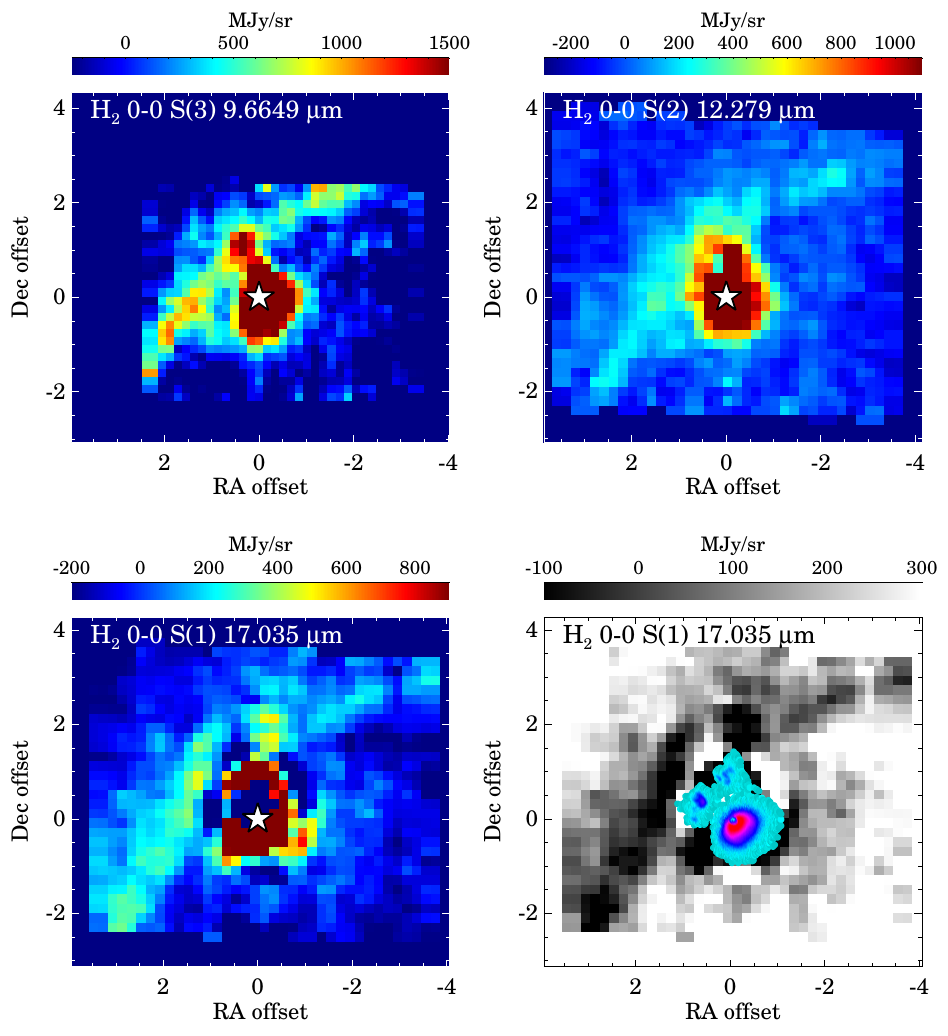}
\caption{
Slightly smoothed images in various  H$_2$ lines showing a central source and an arc to the E and N of DG Tau. The star symbol marks the position of DG Tau. \textit{Top left:} 
H$_2$ S(3) 9.67~$\mu$m.  
\textit{Top right:} 
H$_2$ S(2) 12.28~$\mu$m. \textit{Bottom left:} 
H$_2$ S(1) 17.03~$\mu$m. 
\textit{Bottom right:} Superposition of H$_2$ S(1) emission at 17.03~$\mu$m (gray shades) and (in other colors) a SPHERE scattered-light image \citep{garufi2024} showing a bowl-shaped structure suggested to be created by winds  (see text). 
}
\label{fig:h2arc}
\end{figure}
We map in Fig.~\ref{fig:h2arc} the entire MRS field in the three molecular hydrogen lines of H$_2$ S(1) 17.03~$\mu$m, S(2) 12.28~$\mu$m, and S(3) 9.67~$\mu$m. The images are continuum-subtracted, using as a continuum level a line-poor spectral region next to the lines. Each of the maps shows two emitting features: 

(i) A bright patch about coincident with the disk region and possibly asymmetrically extended to the SW.  Given the limited spatial resolution of the MRS, we do not further discuss the MIR emission; however, we note that its appearance approximately matches the appearance of a cone-like structure seen in scattered light with SPHERE (Fig.~\ref{fig:h2arc}, lower right; \citealt{garufi2024}).

(ii) We find an extended arc running from NW to SE, approximately perpendicularly to the jet axis and displaced from the star in the NE direction by about 1\arcsec. 
The emitting material may reveal some substructures that differ slightly between the three maps.

\subsection{Discussion: Structure and dynamics of winds and jets}\label{structure}

\begin{figure}
    \centering
\includegraphics[width=1.0\linewidth]{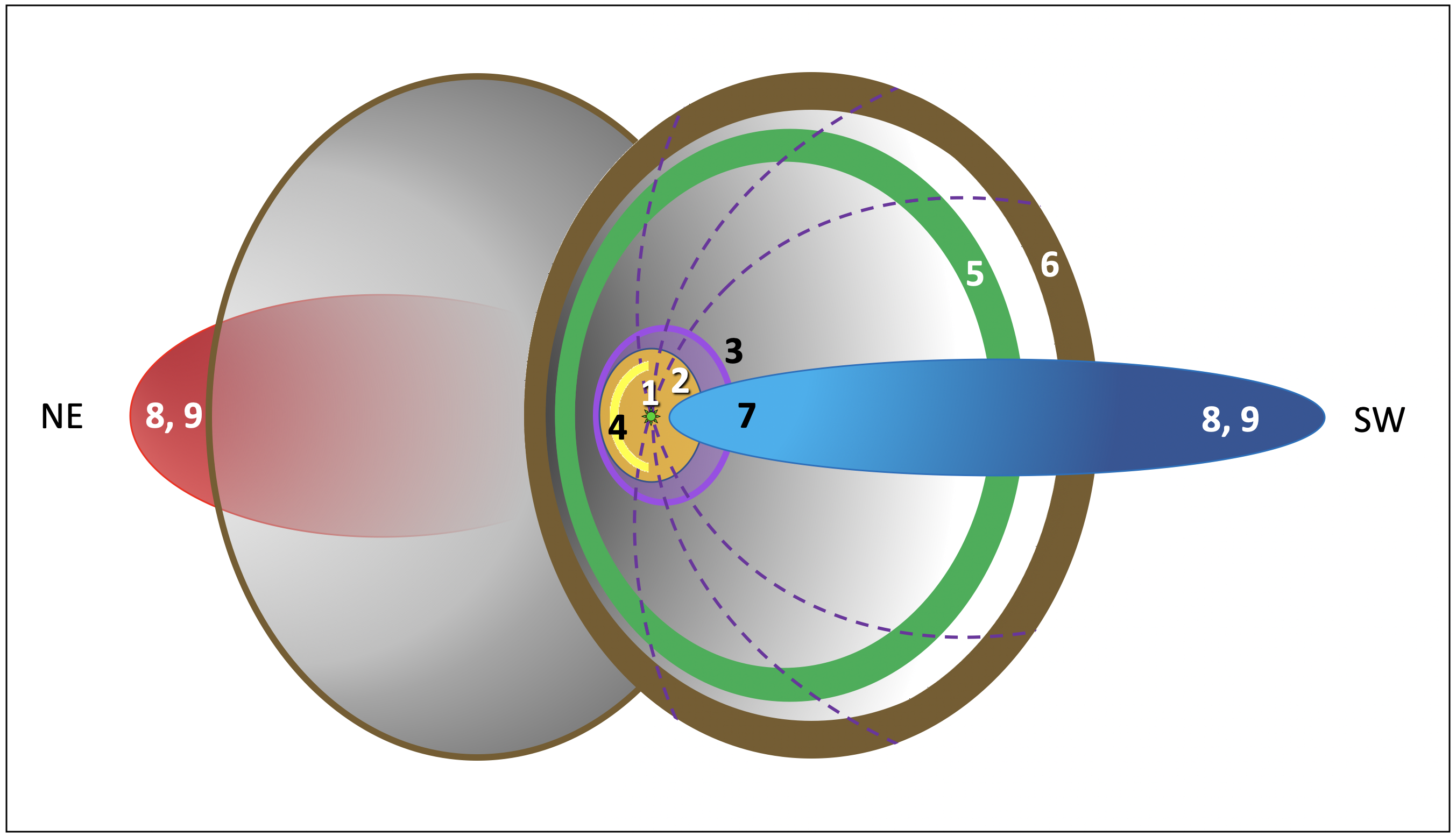}
    \caption{Interpretation of the environment of DG Tau putting features into context. The gray shapes illustrate axisymmetric bowl-shaped outflow structures containing winds and the jets. The gray structure on the left-hand side (pointing toward the NE) is not observationally evident except close to the star in H$_2$ emission \citep{white2014a, white2014b}. The colored and numbered structures correspond to features revealed in Figs.~\ref{fig:overview} and \ref{fig:overview_alt}. 
    1. Bright green star: The star DG Tau. 
     \textbf{2. }Filled orange ellipse: Protoplanetary disk of DG Tau. 
     \textbf{3. }Purple: Inner wind region seen in IR H$_2$ 
        \citep{agraamboage2014}, in H$_2$ UV fluorescence \citep{schneider2013a}, and in scattered 
        light (\citealt{garufi2024}, see Fig.~\ref{fig:h2arc}).     
     \textbf{4. }Yellow arc on disk: Wind seen in the   
        $^{12}$CO~230.538~GHz line with ALMA \citep{guedel2018}.     
     \textbf{5. }Green: Arc in H$_2$ S(1)--S(3) 
        partially observed by MIRI/MRS 
       (Fig.~\ref{fig:h2arc}; visible only toward the SE, E, NE, N, and NW of the star).
     \textbf{6. }Brown: Arc in the $^{12}$CO 230.538~GHz line      
        partially observed by NOEMA \citep{semenov2024} and by ALMA (Figs.~\ref{fig:overview} and \ref{fig:overview_alt}; visible only toward the SE, E, NE, N, and NW of the star). 
     \textbf{7. }Bright blue: inner, collimated, forward (blue) jet (seen, e.g., in X-rays).
     \textbf{8/9. }Blue/red: extended forward (blue) and receding (red) jets, respectively, seen, among others, in X-rays and in  [Ne\,{\sc ii}].  
    }
    \label{fig:sketch}
\end{figure}
\subsubsection{Structure of the winds}  
Figure~\ref{fig:sketch} gives an interpretation of structures seen in our MIRI observations and also by other instruments, as displayed in the composite sketch in Fig.~\ref{fig:overview}. We interpret the immediate environment of DG Tau above the disk plane as a system of winds and jets flowing inside a bowl-shaped structure. Similar structures were described and proposed previously by \citet{kitamura1996}, \citet{agraamboage2011, agraamboage2014}, \citet{white2014a, white2014b};  \citet{delabrosse2024} discussed analogous features for the protostar DG Tau B. In this interpretation, progressively larger, axisymmetric arc-like features are detected at different, decreasing flow velocities. 

The innermost (distances of $\la$1\arcsec) wind structures are seen in fluorescent FUV lines of H$_2$ \citep{schneider2013a} and  in IR H$_2$ lines, 
flowing with a speed of $\sim$($10-30)$~km~s$^{-1}$ 
(\citealt{takami2004, agraamboage2011, agraamboage2014, white2014a, white2014b},  and also identified in our MIRI images in Fig.~\ref{fig:h2arc}), and in NIR scattered light \citep{garufi2024}. 
\citet{takami2004} and \citet{beck2008} propose shock excitation of H$_2$ rather than FUV or X-ray irradiation as the cause of this emission.

FUV C\,{\sc iv} line emission is found inside the H$_2$ cone \citep{schneider2013a, schneider2013b}; on the jet axis itself, X-ray emission from multi-million K plasma is detected \citep{guedel2008, guedel2011, schneider2008, Frank2014}. This suggests  a temperature gradient perpendicularly to the jet axis, from the hottest, $3.7\times 10^6$~K X-ray emitting plasma to the $\sim 10^5$~K C\,{\sc iv} emitting  plasma to the $T \approx 2000$~K H$_2$ line emitting wide cone and on to the cooler cavity walls. These compact structures at $0\farcs 2-0\farcs 3$ (the innermost jet X-ray source, the narrow H$_2$ cavity, and the innermost C\,{\sc iv} feature) all seem to be stationary \citep{guedel2011, schneider2013a, schneider2013b}.

Some of the H$_2$ emission displaced by $\la$1\arcsec\ toward the NE is however redshifted by $\sim$5$~$km~s$^{-1}$ as described by \citet{beck2008}, \citet{agraamboage2014}, and \citet{white2014a, white2014b} and interpreted as part of the red lobe of DG Tau's jet and outflow.
This is supported by filamentary radial structure in the H$_2$ emission and similarly also in the dust continuum (Fig.~\ref{fig:overview} and Fig.~\ref{fig:h2arc}; \citealt{garufi2024}).

At distances of approximately 40--70~au from the star, a molecular wind was found in ALMA CO observations, with a flow speed of up to 3.1~km~s$^{-1}$ \citep{guedel2018}; although part of this feature was, alternatively, interpreted as an accretion stream approaching the disk from behind, it is interesting to note that polarization features align well with the flow (see images in \citealt{bacciotti2018}).

At larger distances, we find the H$_2$ arc-like structure reported in Fig.~\ref{fig:h2arc},  at a projected minimum distance of $\sim$1\arcsec\ from the star.
Higher-resolution millimeter observations of $^{12}$CO (2--1) with NOEMA at 230.54~GHz also reveal a distinct arc to the NE of DG Tau resembling our H$_2$ feature, but at a larger distance of approximately 2\farcs 0--2\farcs 5 revealing blueshift by a few km~s$^{-1}$ \citep{semenov2024}. We checked still higher-resolution ALMA observations published in \citet{guedel2018} to find that this arc varies in size as a function of frequency; at a position angle of $\sim$220\degr\ (toward the NE) the minimal projected distance from DG Tau varies from 1\farcs 4 at a blueshift of 8.1~km~s$^{-1}$ relative to the star, to 2\farcs 1 at a blueshift of 4.2~km~s$^{-1}$, i.e., the emitting material decelerates outward. We note that the H$_2$ arc could be located closer to the star along the wall of the bowl-shaped structure, as depicted in Fig.~\ref{fig:sketch}, but alternatively the H$_2$ emission could also be located closer to the axis, inside the CO emitting surface as proposed by \citet{delabrosse2024} for the protostar DG Tau B.

Similar features, also on larger scales, were reported before: \citet{nakajima1995} described two similar arcs seen in the \textit{I-}band at distances of 2--3\arcsec\ and $\sim$12\arcsec\ from the disk center, interpreted as reflection nebulae and suggested to be the remnants of an infalling envelope or a flattened pseudodisk feeding the accretion disk. Arcs are also seen in an SDSS image in the optical (see Fig.~1 in \citealt{varga2017}) at about 5\arcsec\ and 12\arcsec\ to the NE. 
And at even larger distances of $\sim$2800 au, \citet{kitamura1996} identified an outflow probably related to a flattened disk-like structure as a remnant of the protostellar envelope, with an outflow velocity of 1.5~km~s$^{-1}$. 

The arcs seen in Figs.~\ref{fig:overview} and \ref{fig:h2arc}  are thus surrounding the blue jet and are unrelated to the receding lobe of DG Tau's winds and jets. The material on the SE--NE--NW sides of DG Tau seen in H$_2$, CO, or in scattered light approaches the observer on the nearer, steep wall of the cavity, except for some compact ($\ga$1\arcsec) structure immediately to the NE of the star related to the receding lobe of the outflow \citep{white2014a, white2014b}. Material presumably flowing along the far side of the forward cavity would flow to the SE, but this material is not detected here.

\begin{figure}[t!]
    \centering
\includegraphics[width=1.0\linewidth]{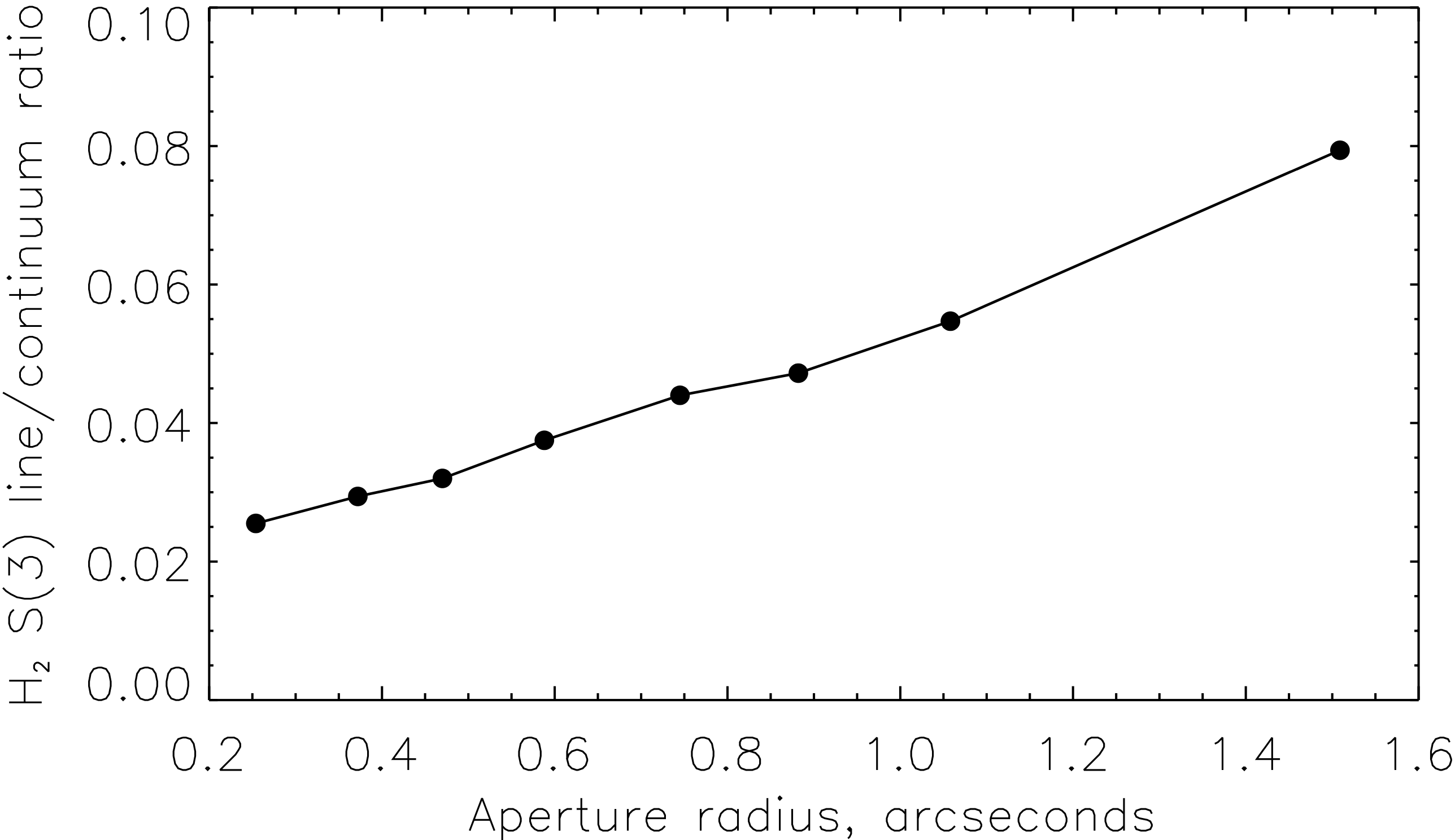}
    \caption{H$_2$ S(3) 9.6649~$\mu$m line-to-continuum ratio as a function of increasing aperture for spectral extraction, centered at the stars position.}
    \label{fig:scatter}
\end{figure}
At this point, we might consider whether H$_2$ emission could, in fact, be scattered light originating in the inner disk that had gone unresolved in our observations. As discussed in \citet{delabrosse2024} for the protostar DG Tau B (located $\sim$50\arcsec\ south of DG Tau), in the scattered-light scenario we would expect a constant line/continuum ratio for the H$_2$ lines (given the same PSF for line and continuum), reflecting the ratio in the central source. Instead, for in-situ radiation, we expect non-constant ratios and an increasing line/continuum ratio with increasing distance from the central source. We analyzed this ratio by increasing the aperture for spectral extraction around the central star from 0\farcs 2 to 1\farcs 5, thus comprising the compact H$_2$ source in Fig.~\ref{fig:h2arc} as well as the nearer part of the arc toward the NE. We find a monotonically increasing line ratio in Fig.~\ref{fig:scatter}, supporting the model of in-situ H$_2$ emission.

\subsubsection{Structure of the jets} 
Next, we attempted to interpret our fine structure line images of [Ne\,{\sc ii}],
[Ne\,{\sc iii}],  [Ni\,{\sc ii}], [Ar\,{\sc ii}], and [Fe\,{\sc ii}] as dynamic structures. We compared our observations with [Fe\,{\sc ii}]~1.644~$\mu$m and  H$_2$~1--0 S(1) 2.12~$\mu$m observations obtained mostly in $2005-2006$ by \citet{agraamboage2011, agraamboage2014} with VLT/SINFONI, and by \citet{white2014a, white2014b} with Gemini North/NIFS. Both groups report features close to the jet axis within $\sim$1\farcs 3 of the central star, a region that is very difficult to access by MIRI MRS given the size of its point-spread function (PSF) and the brightness of the disk.
The above authors agree that the features in the blue and red lobes of the outflow are highly asymmetric. The approaching jet is collimated, with a half-opening angle of $\sim$14$-$15\degr; it shows a number of knots that are ejected every few years, perhaps periodically \citep{agraamboage2011, rodriguez2012, takami2023, pyo2024}. 

\citet{bacciotti2000}, studying spectro-imaging of the blue jet in optical lines of [N\,{\sc ii}], [S\,{\sc ii}], [O\,{\sc i}], and H$\alpha$ within $\sim$1\farcs 5 of the star, noted a velocity gradient toward the jet axis, defining a so-called onion-like velocity structure (see also \citealt{agraamboage2011, agraamboage2014, white2014a, white2014b}). \citet{bacciotti2000} also find a velocity gradient along the (blue) jet axis, lower velocities being found more prominently closer to the star compared to high-velocity features. Indeed, radial velocities initially increase with distance up to 420~km~s$^{-1}$ at distances of $0\farcs 7-3\farcs 5$ from the star in HST observations \citep{maurri2014}. This type of structure has been seen by JWST in many outflows (from Class 0 to Class II sources), as described, for instance, by \citet{caratti2024}, \citet{pascucci2025}, or \citet{vanDishoeck2025}.

In contrast, \citet{agraamboage2011, agraamboage2014} and \citet{white2014a, white2014b} found the DG Tau red jet in $2005-2006$ to resemble an expanding bubble, with radial velocities increasing from $\sim$100~km~s$^{-1}$ to $\sim$180~km~s$^{-1}$ between 0\farcs 7 and 1\farcs 3, while jet propagation to larger distances may be impeded by ambient material left over from the protostellar envelope; no extended, receding jet has been recorded in the optical or IR, although \citet{white2014a} reported a fast knot exactly on the jet axis 0\farcs 84 from the star that may be part of a collimated inner jet, with a radial velocity of 170~km~s$^{-1}$, corresponding to a true velocity along the flow of $\sim$220~km~s$^{-1}$. They suggested that this fast flow is being dispersed by its interaction with the remnant pre-stellar envelope seen in H$_2$ at very low velocities (5--10~km~s$^{-1}$), driving the expansion of the bubble. 

\begin{figure}[h!]
\vskip 0.1cm
\centering
\includegraphics[width=0.39\textwidth]{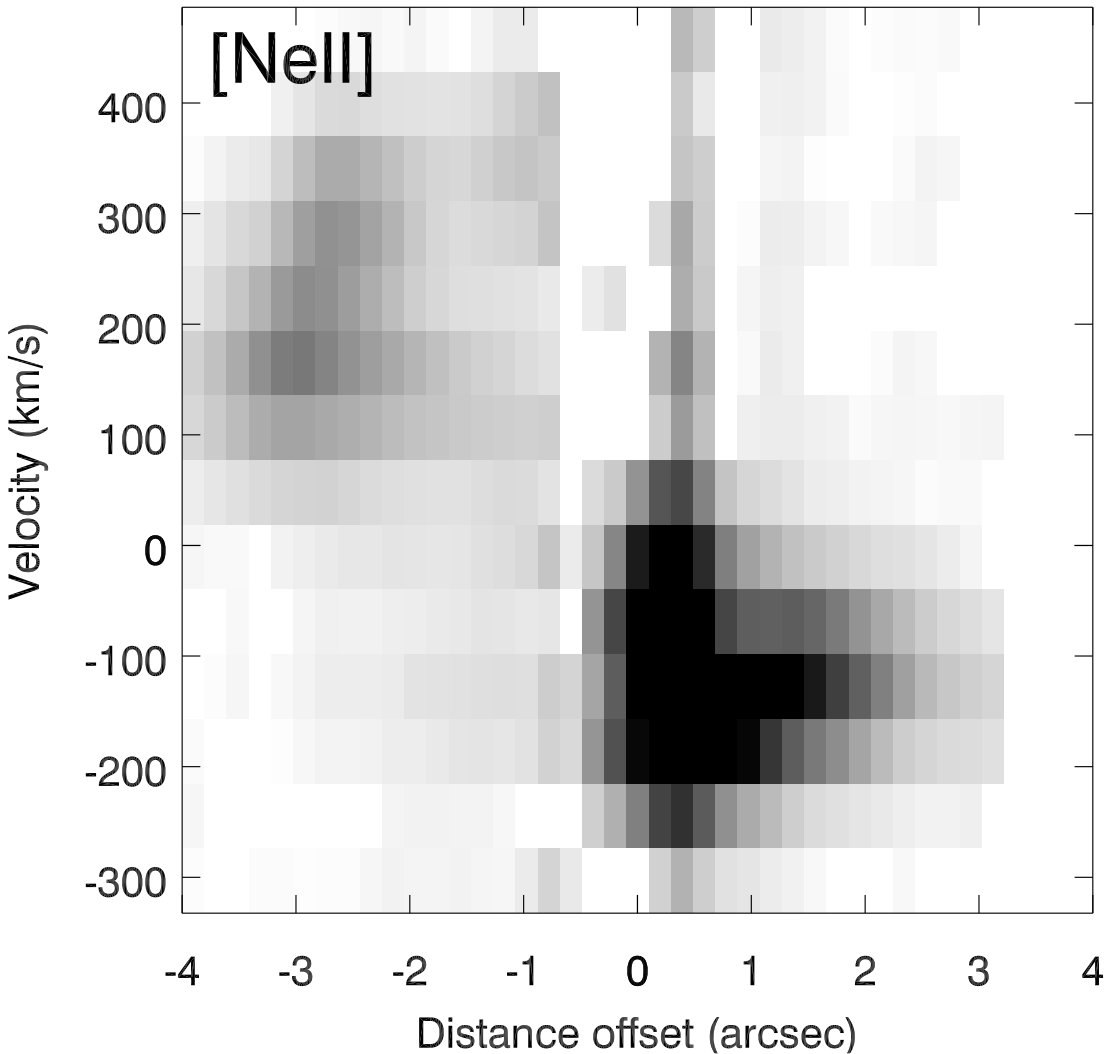}
\includegraphics[width=0.4\textwidth,angle=180,page=1]{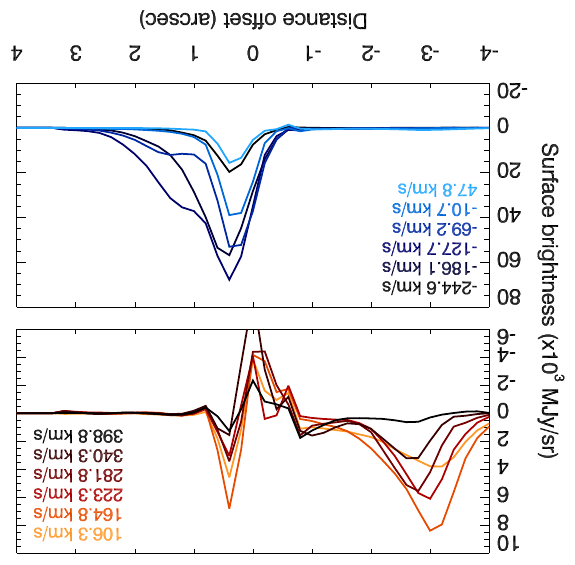}
\caption{\textit{Top}: pv diagram for [Ne\,{\sc ii}]~12.81$\mu$m along the jet axis. Distance is positive to the SW (blue jet). \textit{Bottom}: 1D cuts of the brightness distribution of [Ne\,{\sc ii}]~12.81$\mu$m emission along the jet for different velocity components. The lines represent 1D cuts through the stellar position along the jet axis using bilinear interpolation to obtain surface brightness values (in MJy~sr$^{-1}$) at one pixel spacing. The same values were used for the grayscale panel.}\label{fig:pv}
\end{figure}

In Fig.~\ref{fig:pv} we show a position-velocity (pv) diagram in [Ne\,{\sc ii}] along the jet axis. The blue jet registers at all velocity bins from zero up to $\sim$250~km~s$^{-1}$. In support of \citet{maurri2014}, we find emission from intermediate-velocity, $\sim$50--150~km~s$^{-1}$ gas in the blue jet to be the most extended along the jet axis, dominating the fluxes at $1-3$\arcsec\ (Fig.~\ref{fig:pv} bottom). This could indicate deceleration at $1-2$\arcsec\ although variable emissivity along the jet axis also plays a role, and MIRI MRS cannot resolve profiles in the lateral direction to distinguish between narrower fast and broader slow flows (see, e.g., \citealt{white2014b}). The pv diagram of the blue jet is similar to the diagram presented by \citet{white2014b} for [Fe\,{\sc ii}] (their Fig.~15), albeit at smaller distances. 
  
For the red jet, we identify, as pointed out in Sect.~\ref{morphology} and Fig.~\ref{fig:RGBimages}, 
an onion-like velocity structure. Our observations include a highly collimated jet at radial velocities of over 300~km~s$^{-1}$ out to $\sim$3\farcs 5; this is about three times farther than seen by \citet{white2014a, white2014b}. A similarly extended and apparently collimated jet was recorded in X-rays in $2005-2006$ \citep{guedel2008}. The redshifted [Ne\,{\sc ii}] peak flux in our pv diagram (Fig.~\ref{fig:pv}) shows decreasing velocity with increasing distance, possibly indicating deceleration of expanding material. 
We also note that all three velocity features terminate at similar locations at $(3-4)$\arcsec\ to the NE of the star, suggesting that the faster material is colliding in shocks with slower gas flows.

Next, we investigate whether any of the structures seen in [Fe\,{\sc ii}]~1.644~$\mu$m observations between 2006 and 2009 reported by \citet{white2014a, white2014b} could correspond to structures in our MIRI observations obtained in February 2023. First, \citet{white2014a} find that a bubble-like structure at $0\farcs 7-1\farcs 3$ from the center in the red lobe moves along the outflow axis at a projected velocity of 0\farcs 07 in 1.1~yrs (after a correction for registration uncertainty), translating to a projected proper motion of 38~km~s$^{-1}$ at 125~pc distance, or a deprojected velocity of 61~km~s$^{-1}$ along the outflow axis, and a radial velocity component of 48~km~s$^{-1}$ (which does not need to coincide with the general flow velocity, for example in the case of shocks). If the structure expanded at constant projected velocity, its apex (at --1\farcs 3 in 2005.87) would reach a distance of 1\farcs 3  + 1\farcs 1 = 2\farcs 4 in the MIRI observations.  This is at the base of the brightest parts of the blob region in Fig.~\ref{fig:RGBimages}. If instead the Feb. 2023 distance of the center of the blob region, about 3\farcs 2 from the star, is assumed to correspond to the expanding  structure seen in 2005/2006 at 1\farcs 3, then the proper motion would be $0\farcs 11$~yr$^{-1}-0\farcs 12$~yr$^{-1}$, corresponding to $\sim$68~km~s$^{-1}$ and therefore a radial velocity of $\sim$87~km~s$^{-1}$ for an inclination of 38\degr. This is smaller than the typical radial velocity of the blob by a factor of 1.9; given that the flow speed and the expansion speed of features do not need to be identical, it remains conceivable that the two structures are related. 

If our suggestion holds, then we may be witnessing a momentum driven bubble expansion followed by breakout phase of the fast jet beyond it,  entering a classical jet phase as proposed and modelled by \citet{sutherland2007} for Active Galactic Nuclei (AGN) and applied to DG Tau by \citet{white2014a}. The analytical model worked out by \citet{white2014a} for the expansion of a bubble that is driven by momentum transfer from a fast jet would, however, expand by no more than 0\farcs 2 between 2006 and 2023, much less than suggested by our observations. We also note that X-ray observations obtained at the time of White et al.'s [Fe\,{\sc ii}]~1.644~$\mu$m observations (late 2005 to early 2006) showed a receding X-ray jet source out to $\sim$4$\arcsec-5\arcsec$, suggesting that the red jet would have broken through the ambient material already earlier.

Second, the bright feature observed in 2005.86 at 0\farcs 84 exactly on the outflow axis had a radial velocity of 170~km~s$^{-1}$ (range $\sim 140-200$~km~s$^{-1}$, \citealt{white2014a}), corresponding to a projected distance of $\sim$0\farcs 84 + 3\farcs 86 = 4\farcs 7 in 2023.14 in the absence of deceleration. The apex of the fast jet recorded by MIRI in [Ne\,{\sc ii}] is seen at 3\farcs 5 but at higher radial velocities of $\ga$300~km~s$^{-1}$ (1.8 times faster than in \citealt{white2014a}). But again, the coincidence of these values well within a factor of 2 may suggest continuous jet expansion between 2006 and 2023.

We also highlight a small but bright feature at a distance of   $\sim$1\arcsec\ on the red jet axis, i.e., in projection just outside the ALMA CO gas disk, seen in the maps of  [Ne\,{\sc ii}] and [Ne\,{\sc iii}], see Fig.~\ref{fig:ratio7} and \ref{fig:ratio8}.
In the fine structure lines, this feature shows high velocities of 250--400~km~s$^{-1}$ (see also pv diagram in Fig.~\ref{fig:pv}) while in H$_2$ the radial velocity is less than 50~km~s$^{-1}$. These features are reminiscent of what \citet{white2014a} reported from [Fe\,{\sc ii}] 1.644~$\mu$m and H$_2$~2.12$~\mu$m observations at the same location. It is difficult to judge if some of these features are related or perhaps analogous features due to repeated ejection of jet knots, given the limited access to distances $<$1\arcsec\ in the MIRI MRS and a time gap between the two observations of about 16 years. 

To conclude, there is some (limited) evidence that provides a connection between the jet and wide-angle wind flows described in detail by \citet{white2014a, white2014b} and the larger structures we have imaged with MIRI. This is supported by the measured kinematic properties of the structures in the outflows and jets.

\subsection{The cause of neon ionization}\label{cause}

Comprehensive observational surveys, especially with  \textit{Spitzer}, have detected many [Ne\,{\sc ii}]~12.81~$\mu$m lines in CTTS (and also Herbig stars) but only very few [Ne\,{\sc iii}]~15.55~$\mu$m lines (e.g., \citealt{flaccomio2009, guedel2010, baldovin2011, sacco2012, pascucci2020, akhila2025}). \citet{Espaillat2013} studied flux ratios $f =$ {\rm flux([Ne\,{\sc iii}]~15.55~$\mu$m)/{\rm flux([Ne\,{\sc ii}]~12.81~$\mu$m)}}  for a small sample of objects with detected [Ne\,{\sc iii}]~15.55~$\mu$m lines and found ratios of $f \approx 0.05-0.35$ with one exception that showed $f$ somewhat larger than unity. For a sample of 56 transitional disks, \citet{Szulagyi2012} reported 32 [Ne\,{\sc ii}]~12.81~$\mu$m detections but found [Ne\,{\sc iii}]~15.55~$\mu$m in only three of the objects, concluding that $f$ is typically significantly below unity.

We emphasize that the sources in the cited surveys were not spatially resolved; the comparison with models assumed that the neon emission comes from a disk atmosphere. The situation for DG Tau is, as already summarized in Sect.~\ref{structure}, more complex: While DG Tau is surrounded by a standard protoplanetary disk with a high accretion rate, various winds escape from the disk, and two prominent, partly collimated jets are present that we identify as strong sources of emission from ionized neon, argon, iron, and nickel. 
X-rays originate from the star with $L_{\rm X} = (0.7-1.1)\times 10^{30}$~erg~s$^{-1}$ in a hard spectrum, while the softer stellar X-rays are strongly absorbed; but a weakly absorbed soft spectrum is emitted by the inner jet within $\approx 40$~au from the star (Figs.~\ref{fig:overview} and \ref{fig:overview_alt}, and Fig. 5 in \citealt{Frank2014}), with $L_{\rm X} \approx 10^{29}$~erg~s$^{-1}$ \citep{guedel2008}. The strong X-ray absorption of the star is not accompanied by commensurably strong visual extinction based on standard gas-to-dust ratios \citep{Vuong2003}, suggesting that the absorber is dust-depleted gas.

\citet{maurri2014} found that the ionization degree in the blue jet is highest, up to 70\%, at distances of $2\arcsec-4\arcsec$ from the star while it is much lower inside 1\arcsec. This is supported by \citet{giannini2019} from spatially unresolved high-resolution line spectroscopy.
The question then is where and how the ionization is produced.
Generally, ionization could take place close to the star; for instance, in a recollimation shock \citep{bonito2011} or via photoionization due to irradiation by extreme UV (EUV) or X-rays from the star or the jet itself \citep{guedel2008}. The ionization degree will only gradually decrease with distance because of slow recombination. 
\citet{guedel2008} studied plasma cooling by radiative cooling and  cooling by expansion of the jet, concluding that the former alone explains the observed jet lengths with hot-plasma densities of $\sim 10^5$~cm$^{-3}$ and volume filling factors on the order of $10^{-6}$ (although for lower densities cooling is predominantly adiabatic). 
\citet{schneider2008} also presented a simple radiative cooling model for the hot plasma flowing in the jet with a velocity of 300~km~s$^{-1}$ behind a shock, based on X-ray observations \citep{guedel2007, guedel2008}; they also concluded that the hot-plasma filling factor is small, but the plasma radiative cooling time is long, resulting in a cooling length of at least 100~au. Considering hydrogen alone, \citet{maurri2014} estimated a recombination time, i.e., a significant decay of the ionization fraction, of $6\times 10^8$~s in densities declining from $10^5$~cm$^{-3}$ to $\sim$10$^3$~cm$^{-3}$ beyond 1\arcsec\, resulting in a scale length of the ionized jet of $\sim$6\arcsec\ in projection at a distance of 125~pc. This agrees with our MIRI/MRS observations as far as the field of view permits detection of emission from ionized neon, argon,  iron, or nickel (see in particular Fig.~\ref{fig:bluered}--bottom-right for Fe\,{\sc ii}), and incidentally it also agrees with the length of the X-ray jet \citep{guedel2008}. One possibility is therefore that ionization and heating to millions of K occurs within an arcsecond from the star. However, continued \, (re-)ionization by high-velocity shocks could also take place along the jet, either due to shocks or due to irradiation. \citet{liu2016} put [Ne\,{\sc iii}]~$\lambda$3869 observations of the DG Tau jet into context with photoionization by either stellar X-rays or X-rays forming in the jet. We discuss the possibility of the former mechanism below. However, heating to X-ray-emitting temperatures and photoionization of cold material by the jet X-ray sources should not be treated as separate mechanisms as they naturally occur in  concert. We therefore propose a self-consistent approach below that includes both in the same model. A further alternative would be photoionization by external hot stars, but this can be excluded as the Taurus star-forming region is devoid of such massive stars. Thus, we considered three possible causes for the ionization of neon (or analogously other metals): photoionization by X-rays or EUV radiation originating in the stellar corona, or ionization via shock heating that naturally involves both collisional ionization and photoionization.

\subsubsection{Photoionization by the star}
Neon has high first and second ionization potentials, namely 21.56~eV for neutral Ne and 41.0~eV for Ne$^+$. Photoionization therefore requires high-energy photons with wavelengths less than 57.5~nm and 30.2~nm, respectively (mid-to-short wavelength EUV or X-rays). The fine-structure lines of neon are predominantly excited by collisions with electrons (rather than H atoms; \citealt{glassgold2007}).

DG Tau's X-ray luminosity is $L_{\rm X} = 10^{30}$~erg~s$^{-1}$ \citep{guedel2008}. For a stellar radius of $R_{\odot} = 2.6R_{\odot}$ (from the bolometric luminosity and the effective temperature $T_{\rm eff}$ given in \citealt{garufi2024}) we find an average X-ray surface flux of $F_{\rm X} \approx 2.5\times 10^6$~erg~cm$^{-2}$~s$^{-1}$ and from this estimate an EUV-to-X-ray flux or luminosity ratio of 1.0 for EUV radiation between 10 and 36~nm (Eq.~20 in \citealt{johnstone2021}). The range of 36-91~nm adds $\sim 29$\% to it (Eq.~22 of \citealt{johnstone2021}), i.e., we adopt $L_{\rm EUV} = 1.3\times 10^{30}$~erg~s$^{-1}$. 

\subsubsection{Stellar EUV photoionization} 
EUV photons are strongly absorbed shortward of the H Lyman edge at 91.2~nm. Photoionization is therefore possible only if the foreground gas is highly ionized and thus transparent to EUV radiation. The ionization fraction in the jet is indeed very high, up to $\sim$(70--80)\% \citep{maurri2014}. We calculate the Str\"omgren radius within which neutral hydrogen gets fully ionized to a H\,{\sc ii} region as
\begin{equation}\label{stroemgren}
    R_{\rm S} = \left(\frac{3\Phi_{\rm EUV}}{4\pi \alpha n_0^2}\right)^{1/3},
\end{equation}
where $\alpha \approx 3\times 10^{-13}$~cm$^3$~s$^{-1}$ is the recombination coefficient of the dominant hydrogen gas in the typical jet temperature range of $(1-2)\times 10^4$~K \citep{nahar2021}, and $n_0$ is the gas number density in the jet. For the latter, we adopt $n_0 = (10^4-10^5)$~cm$^{-3}$ \citep{guedel2008, maurri2014, agraamboage2014}. 
From $L_{\rm EUV}$ and an intermediate EUV photon energy of $6.6\times 10^{-11}$~erg (corresponding to 30~nm wavelength), we find the photon luminosity $\Phi_{\rm EUV} = 2\times 10^{40}$~photons per second and
obtain  $R_{\rm S} \approx (1-5)\times 10^{14}~{\rm cm} \la 40~{\rm au}$. This should be considered an upper limit in view of even higher densities (up to $10^6$~cm$^{-3}$) in cool ($< 10^4$~K) low-velocity regions of the outflow \citep{giannini2019}, especially in the innermost 0\farcs 5 of the star \citep{maurri2014}.
The maximum ionization distance is thus much smaller than the observed ionized-jet length. These results mirror estimates for ionizing radii for protoplanetary disk atmospheres in \citet{hollenbach2009}.
We conclude that ionization along the jet out to $\approx (400-600)$~au by EUV photons from DG Tau is not a valid model.

\subsubsection{Stellar X-ray photoionization} 
Turning to X-rays from the star, we make two estimates. 
First, irrespective of the general ionization degree of the jet, we consider the maximum X-ray absorption along the jet. Again assuming a gas number density of $n_{\rm H} = (10^4 - 10^5)$~cm$^{-3}$ and considering the extreme case of 100\%\ neutral material along the line of sight to the neon line emitting source, we find the hydrogen absorption column density as $N_{\rm H} \approx n_{\rm H} \ell$ where $\ell$ is the length of the jet. The blue neon jet reaches out to 3\arcsec\ in projection, which corresponds to $\ell \approx 600$~au along the true jet axis. Therefore, $N_{\rm H} = (0.9-9)\times 10^{20}$~cm$^{-2}$. This still allows X-rays even below 1~keV to reach the most distant regions of the jet.

In photoionization equilibrium, ionizations and recombinations in an irradiated unit volume balance,
\begin{equation}
    \frac{\Phi_{\rm X}}{4\pi d^2}n_{\rm Ne^0} \sigma_{\rm Ne}
    = \alpha_{\rm Ne^+} n_{\rm e} n_{\rm Ne^+},
\end{equation}
where  $\Phi_{\rm X} = L_{\rm X}/\epsilon \approx 6.3\times 10^{38}$~photons~s$^{-1}$ is the stellar X-ray photon luminosity and $\epsilon \approx 1~{\rm keV} = 1.6\times 10^{-9}$~erg is the energy of a typical stellar X-ray photon.
The distance from the star is denoted by $d$, and $n_{\rm Ne^0}$ and $n_{\rm Ne^+}$ are the number densities of neutral and singly ionized neon, $\sigma_{\rm Ne} \approx 0.24\times 10^{-18}$~cm$^{-2}$ is the ionization cross section of neutral neon (\citealt{marr1976} for 1~keV photon energy), $\alpha_{\rm Ne^+} \approx 2.2\times 10^{-13}$~cm$^{3}$~s$^{-1}$ is the recombination rate coefficient for Ne$^+$ at $10^4$~K (from Eqs.~4 and 5 and Table 2 in \citealt{shull1982}) and $n_{\rm e} \approx 10^4$~cm$^{-3}$ is the electron density as observed \citep{maurri2014}. We find, at an intermediate distance of 300~au along the jet, $n_{\rm Ne^+}/n_{\rm Ne^0} \approx 2.7/n_{\rm e} \approx 2.7\times 10^{-4}$, which leaves neon largely neutral; at 600~au where we still see strong [Ne\,{\sc ii}] emission, the fraction is $\sim 7\times 10^{-5}$. This estimate is simplified because K-shell ionization with hard photons also produces Ne$^{2+}$ that can form Ne$^{+}$ via charge exchange (which we ignore for an order-of-magnitude estimate). However, for the same reason as given above for the Str\"omgren radius estimate, higher densities in the innermost region again make our estimate an upper limit. On the other hand,
given the observed high ionization degree of the jet gas \citep{maurri2014}, the ionization degree of neon is expected to be even higher than the ionization degree of hydrogen (see the argument in \citealt{glassgold2007} after their Eqs. 2-13) if the gas is photoionized. Thus, we reject any model where Ne is photoionized by stellar X-rays.

\subsubsection{Radiative cooling of thermally heated shocks}
Shocks at sufficiently high velocity can heat the gas up to millions of K; there is evidence of shock heating from spatially resolved observations and numerical simulations (see discussions in \citealt{maurri2014}, \citealt{liu2016}, \citealt{massaglia2005}) and specifically also from the presence of X-ray emitting plasma in the jets \citep{guedel2008}. To thermally produce Ne$^{+}$ and Ne$^{2+}$ ions via collisional ionization, temperatures between $\sim 1.2\times 10^4$~K and $\sim 5\times 10^5$~K are required if the gas is in collisional ionization equilibrium (CIE; \citealt{mazzotta1998}). We propose that the observed [Ne\,{\sc ii}] and [Ne\,{\sc iii}] jet emission is related to cooling, radiative shocks  that are initially seen as an X-ray emitting plasma jet at $\sim$5~MK.

We do not consider the buildup of a new shock (see, e.g., \citealt{gnat2011}) but assume, for a plausibility estimate,  a simple 1D stationary shock in which a continuous gas flow enters shock front where the plasma is heated up to a temperature $T_s$ given by the shock velocity, $v_s$,
\begin{equation}
    T_s = \frac{3\bar{m}}{16k}v_s^2
\end{equation}
\citep{mckee1980} where $\bar{m}$ is the mean mass of a particle and  $k$ is the Boltzmann constant. For ionized gas, we adopt $\bar{m} = 0.61\times 1.67\times 10^{-24}$~g $\approx 10^{-24}$~g. For $v_s = 400$~km~s$^{-1}$ and $v_s = 600$~km~s$^{-1}$, $T_s \approx 2.2$~MK and 5~MK, respectively. The former value is incompatible with the X-ray-derived average temperature, also because this temperature is an average over the entire shock-heated, X-ray-emitting plasma volume which is not well resolved in X-ray observations. Given that this average is $\approx 3.7-4.4\times 10^6$~K \citep{guedel2008}, we adopt $T_s = 5$~MK as a reasonable temperature estimate at the shock jump and therefore use $v_s = 600$~km~s$^{-1}$ as a standard. Following the Rankine-Hugoniot jump conditions for a compressible gas, the density increases by a factor of 4 at the shock jump, while the velocity decreases by a factor of 4 to 150~km~s$^{-1}$. This should not be mistaken for the observed velocity in the shocked gas because the shock front itself may move along the jet, in which case the two velocities must be co-added.
Given the observed densities on the order of $\ga 10^4$~cm$^{-3}$ in the jet \citep{maurri2014, giannini2019}, we adopt a pre-shock density of $10^3$~cm$^{-3}$ which increases in the shock by a factor of 4 following the Rankine-Hugoniot jump conditions for a compressible gas, and further increases as the shock gas cools.

\begin{figure*}[t!]
\centering
\vbox{
\vskip -15.9cm
\includegraphics[width=1.05\textwidth,angle=0,trim={0 0 0 18 cm},clip]{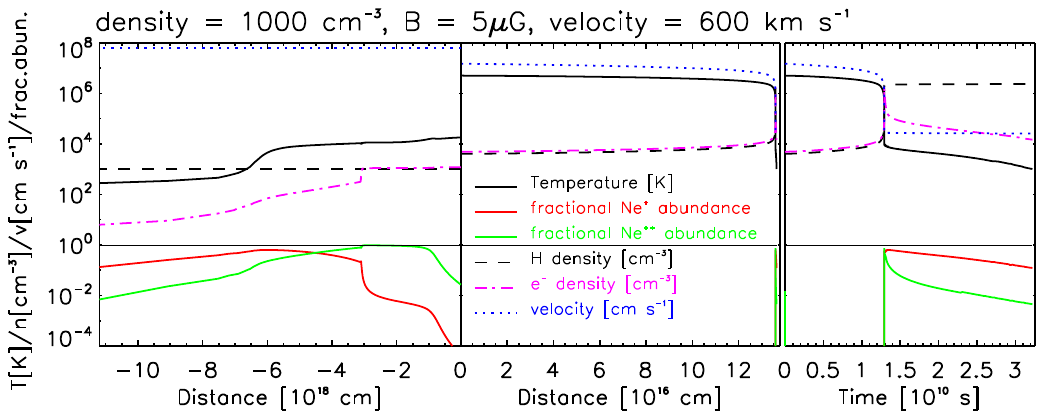}\\
\vskip -18cm
\includegraphics[width=1.05\textwidth,angle=0,trim={0 0 0 18 cm},clip]{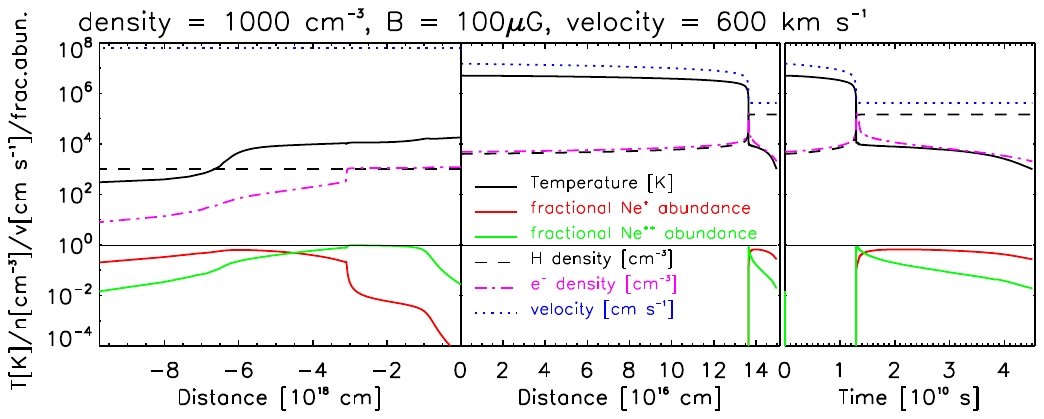}
}
\vskip -2.4cm
\caption{Models from the \citet{allen2008} shock model library. \textit{Top:} For a pre-shock magnetic field strength of 5$\mu$G.
\textit{Bottom:} For 100$\mu$G. 
\textit{Left panels}: Precursor region parameters as a function of distance from the shock jump; the latter is at distance = 0. 
\textit{Middle panels:} Shock region parameters as a function of distance from the shock jump.
\textit{Right panels:} Shock region parameters as a function of time since shock front passage. The individual curves are defined in the middle panels. -- The black horizontal line at $10^0$ marks the unity level for the fractional abundances.}\label{fig:shockmodels}
\end{figure*}
The shocked gas 
cools radiatively behind the shock and therefore each ionized element recombines to lower ionization stages as time proceeds, or as the gas moves away from the shock in the downstream direction. Line radiation in the cooling shock was modelled by, for example, \citet{hollenbach1989} and, including [Ne\,{\sc ii}], \citet{hollenbach2009}.

The IR neon line luminosity integrated along the entire stationary shock, or equivalently, the time-integrated neon luminosity of a moving slab in the jet during its entire cooling process then corresponds to the observed total neon line luminosity. 

It is, however, important to recognize that CIE conditions do not hold during the entire shock cooling process because once the gas temperature cools to $\la 10^6$~K, the time scale of cooling becomes shorter than the recombination time scale of individual ions, meaning that the gas is over-ionized compared to CIE conditions; the gas is cooling under Nonequilibrium Ionization (NEI) conditions. This itself changes the radiation processes and therefore modifies the cooling function. For details on the NEI shock model, we refer the reader to Appendix~\ref{NEI}.

\begin{figure}[t!]
\centering
\vskip -3.15cm
\hbox{\hskip -1cm\includegraphics[width=0.59\textwidth,angle=0,trim={0 0 0 18 cm},clip]{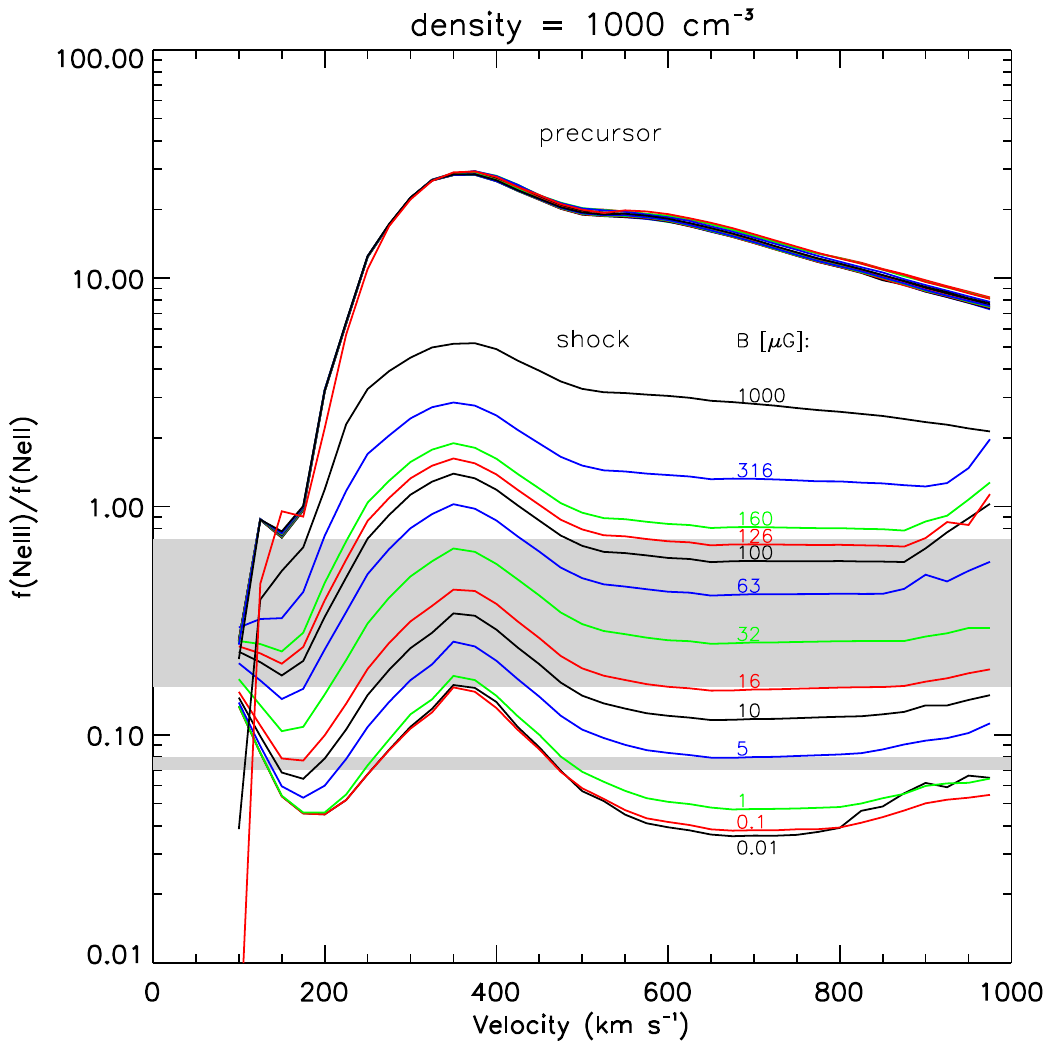}}
\vskip -1.7cm
\caption{
[Ne\,{\sc iii}]~15.55~$\mu$m/[Ne\,{\sc ii}]~12.81~$\mu$m
flux ratios in 1D shock models for a precursor density of 1000~cm$^{-3}$, as a function of shock velocity for various models with different magnetic field strengths from $B = 0.01~\mu$G to 1000$~\mu$G.  The precursor ratios are shown by the nearly coincident curves for all $B$ values at the top,
while the lower curves are for the shock region. The gray areas mark the range of observed flux ratios, the upper broad region being for the red jet and the lower, narrow bar for the blue jet. A shock velocity of $\sim$~600~km~s$^{-1}$ is compatible with the X-ray derived plasma temperature behind the shock, and also the maximum measured flow velocities in the lines (Fig.~\ref{fig:fullspectrum}).}\label{fig:neonratios}
\end{figure}

We use the shock model library of \citet{allen2008} using the \texttt{MAPPINGS III} code, an update of \texttt{MAPPINGS II} \citep{sutherland1993}, to estimate 
[Ne\,{\sc iii}]~15.55~$\mu$m/[Ne\,{\sc ii}]~12.81~$\mu$m flux ratios. 
The \texttt{MAPPINGS} models also include a magnetic-field strength parameter that essentially determines the compressibility of the gas throughout the shock (see Appendix~\ref{NEI}).
We apply the model set using a precursor density of 1000~cm$^{-3}$ with magnetic field strengths between 0.01~$\mu$G and 1000~$\mu$G (models `S' in Table 2 of \citealt{allen2008}) and shock velocities between 100~km~s$^{-1}$ and 1000~km~s$^{-1}$, focusing on $v = 600$~km~s$^{-1}$. 

Figure~\ref{fig:shockmodels} shows two simulations for the two magnetic field strengths  of 5~$\mu$G and 100~$\mu$G. The left panels show the precursor region as a function of distance (negative) from the shock jump which is at distance = 0. The middle panels show the same for the shock region; the right panels show again the shock region but here as a function of time since a gas element has passed the shock front. These latter plots reveal the late phase of the shock better because the gas velocity is decreasing continuously. The individual curves refer to  the temperature (solid black), the fractional Ne$^+$ and Ne$^{2+}$ abundances relative to all neon (solid red and green, respectively), the total hydrogen and electron densities (dashed black and dot-dashed purple, respectively), and the velocity relative to the shock front (dotted blue).

The figures reveal the long phase of the high-temperature plasma regime before a rapid drop from $T \le 10^6$~K to $\sim$10$^4$~K followed by the further cooling photoabsorption/recombination zone, as described in Appendix~\ref{NEI}. The velocity drops toward the most distant region of the shock while the densities increase, depending on the magnetic field, to $10^5-10^7$~cm$^{-3}$. At least the former value is compatible with observations \citep{maurri2014}. The fractional abundances of Ne$^{+}$ and Ne$^{2+}$ reach values close to unity in the nonequilibrium and photoabsorption/recombination regions. Under CIE conditions, these ions would be prominently present in substantial amounts ($\ga$10\%) only for $2\times 10^4$~K~$\la T \la 10^5$~K and $4.5\times 10^4$~K~$\la T \la 2\times 10^5$~K, respectively (see \citealt{mazzotta1998}), while under NEI conditions they emerge only with a delay, but do remain present, especially Ne$^{+}$, for extended periods of time at $T \approx 10^4$~K and below before recombining. The magnetic field has the effect of extending the relative length of the photoabsorption and recombination zone with regard to the hot plasma zone. This is a consequence of magnetic pressure support that counteracts compression of the plasma and strong deceleration of the flow. Some deceleration is observationally supported; in the blue jet, our [Ne\,{\sc ii}] images suggest that intermediate velocities reach to larger distances from the star than the highest velocities; in the red jet, the most distant gas is of intermediate or low velocity (Figs.~\ref{fig:RGBimages}, \ref{fig:pv}, and \ref{fig:ratio1}--\ref{fig:ratio8}).

Figure~\ref{fig:neonratios} shows the resulting, integrated 
flux ratios of [Ne\,{\sc iii}]~15.55~$\mu$m/[Ne\,{\sc ii}]~12.81~$\mu$m as a function of the shock velocity, for various magnetic field strengths from 0.01~$\mu$G to 1000~$\mu$G, both for the shock (in the lower part of the figure) and the precursor (nearly coincident curves in the upper part of the figure).
The gray areas mark the approximate, observed flux ratio ranges (upper area for red jet, lower area for blue jet, see Figs.~\ref{fig:ratio1}--\ref{fig:ratio8}). We see that (i) for the red jet, many shock models produce the observed flux ratios for a shock velocity of $\approx 600$~km~s$^{-1}$ as long as the magnetic field strength is less than $100-130~\mu$G.  
The range of magnetic field strengths for which the model [Ne\,{\sc iii}]/[Ne\,{\sc ii}] flux ratios agree with the range of observed ratios is roughly $5-120$~$\mu$G. 
This is a good match with  $\sim$20~$\mu$G reported for a knot
observed with LOFAR and estimated based on the Razin effect for synchrotron emission, albeit at a larger projected distance of 3000~au \citep{feeney2019}. 
(ii) The blue jet shows considerably lower ratios that are represented by the lowest magnetic field strengths of only a few microGauss ($\mu$G). This range of agreement motivates the choice of models presented in Fig.~\ref{fig:shockmodels}.
(iii) The modelled precursor Ne flux ratios, reaching values of $\sim$20, grossly contradict the observed ratios $< 1$ in the entire jet, however. 

The discrepancy with the precursor ratios is perhaps not surprising.  The shock models assume pristine, cool material entering the shock front; the precursor temperature and ionization fractions are self-consistently calculated in photoionization equilibrium with the high-energy emission of the shocked material. This is not supported by the observations. The X-ray jet is most intense only a few tens of au from the star ($\sim$0\farcs 15 in projection; see 
\citealt{schneider2008, guedel2011, Frank2014}). Initial heating to $\sim 5\times 10^6$~K therefore takes place immediately above the disk, which leaves no space for an idealized, extended precursor. It is also clear that further shocks exist along the jet \citep{maurri2014, liu2016}, and these would be fed by pre-processed, shocked and cooling material, again not leaving space for a standard precursor. We therefore suggest not to consider precursor models for the DG Tau jet shocks. 

The presented 1D models are discussed here only as a rough approximation to assess if shocks are plausible to explain the observed neon flux ratios. The models have severe limitations with regard to jet shocks, however. First, the assumption of 1D prevents shocked gas from expanding laterally, which would be expected given the overpressure in the high-temperature radiation zone (increase of density by a factor of $\sim$4 and of temperature by a factor of $\sim$500 for fully compressible gas). Lateral expansion is evident in the DG Tau jet that shows a half-opening angle of at least $14\degr-15\degr$ \citep{agraamboage2011, white2014b} already within 1\arcsec\ of the star. The same holds for the X-ray jet \citep{guedel2008}. Such expansion prevents the gas from attaining extreme densities during the late cooling phase, but also extends the emission region further downstream because the lowered density increases the cooling and recombination times, while cooling due to expansion will become effective especially for the hot, dense phase \citep{guedel2008}.

The shock velocity of $\sim$600~km~s$^{-1}$ adopted in our simple 1D model also deserves some critical review. Shock velocities inferred from optical line ratios are often $<$100~km~s$^{-1}$ even though the velocity of the jet flow relative to the star may reach many 100~km~s$^{-1}$ (e.g., \citealt{hartigan1994}). Specifically for DG Tau, \citet{lavalley2000} find shock velocities of 70--100~km~s$^{-1}$. While we note that the lines used for such estimates are low-excitation and low-ionization lines of O, N, S, H$\alpha$, and H$\beta$, jets show a range of velocities in an ``onion-like'' arrangement, the fastest parts of the jet being closest to the jet axis \citep{bacciotti2000}. Still, shock velocities may be lower than the jet flow velocity unless the shock is stationary; DG Tau hosts at least one stationary shock close to the star \citep{schneider2008, guedel2011, schneider2013b}. The question is therefore which part of the jet is fast enough to produce shock velocities that can explain our observations. This issue was previously addressed by \citet{guenther2009} who concluded that a small fraction of the jet flow reaches sufficiently high velocities to explain the observed X-rays via shock heating, while the slower bulk flow does not generate X-ray emission (see also \citealt{schneider2008}). This is compatible with the distribution of velocities in the jet as seen in the atomic lines (Fig.~\ref{fig:fullspectrum}). However, a deeper analysis would require higher spatial resolution both in X-rays and in the atomic lines discussed in this paper, together with higher-dimensional models that allow for multiple shocks at different velocities. Such models would require  observational input details presently not at hand.
The main argument in support of the adopted temperature therefore remains the X-ray measured peak temperature of $\sim$5$\times 10^6$~K and the observed maximum velocities in the atomic lines; these velocities may be surpassed by flows at still higher velocities containing only a small fraction of the jet mass. 

While other lines from ionized species in the jet are available, we refrain from
studying line ratios between different elements. Doing so would require accurate knowledge of the element abundance ratios in the jet, which we do not have. For example, fractionation in the jet heating and acceleration process may alter the
element ratios from the initial values in the disk. Neon in particular has been suspected to be significantly more abundant in Sun-like stars than standard estimates based, for example, on solar energetic particles or hot-star spectroscopy suggest (e.g., \citealt{anders1989, asplund2005}); such elevated neon abundances as derived from stellar X-ray measurements \citep{drake2005} have been proposed to also be used in shock models \citep{gnat2009}, which is not the case for the \texttt{MAPPINGS III} models of \citet{allen2008} used here.
Iron could have been released by evaporating dust, which would require information on the dust abundance at the base of the jet (we refer to studies of the depletion of Fe onto grains at the base of the DG Tau jet, by \citealt{agraamboage2011} and \citealt{giannini2019}; such depletion can alter the Fe abundances by up to 1.5 orders of magnitude in DG Tau).

The jet cooling lengths of the models are on the order of $10^{17}$~cm; this is an order of magnitude more than suggested by observations. Because $n_{\rm pre}t$ is approximately invariant (see above), a pre-shock density increase to $10^4$~cm$^{-3}$ would bring the jet length into agreement, but we wish to avoid optical depth effects in the late cooling phase of the jet and again point to the limitations introduced by 1D, for example due to the lack of lateral expansion and consequent additional cooling \citep{guedel2008}.

\section{Conclusions}
\label{sec:conclusion}
We comprehensively analyzed and interpreted the content of the MIR images and spectra of the transition Class I/Class II object DG Tau obtained by MIRI/MRS. DG Tau is widely recognized as a key young stellar object because it maintains some properties of Class I protostars, but without the extinction and absorption by a surrounding molecular envelope that typically prevent access to the immediate stellar environment at optical, IR, and X-ray wavelengths in normal Class I protostars. DG Tau is surrounded by a young protoplanetary disk, but there are also remnants of the molecular envelope from which the disk accreted material. We studied the disk mineralogy, disk gas composition, and the jet dynamics and ionization of the system and put them into context of earlier observations, which include observations at other wavelengths. The main results of our study can be summarized as follows:
\begin{itemize}
    \item The continuum flux in the MIR is very high, reaching levels of $10-15$~Jy at wavelengths of $15-25$~$\mu$m. This may be a signature of elevated UV irradiation due to strong mass accretion from the disk onto the star.
    \item The spectrum shows several silicate features that can be decomposed into contributions of different minerals in the protoplanetary disk. We found a predominance of amorphous Mg$_2$SiO$_4$ and MgSiO$_3$ (up to a total of 95\%), but also a significant presence of crystalline forsterite (up to 14\%). The dominant fraction of the dust has grown to sizes on the order of 2~$\mu$m despite the very early evolutionary stage of this disk. This indicates that dust growth occurs on short time scales of a few 10$^5$~yrs.
    \item The gas disk reveals itself in a spectrum very rich in H$_2$O and CO lines, lines of OH, H~{\sc i}, and H$_2$ (the latter likely from winds) but displays no other molecular lines. Our analysis of the CO and H$_2$O lines indicates very high temperatures of $600-2000$~K, obviously related to the innermost disk that we suggest is strongly irradiated by UV irradiation from accretion hotspots. We note that DG Tau is subject to a high accretion rate of $\sim$10$^{-7}~M_{\odot}$~yr$^{-1}$.
    \item The emission of H$_2$ is partly extended, approximately co-spatial with millimetric CO emission. We interpret the elliptically shaped traces of emission as regions of a conical outflow wall left over from the Class-I envelope stage; its radial velocity decreases with increasing distance from the star, suggesting a wind decelerating  outward, oriented toward the observer, and shocking at the cavity walls.
    \item DG Tau reveals a complex bipolar jet, the two jets being largely different from each other. Mapped in lines of singly or doubly ionized metals, the blue jet is collimated, while the red jet is complex, showing a structure in which the gas flow close to the axis is the fastest and most collimated, while slower components wrap around it with larger opening angles. 
    The redshifted jet, so far not described at scales of several arcseconds, may be related to features within $\sim$1\arcsec\ of the star observed about 16~yrs ago.
    \item The prominent spectral lines of [Ne\,{\sc ii}], [Ne\,{\sc iii}], 
    [Ar\,{\sc ii}], [Fe\,{\sc ii}], and [Ni\,{\sc ii}] reveal radial velocities of up to at least 400--450~km~s$^{-1}$, corresponding to flow velocities along the jet of up to at least 500--570~km~s$^{-1}$. The presence of ions such as Fe 
    in the jet may be interesting for future studies because the atoms most likely were part of dust before jet launching and heating, and may therefore provide information on dust composition.
    \item We studied the origin of the abundant Ne$^{+}$, Ne$^{2+}$, Ar$^{+}$,  Fe$^{+}$, and Ni$^{+}$ ions, focusing on neon ions. Sufficient photoionization by the stellar EUV or X-rays is not feasible given the distance from the stellar source. We propose that the ions are related to cooling shocks suggested by strong, spatially resolved X-ray jets of DG Tau. The 1D models using nonequilibrium ionization evolution during the cooling process provide a good description of the line ratios for shock velocities around 600~km~s$^{-1}$ with at best moderate magnetic fields that provide some support of the plasma against contraction during the late cooling phase. These models are simplified but give a plausibility assessment for linking the high-energy jet with the ionic jet 
    observed in the MIR. The models do not account for lateral expansion due to overpressure in the hot region just behind the shock front, and also do not consider multiple shocks in the jet either in parallel or in sequence behind each other, given missing information on spatial scales not accessible by present-day observations. 
\end{itemize}
It is evident that the strong emission originating from ions is not primarily related to X-ray irradiated disk surfaces (e.g., \citealt{glassgold2007}), but originates from gas in the energetic jets, most likely related to strong shocks of fast material, confirming the assertion in \citet{guedel2010} that may analogously hold for other T Tauri stars with strong MIR neon, argon, and iron line emission.

DG Tau deserves further scrutiny, specifically in the following directions: (i) it is surprising that its inner disk does not reveal any emission from molecules unrelated to CO, hydrogen, H$_2$O, and OH (but seen in other CTTS disks, at least at the sensitivity level achieved with our observations). One possibility is that other molecules reside below the optical depth unity surface defined by the dust, which may be related to strong disk flaring and insufficient dust settling (e.g., \citealt{vanDishoeck2025} for young disks); (ii) higher spatial resolution in lines of metal ions would be helpful to better characterize the locations and the number of shocks; and (iii) such observations could better inform more realistic 3D simulations of jet shock heating and cooling, including sequences of shocks along the jet flow.

\vspace{5mm}

\begin{acknowledgements} 
We thank Mark Allen for giving us access to the numerical shock model library presented in \citet{allen2008}. Antonio Garufi provided the SPHERE image of DG Tau shown in Fig.~\ref{fig:overview} and Fig.~\ref{fig:h2arc} \citep{garufi2024}. Cosima Eibensteiner prepared the ALMA data cubes for CO observations (published in \citealt{guedel2018} and sketched in Fig.~\ref{fig:overview}). 
We thank an anonymous referee who made many significant suggestions that improved the text and the presentation of our results. 
We furthermore acknowledge helpful discussions with Gerhard Hensler and Dieter Breitschwerdt on plasma cooling, and with Almudena Alonso-Herrero on neon line emission.

This work is based on observations made with the NASA/ESA/CSA \textit{James Webb} Space Telescope. The data were obtained from the Mikulski Archive for Space Telescopes at the Space Telescope Science Institute, which is operated by the Association of Universities for Research in Astronomy, Inc., under NASA contract NAS 5-03127 for JWST. 
These observations are associated with program 1282. The following National and International Funding Agencies funded and supported the MIRI development: 
NASA; ESA; Belgian Science Policy Office (BELSPO); Centre Nationale d'\'Etudes Spatiales (CNES); Danish National Space Centre; Deutsches Zentrum f\"ur Luft- und Raumfahrt (DLR); Enterprise Ireland; 
Ministerio De Econom\'ia y Competividad; 
Netherlands Research School for Astronomy (NOVA); 
Netherlands Organisation for Scientific Research (NWO); 
Science and Technology Facilities Council; Swiss Space Office; 
Swedish National Space Agency; and UK Space Agency.

This work was also supported by ADVANCED 149943 and 2023-1.2.4-T\'ET-2023-00056 grants, which have been implemented with the support provided by the Ministry of Culture and Innovation of Hungary from the National Research, Development and Innovation Fund, financed under the NKKP ADVANCED and 2023-1.2.4-T\'ET-2023 funding schemes. This work received funding from the Hungarian NKFIH OTKA project no. K-147380. 
This research was supported in part by grant NSF PHY-2309135 to the Kavli Institute for Theoretical Physics (KITP).
T.K. acknowledges support from STFC Grant ST/Y002415/1.
A.C.G. acknowledges support from PRIN-MUR 2022 20228JPA3A ``The path to star and planet formation in the JWST era (PATH)'' funded by NextGeneration EU and by INAF-GoG 2022 ``NIR-dark Accretion Outbursts in Massive Young stellar objects (NAOMY)'' and Large Grant INAF-2024 ``Spectral Key features of Young stellar objects: Wind-Accretion LinKs Explored in the infraRed (SKYWALKER)''.
Leiden is supported by funding from the  ERC Advanced Grant 
No. 291141 MOLDISK. A.M.A. and E.v.D. acknowledge support from grant TOP-1 614.001.751 from the Dutch Research Council (NWO).

SAOImageDS9 is fully funded by the Chandra X-Ray Science Center (CXC) 
and is licensed in part under the GNU General Public License, version 3.
JS9 is funded by the Smithsonian Institution, the Chandra X-Ray Science Center (NAS8-03060), and NASA's Universe of Learning (STScI-509913). JS9 is distributed under the terms of the MIT License.
\end{acknowledgements} 

\bibliographystyle{aa}
\bibliography{references}

\begin{appendix}

\section{Maps of atomic emission in different velocity channels}\label{appendixfigures}
\begin{figure*}
\centering
\includegraphics[width=0.96\textwidth]{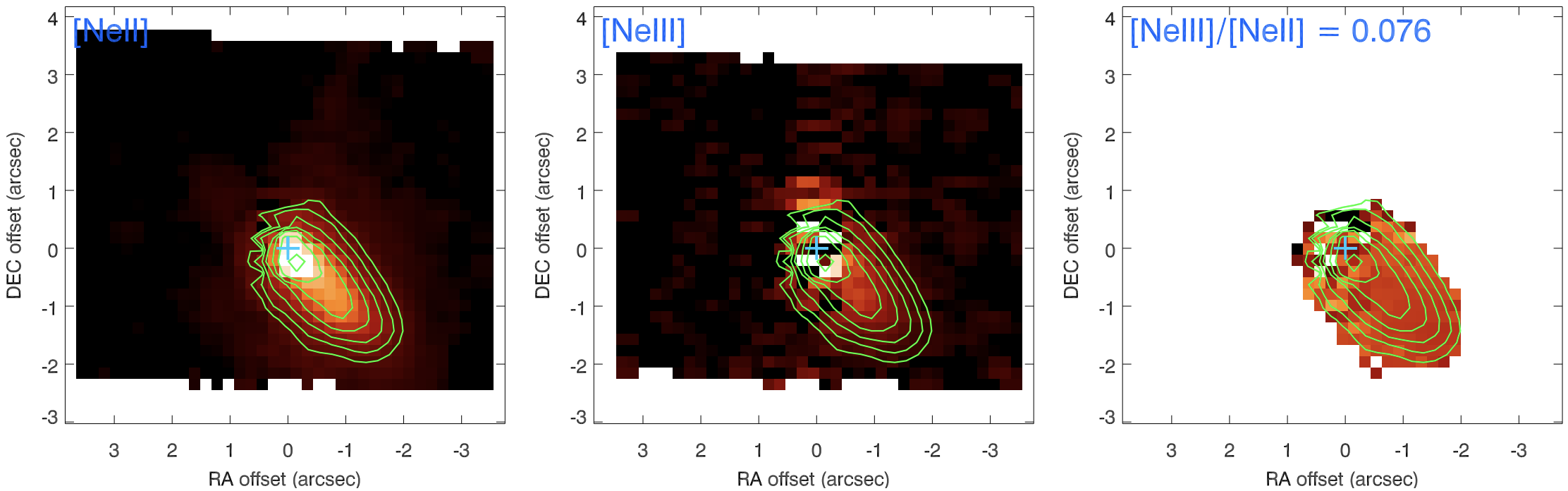}
\caption{\textit{Left:} [Ne~{\sc ii}] flux image for blue jet at 12.80875~$\mu$m and 12.81125~$\mu$m (average), corresponding to an average radial velocity of  $\sim-98$~km~s$^{-1}$ (range: $\sim-128$~km~s$^{-1}$ to $\sim-69$~km~s$^{-1}$).  \textit{Middle:} [Ne~{\sc iii}] flux image at 15.54875~$\mu$m and 15.55125~$\mu$m (average), corresponding to an average radial velocity of  $\sim-113$~km~s$^{-1}$ (range: $\sim-137$~km~s$^{-1}$ to $\sim-89$~km~s$^{-1}$). \textit{Right:}  [Ne~{\sc iii}]/[Ne~{\sc ii}] flux ratio. Areas where the [Ne~{\sc ii]} flux is low are masked. The value at the top of the right panel gives an average flux ratio over the inner contours but excluding the central PSF region from the star+disk. Contours are from the [Ne~{\sc ii}] figure and are overplotted on all three panels to guide the eye.
}\label{fig:ratio1}
\end{figure*}

\begin{figure*}
\centering
\includegraphics[width=0.96\textwidth]{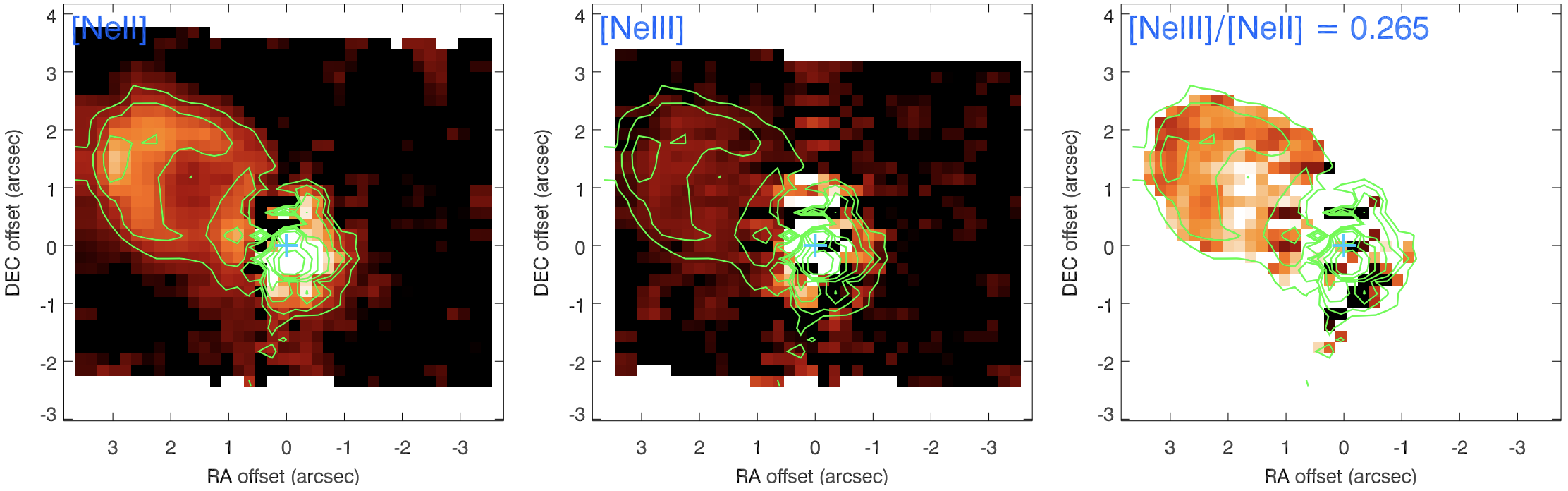}
\caption{Similar to Fig.~\ref{fig:ratio1} but for the arc region of the red jet. 
\textit{Left}: [Ne~{\sc ii}] at 12.81625~$\mu$m, corresponding to a radial velocity of $\sim+48$~km~s$^{-1}$. 
\textit{Middle:} [Ne~{\sc iii}] at 15.55875~$\mu$m corresponding to $\sim+56$~km~s$^{-1}$.}\label{fig:ratio2}
\end{figure*}

\begin{figure*}
\centering
\includegraphics[width=0.96\textwidth]{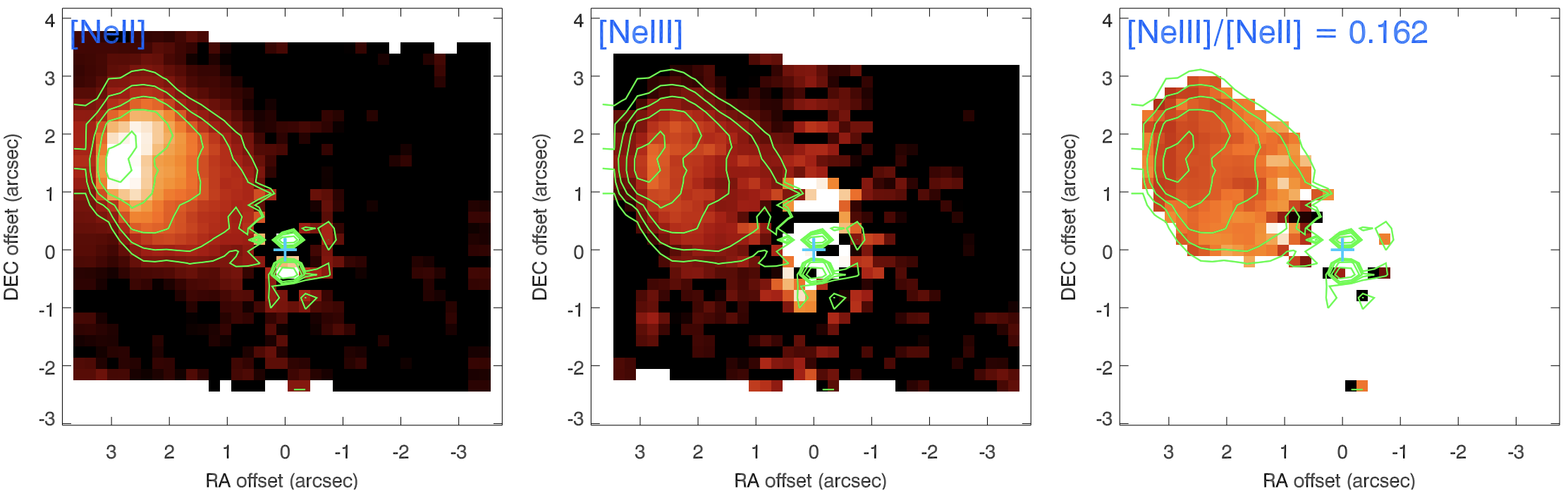}
\caption{Similar to Fig.~\ref{fig:ratio1} but for the arc region of the red jet.
\textit{Left}: [Ne~{\sc ii}] at  12.81875~$\mu$m, corresponding to a radial velocity of $\sim
+106$~km~s$^{-1}$. \textit{Middle:} [Ne~{\sc iii}] at 15.56125~$\mu$m ~$\mu$m corresponding to $\sim+104$~km~s$^{-1}$.}\label{fig:ratio3}
\end{figure*}

\begin{figure*}
\centering
\includegraphics[width=0.96\textwidth]{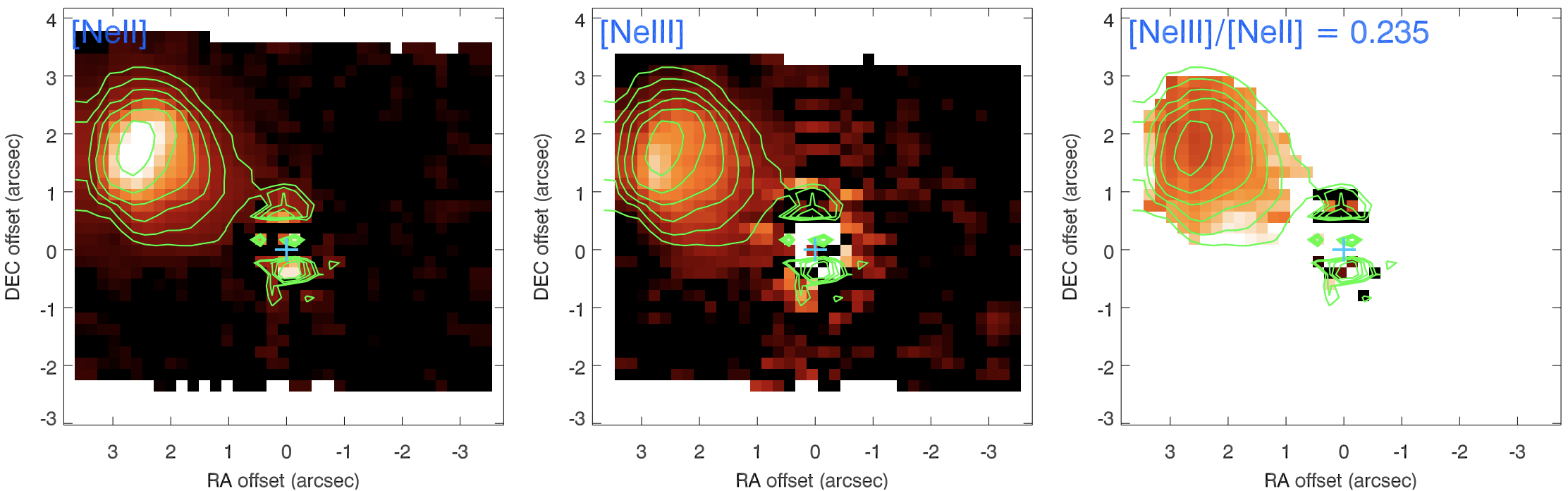}
\caption{Similar to Fig.~\ref{fig:ratio1} but for the blob region of the red jet. \textit{Left}: [Ne~{\sc ii}] at  12.82125~$\mu$m, corresponding to a radial velocity of $\sim+165$~km~s$^{-1}$. 
\textit{Middle:} [Ne~{\sc iii}] at 15.56375~$\mu$m corresponding to $\sim+152$~km~s$^{-1}$.}\label{fig:ratio4}
\end{figure*}

\begin{figure*}
\centering
\includegraphics[width=0.96\textwidth]{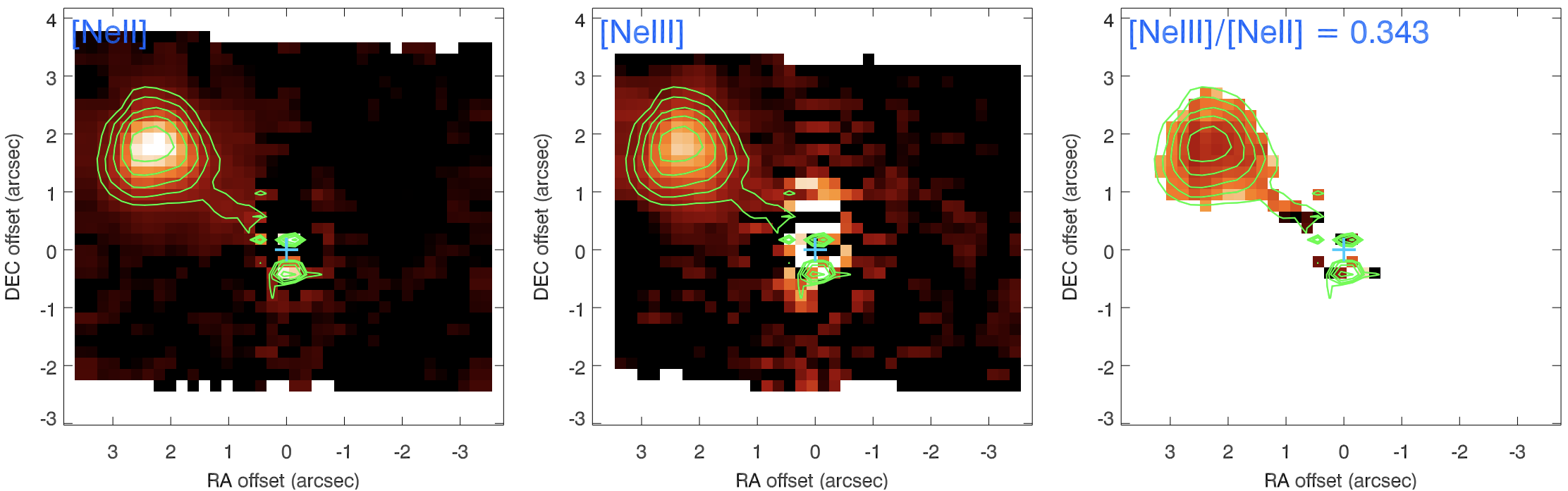}
\caption{Similar to Fig.~\ref{fig:ratio1} but for the collimated jet region of the red jet. 
\textit{Left}: [Ne~{\sc ii}] at  12.82375~$\mu$m and 12.82625~$\mu$m (average), corresponding to a radial velocity of $\sim+253$~km~s$^{-1}$ (range $\sim+223$~km~s$^{-1}-\sim+282$~km~s$^{-1}$). 
\textit{Middle:} [Ne~{\sc iii}] at 15.56875~$\mu$m  corresponding to $\sim+249$~km~s$^{-1}$.}\label{fig:ratio5}
\end{figure*}

\begin{figure*}
\centering
\includegraphics[width=0.96\textwidth]{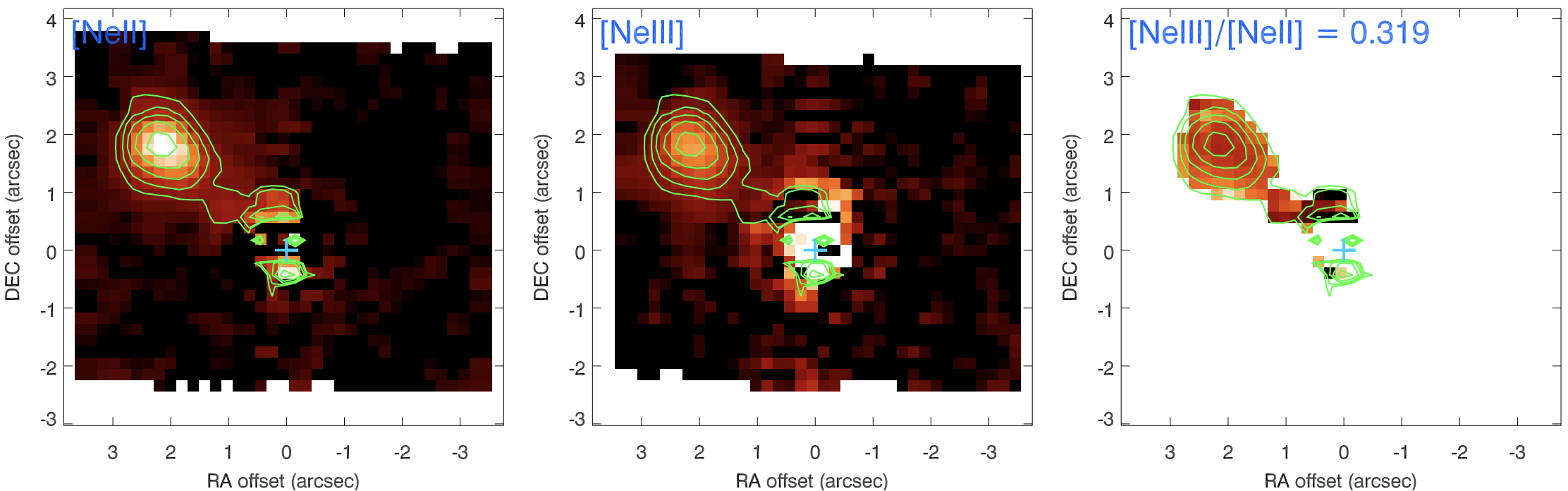}
\caption{Similar to Fig.~\ref{fig:ratio1} but for the collimated jet region of the red jet. \textit{Left}: [Ne~{\sc ii}] at  12.82625~$\mu$m, corresponding to a radial velocity of $\sim+282$~km~s$^{-1}$. \textit{Middle:} [Ne~{\sc iii}] at 15.57125~$\mu$m corresponding to $\sim+297$~km~s$^{-1}$.
}\label{fig:ratio6}
\end{figure*}

\begin{figure*}
\centering
\includegraphics[width=0.96\textwidth]{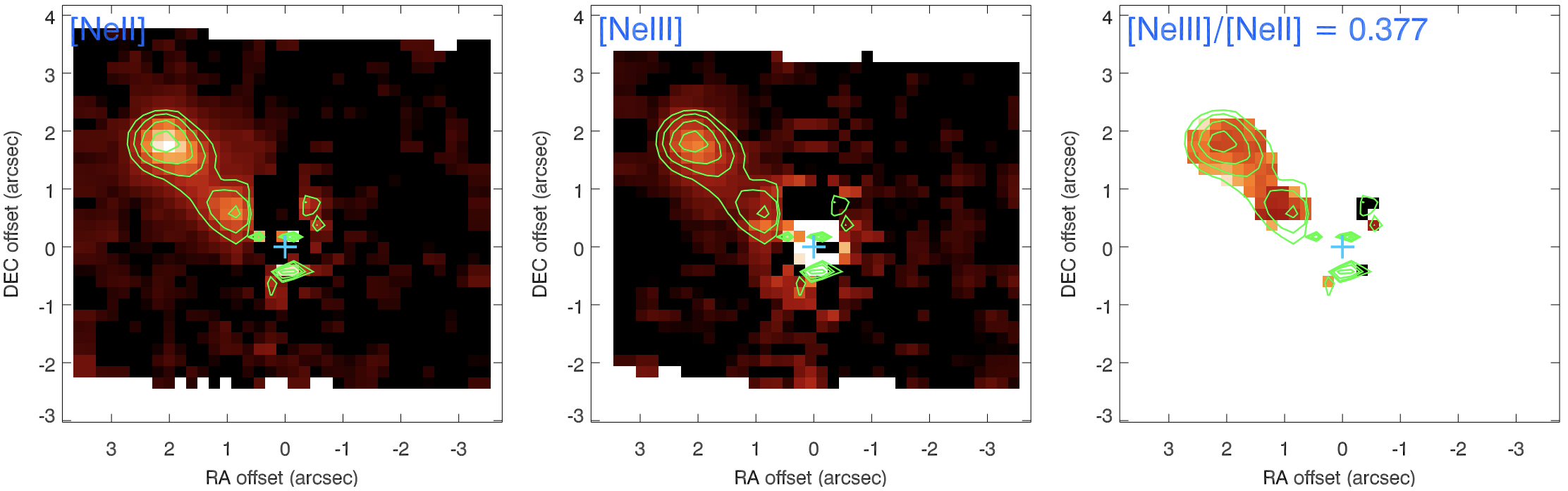}
\caption{Similar to Fig.~\ref{fig:ratio1} but for the collimated jet region of the red jet. 
\textit{Left}: [Ne~{\sc ii}] at  12.82875~$\mu$m, corresponding to a radial velocity of $\sim+341$~km~s$^{-1}$. 
\textit{Middle:} [Ne~{\sc iii}] at 15.57375~$\mu$m corresponding to $\sim+345$~km~s$^{-1}$.}\label{fig:ratio7}
\end{figure*}

\begin{figure*}
\centering
\includegraphics[width=0.96\textwidth]{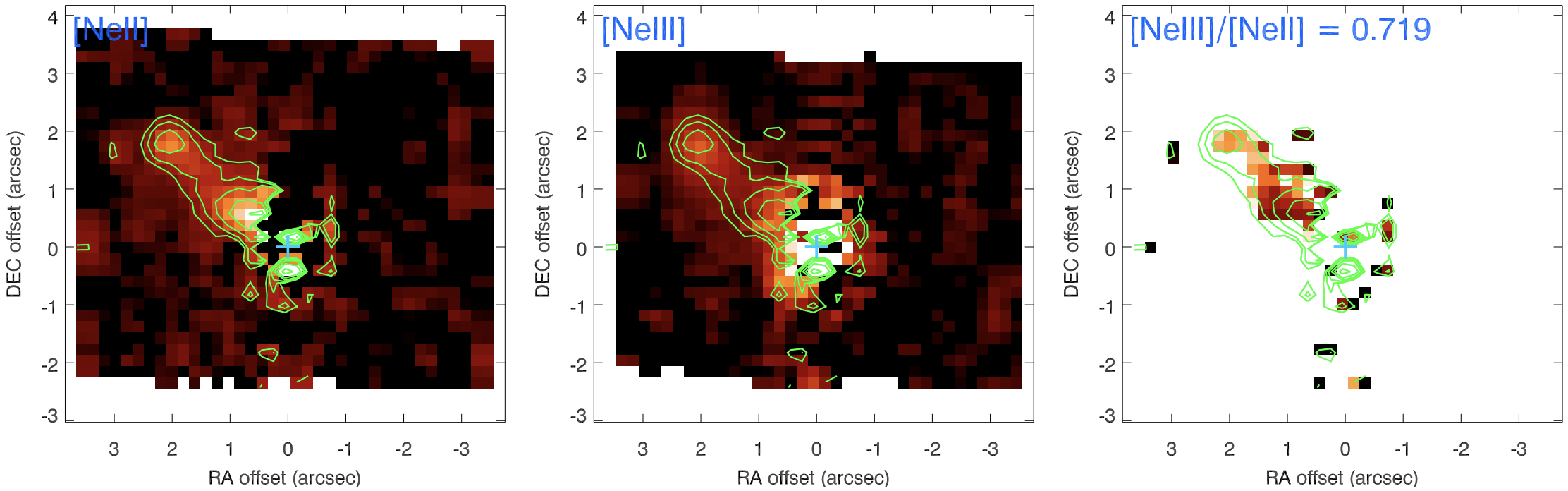}
\caption{Similar to Fig.~\ref{fig:ratio1} but for the collimated jet region of the red jet. \textit{Left}: [Ne~{\sc ii}] at  12.83125~$\mu$m, corresponding to a radial velocity of $\sim+399$~km~s$^{-1}$. \textit{Middle:} [Ne~{\sc iii}] at 15.57625~$\mu$m corresponding to $\sim+393$~km~s$^{-1}$.}\label{fig:ratio8}
\end{figure*}

\begin{figure*}
\centering
\begin{center}
\hbox{
\includegraphics[width=0.25\textwidth, angle=180]{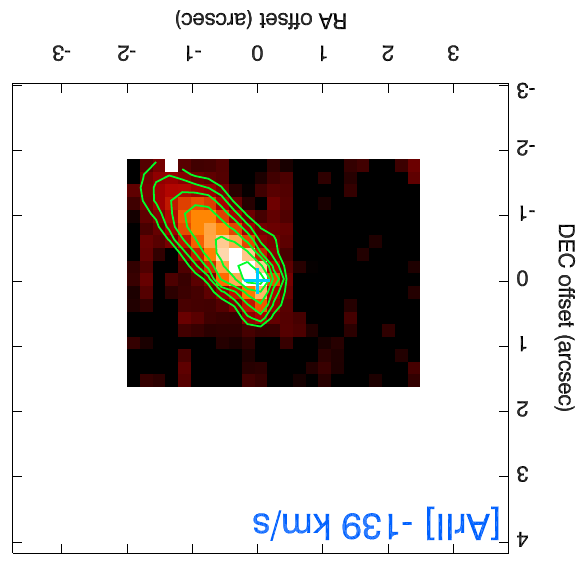}\hskip 1cm
\includegraphics[width=0.25\textwidth, angle=180]{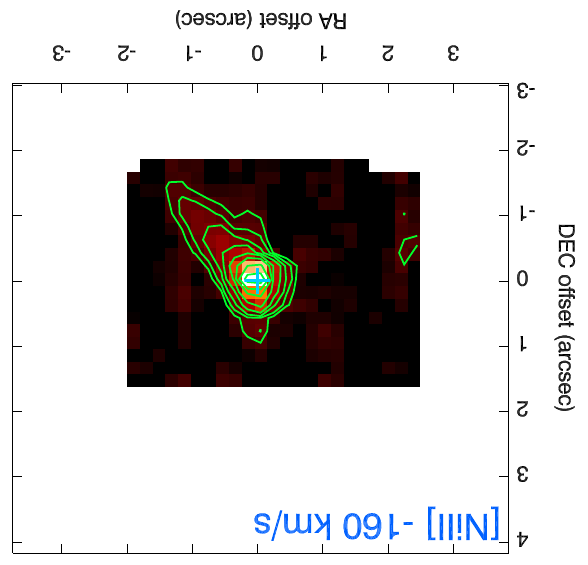}\hskip 1cm
\includegraphics[width=0.25\textwidth, angle=180]{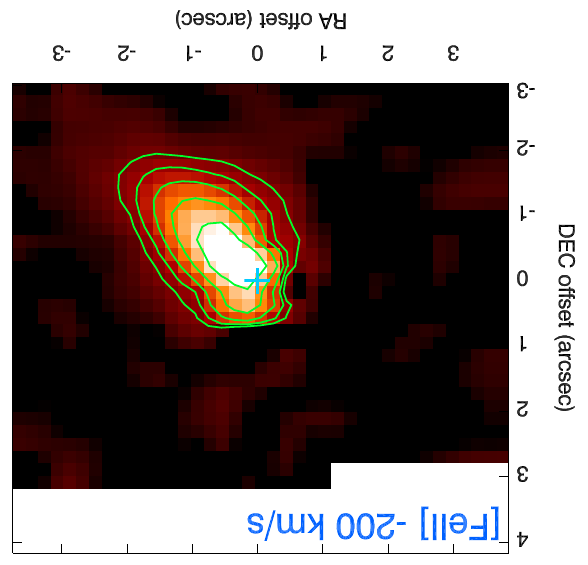}
}
\hbox{
\includegraphics[width=0.25\textwidth, angle=180]{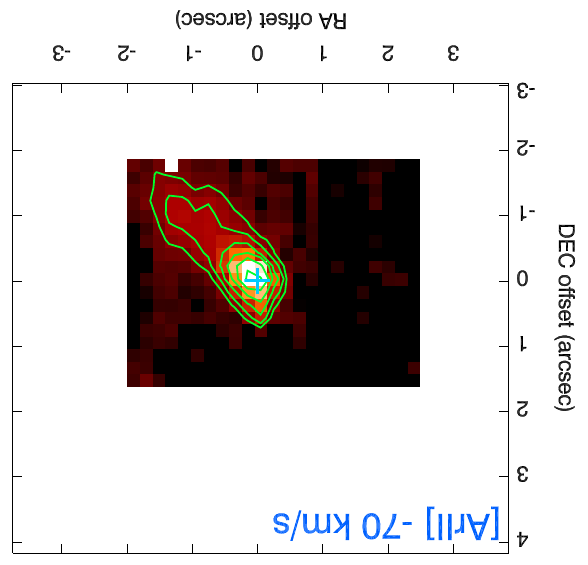}\hskip 1cm
\includegraphics[width=0.25\textwidth, angle=180]{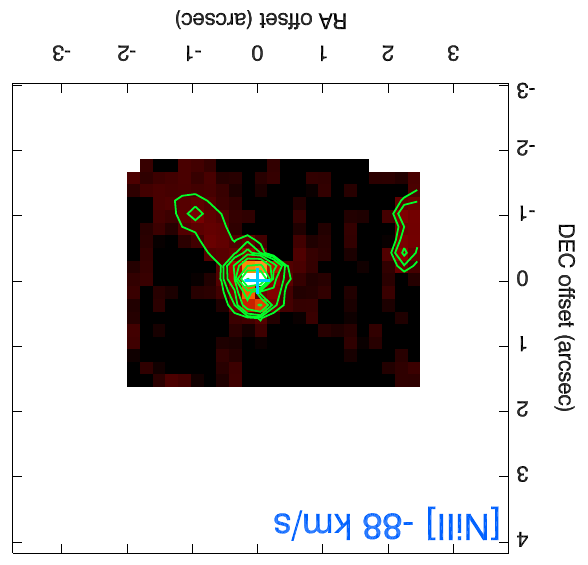}\hskip 1cm
\includegraphics[width=0.25\textwidth, angle=180]{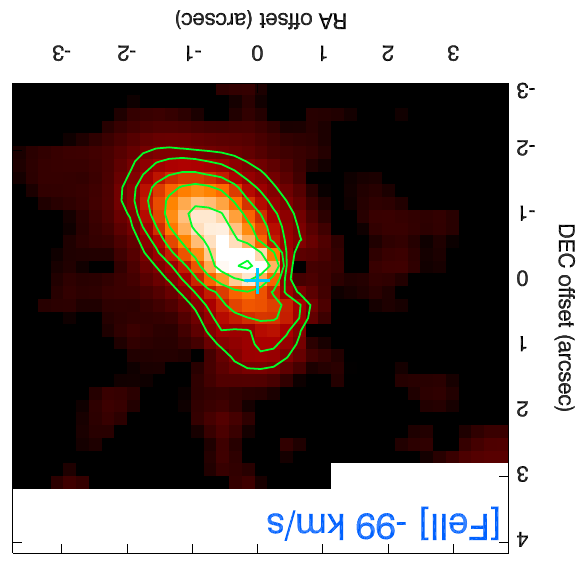}

}
\hbox{
\includegraphics[width=0.25\textwidth, angle=180]{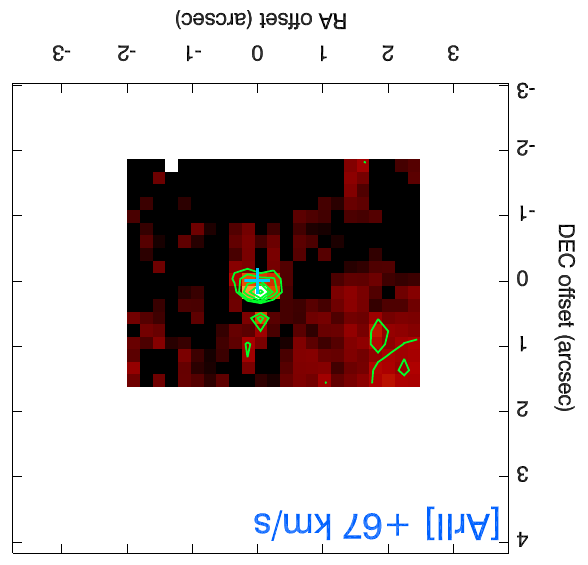}\hskip 1cm
\includegraphics[width=0.25\textwidth, angle=180]{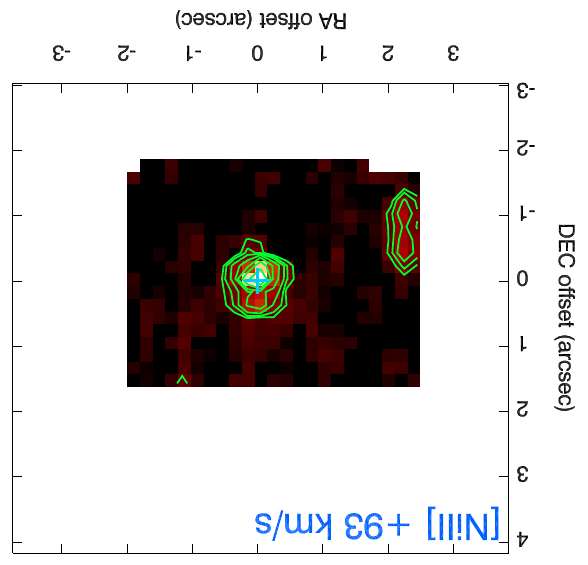}\hskip 1cm
\includegraphics[width=0.25\textwidth, angle=180]{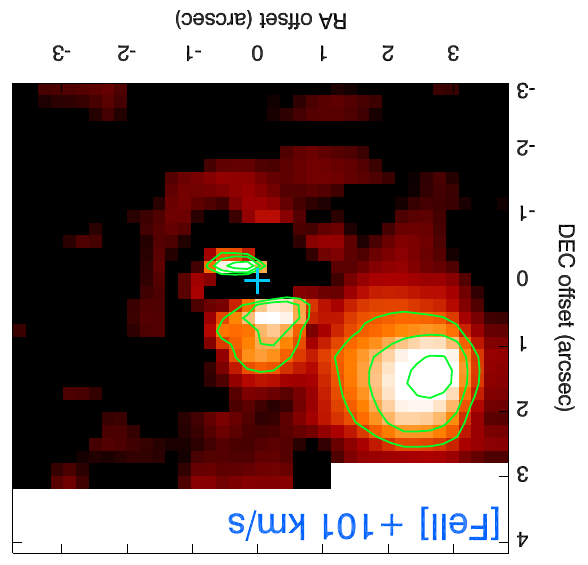}

}
\hbox{
\includegraphics[width=0.25\textwidth, angle=180]{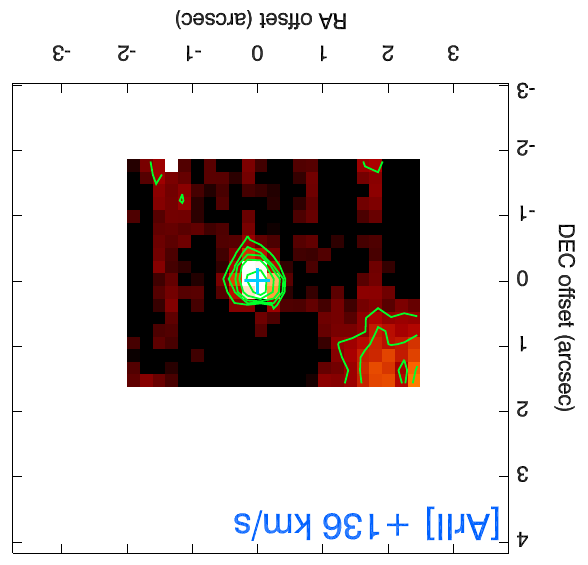}\hskip 1cm
\includegraphics[width=0.25\textwidth, angle=180]{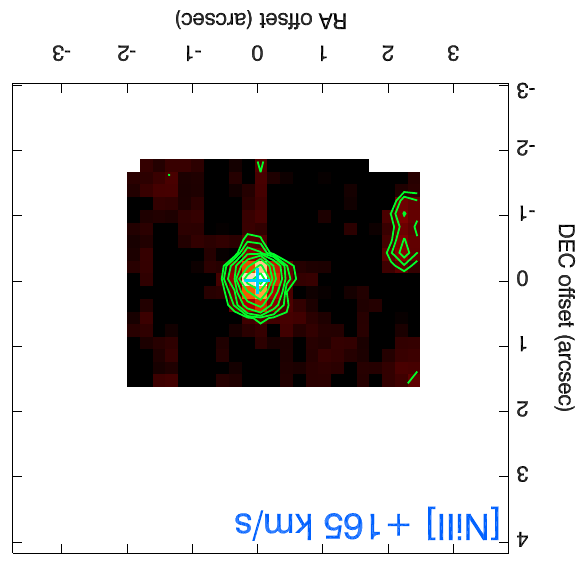}\hskip 1cm
\includegraphics[width=0.25\textwidth, angle=180]{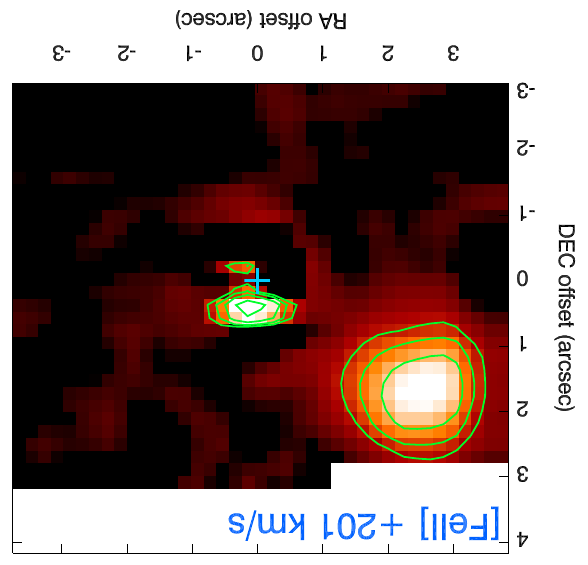}
}
\hbox{
\includegraphics[width=0.25\textwidth, angle=180]{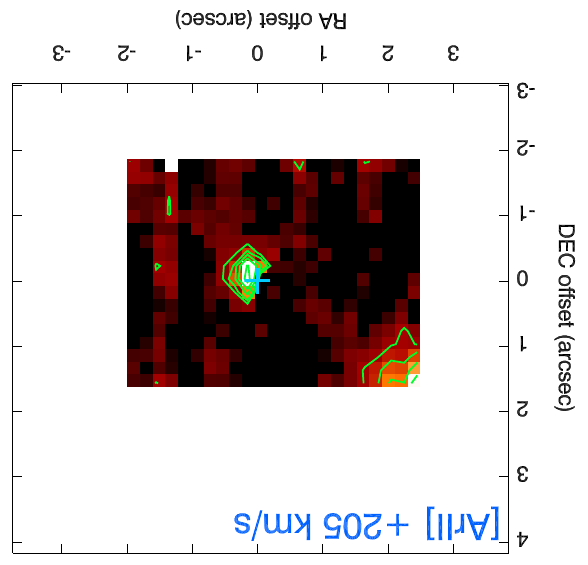}\hskip 1cm
\includegraphics[width=0.25\textwidth, angle=180]{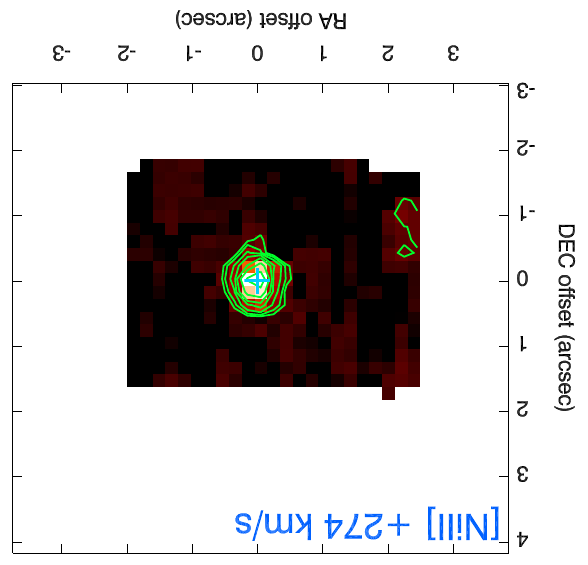}\hskip 1cm
\includegraphics[width=0.25\textwidth, angle=180]{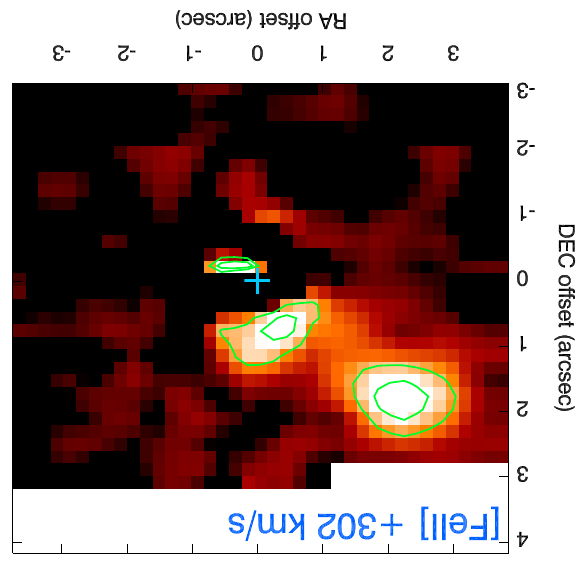}
}
\end{center}
\vskip -.6cm
\caption{\textit{Left:} [Ar~{\sc ii}]~6.98~$\mu$m flux images for different wavelengths resp. radial velocities: from top to bottom,  
6.9824~$\mu$m resp. \nobreak{--139~km~s$^{-1}$}; 
6.9840~$\mu$m resp. --70~km~s$^{-1}$;
6.9872~$\mu$m resp. +67~km~s$^{-1}$;
6.9888~$\mu$m resp. +136~km~s$^{-1}$;
6.9904~$\mu$m resp. +205~km~s$^{-1}$.  \textit{Middle:} Similar for [Ni~{\sc ii}]~6.64~$\mu$m, on the same spatial scale as [Ar~{\sc ii}]; from top to bottom:
6.6328~$\mu$m resp. --160~km~s$^{-1}$;
6.6344~$\mu$m resp. --88~km~s$^{-1}$; 
6.6384~$\mu$m resp. +93~km~s$^{-1}$; 
6.6400~$\mu$m resp. +165~km~s$^{-1}$; 
6.6424~$\mu$m resp. +274~km~s$^{-1}$. 
\textit{Right:} Similar for [Fe~{\sc ii}]~17.94~$\mu$m, on the same spatial scale as [Ar~{\sc ii}]; from top to bottom:
17.925~$\mu$m resp. --200~km~s$^{-1}$;
17.931~$\mu$m resp.  --99~km~s$^{-1}$; 
17.943~$\mu$m resp. +101~km~s$^{-1}$; 
17.949~$\mu$m resp. +201~km~s$^{-1}$; 
17.955~$\mu$m resp. +302~km~s$^{-1}$.  Note that the three selected velocities in any given row are not the same.}\label{fig:arfe}
\end{figure*}

\begin{figure*}
\hbox{
\includegraphics[width=0.35\textwidth, angle=180]{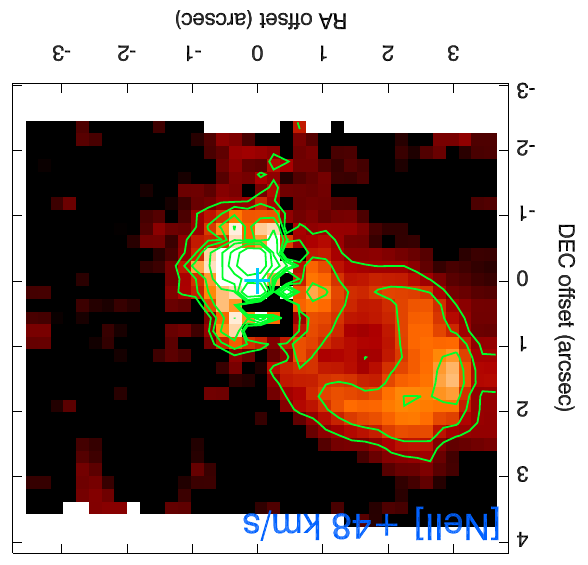}\hskip 1cm
\includegraphics[width=0.35\textwidth, angle=180]{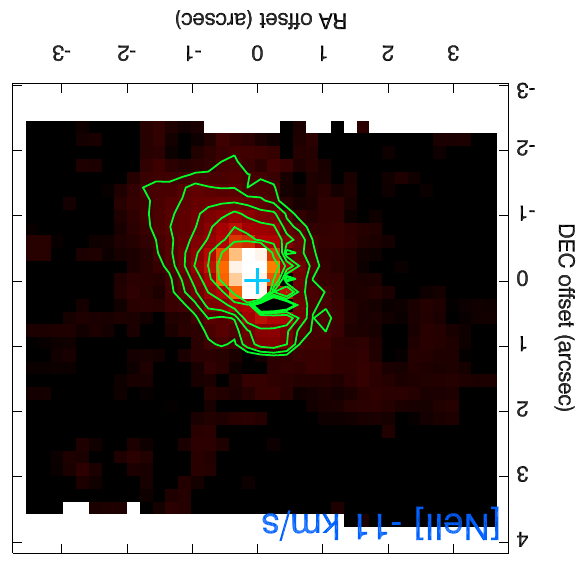}
}
\hbox{
\includegraphics[width=0.35\textwidth, angle=180]{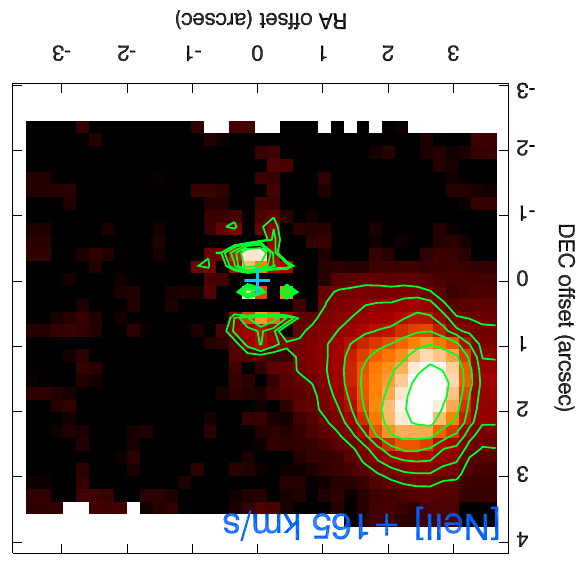}\hskip 1cm
\includegraphics[width=0.35\textwidth, angle=180]{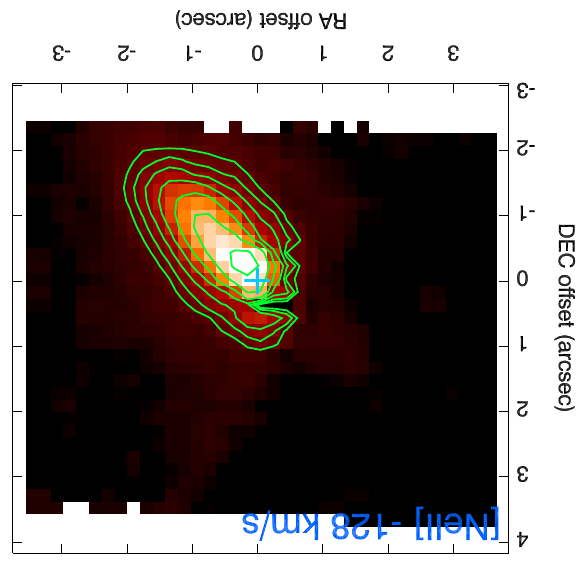}
}
\hbox{
\includegraphics[width=0.35\textwidth, angle=180]{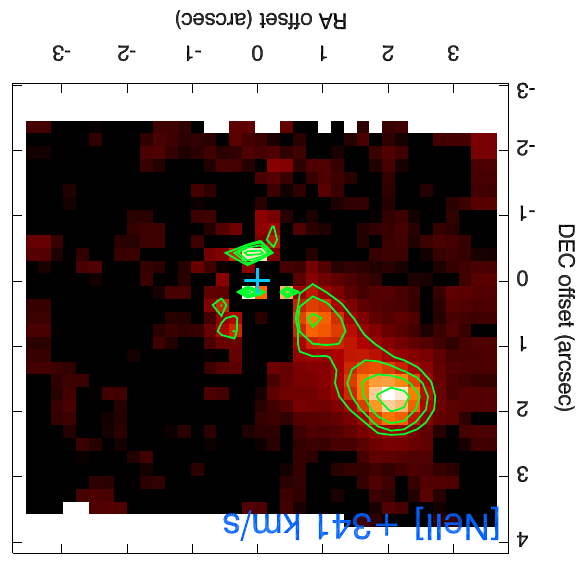}\hskip 1cm
\includegraphics[width=0.35\textwidth, angle=180]{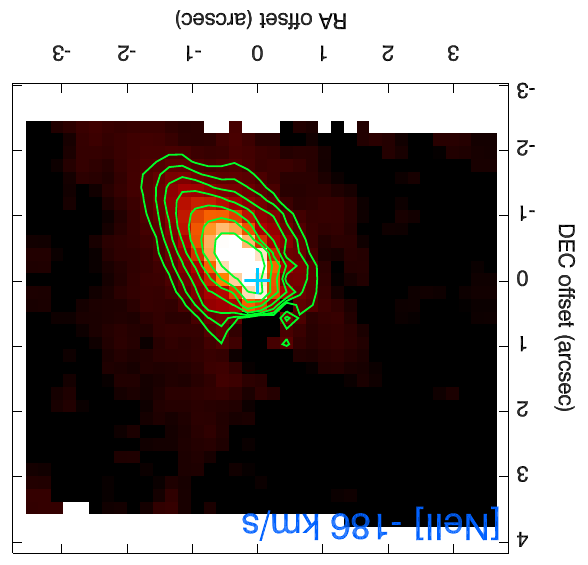}
}
\caption{Pairs of flux images for [Ne~{\sc ii}]~12.81~$\mu$m at low, intermediate, and high velocities, corresponding to the images superimposed in Fig.~\ref{fig:RGBimages}.
The left column shows positive velocities (redshifted line) and the right column negative velocities (blueshifted line).
\textit{Top row:} 
Low-velocity jet, shown in blue in Fig.~\ref{fig:RGBimages}: \textit{left:} 12.81625~$\mu$m resp. +48~km~s$^{-1}$; \textit{right}: 12.813755~$\mu$m resp. -11~km~s$^{-1}$. \textit{Middle row:} 
Intermediate-velocity jet, shown in green in Fig.~\ref{fig:RGBimages}: 
\textit{left:} 12.82125~$\mu$m resp. +165~km~s$^{-1}$; \textit{right}: 12.80875~$\mu$m resp. -128~km~s$^{-1}$. 
\textit{Bottom row:} 
High-velocity jet, shown in red in Fig.~\ref{fig:RGBimages}: 
\textit{left:} 12.82875~$\mu$m resp. +340~km~s$^{-1}$; \textit{right}: 12.80625~$\mu$m resp. -186~km~s$^{-1}$. Note that the absolute values of the velocities in the left column do not correspond to those in the right column but were chosen such that they optimally highlight different morphological features described in the main text.
}\label{fig:neIIvelocity}
\end{figure*}

This section shows channel maps for lines of [Ne~{\sc ii}], [Ne~{\sc iii}], [Ar~{\sc ii}], [Ni~{\sc ii}], and [Fe~{\sc ii}]  at different radial velocities relative to DG Tau. 
Figures~\ref{fig:ratio1}--\ref{fig:ratio8} show pairs of [Ne~{\sc ii}] and [Ne~{\sc iii}] images at similar radial velocities together with the flux ratio images. If the velocities given by the spectral bins were different by more than 15~km~s$^{-1}$, we also used averages of two spectral bins to obtain closer velocity agreement. Figure~\ref{fig:arfe} shows a series of images taken at significantly different radial velocities in the [Ar~{\sc ii}], [Fe~{\sc ii}], and [Ni~{\sc ii}] lines. And Fig.~\ref{fig:neIIvelocity} compares [Ne~{\sc ii}] images taken at low, intermediate, and high velocities both in the blue and red jets; these data were used to produce the RGB images shown in Fig.~\ref{fig:RGBimages}.

\section{Nonequilibrium Ionization behavior of cooling shocks}\label{NEI}

It is well known that CIE conditions are violated during some stages of shock cooling, and Nonequilibrium Ionization (NEI) models must be applied when the plasma cooling time becomes shorter than the recombination time scales (e.g., \citealt{allen2008, gnat2009}). 

Furthermore, photoionization of  both the shocked gas and the gas 
upstream of the shock (i.e., the precursor) by the emitted radiation from the shock itself (``self-radiation'') must be considered; the precursor will adjust to photoionization equilibrium and could become a strong line emitter itself \citep{gnat2009, allen2008}. The relevant stages in a shock are the following \citep{sutherland1993, dopita1996, gnat2007, gnat2009, allen2008}:
\begin{itemize}
\item The short ionization region right behind the shock rapidly ionizes and heats the inflowing pre-shock gas to $T_s$ with ion abundance fractions lagging behind the temperature increase, i.e., the gas is initially under-ionized. 
\item In the high-temperature radiation zone at $T_s \gg 10^6$~K, cooling via radiation of EUV and X-rays is slow, and the gas rapidly adjusts to CIE conditions.
In the example shown in Fig.~\ref{fig:shockmodels}, this region would range between a distance of $\sim 0$ and about $13.5\times 10^{16}$~cm or a time up to $\sim 1.3\times 10^{10}$~s.
\item The nonequilibrium cooling zone sets in once $T \la 10^6$~K when cooling becomes very rapid and shorter than individual recombination time scales; the gas is out of collisional ionization  equilibrium, and recombination lags behind the temperature development under CIE conditions. The NEI cooling efficiency is smaller than for CIE because the gas is over\-ionized while cooling, which results in less efficient radiation. (At a location of $\sim 13.5\times 10^{16}$~cm or time of $\sim 1.3\times 10^{10}$~s in Fig.~\ref{fig:shockmodels}).
\item A super-cooled zone develops when $T$ reaches $\sim$10$^4$~K as photoionization sets in, but the ionization state is still too high (due to recombination lag) to efficiently absorb photons (right after the rapid temperature decline in Fig.~\ref{fig:shockmodels}, but not resolved in these plots).
\item The photoabsorption and recombination zone develops below $10^4$~K in which the shock-emitted EUV photons develop a H\,{\sc ii} layer and X-rays continue ionizing some of the recombining, cooling gas. (After  $13.5\times 10^{16}$~cm or $\sim 1.3\times 10^{10}$~s in Fig.~\ref{fig:shockmodels}).
\end{itemize}
Models are complemented with a magnetic-field strength parameter that essentially determines the compressibility of the gas; the extreme cases are an incompressible fluid for high $B$ (``isochoric cooling'' with constant hydrogen density), in which the magnetic pressure counteracts compression of the cooling gas,
and a fully compressible fluid for zero $B$ (``isobaric cooling'' with constant pressure). Below $10^5$~K, isobaric shocks cool rapidly because the gas density increases inversely to $T$. Isochoric jets, instead, show an extended photoabsorption and recombination zone at $10^4$~K and below.  Conservation conditions require the gas velocity to scale roughly inversely with the particle density, which at the end of the cooling phase slows down the jet gas relative to the shock front, in contrast to the isochoric case.

To first order and for pre-shock densities up to $n_{\rm pre} = 10^3$~cm$^{-3}$, the ratios between cooling time and recombination times are independent of the pre-shock gas density because both processes scale with density. Also, all ion fractions are density-independent because all relevant reactions including bremsstrahlung are ``two-body'' interactions, i.e., the $n^2$ drops out for ratios. The parameter $n_{\rm pre}t$ is invariant for changing density ($t$ is the time since a gas parcel entered the shock). Only for higher pre-shock densities, collisional de-excitation begins to play a role in the recombination zone as well.

\section{Overview of the DG Tau star-disk-winds-jet system}

Figure~\ref{fig:overview_alt} gives an alternative representation of Fig.~\ref{fig:overview}, separating features within $\sim$1\arcsec\ of the star
from the large-scale features (plus the disk and the $^{12}$CO arc at the disk edge). 

\begin{figure}[h!]
    \centering
\includegraphics[width=1.0\linewidth]{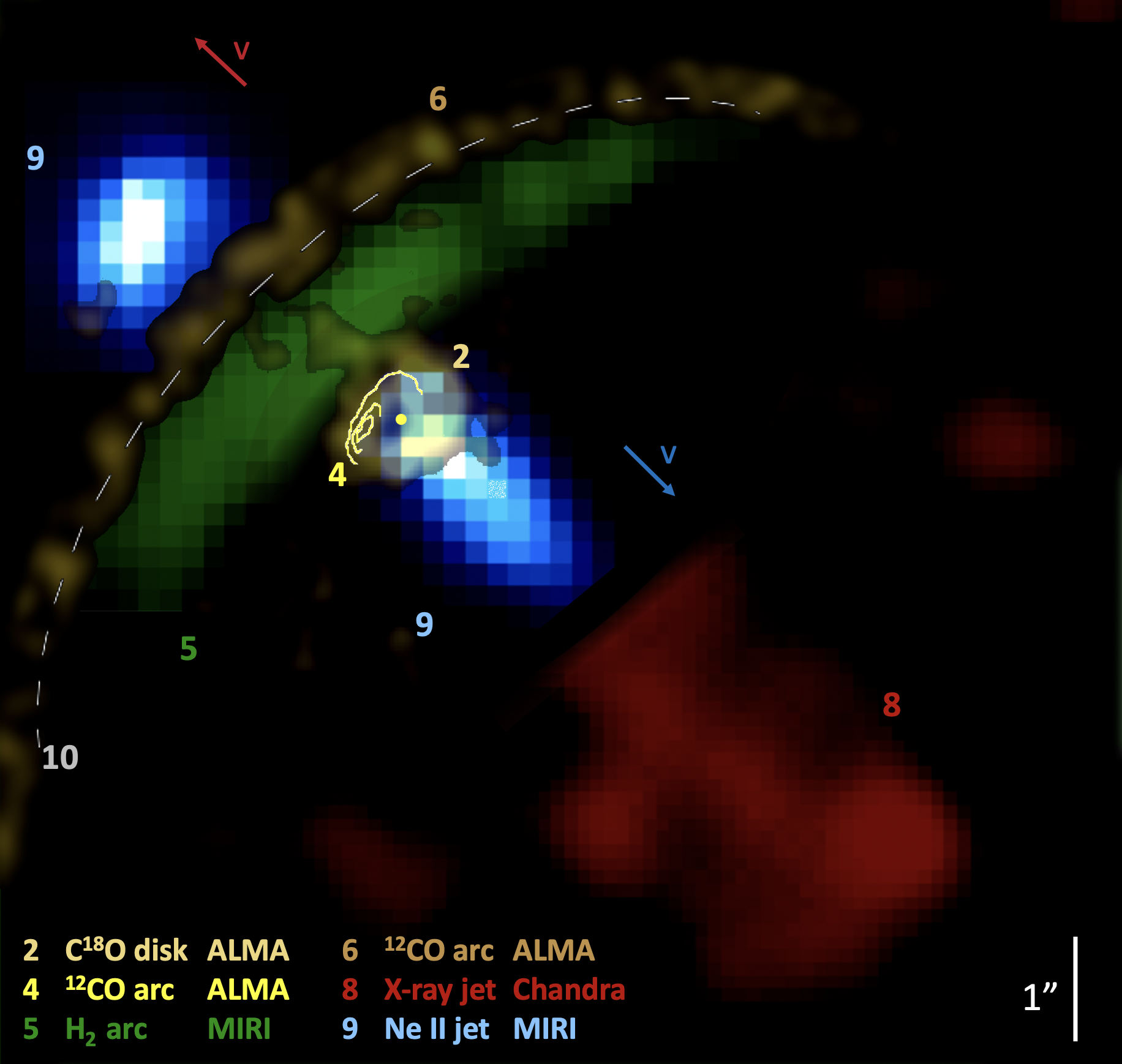}
\includegraphics[width=1.0\linewidth]{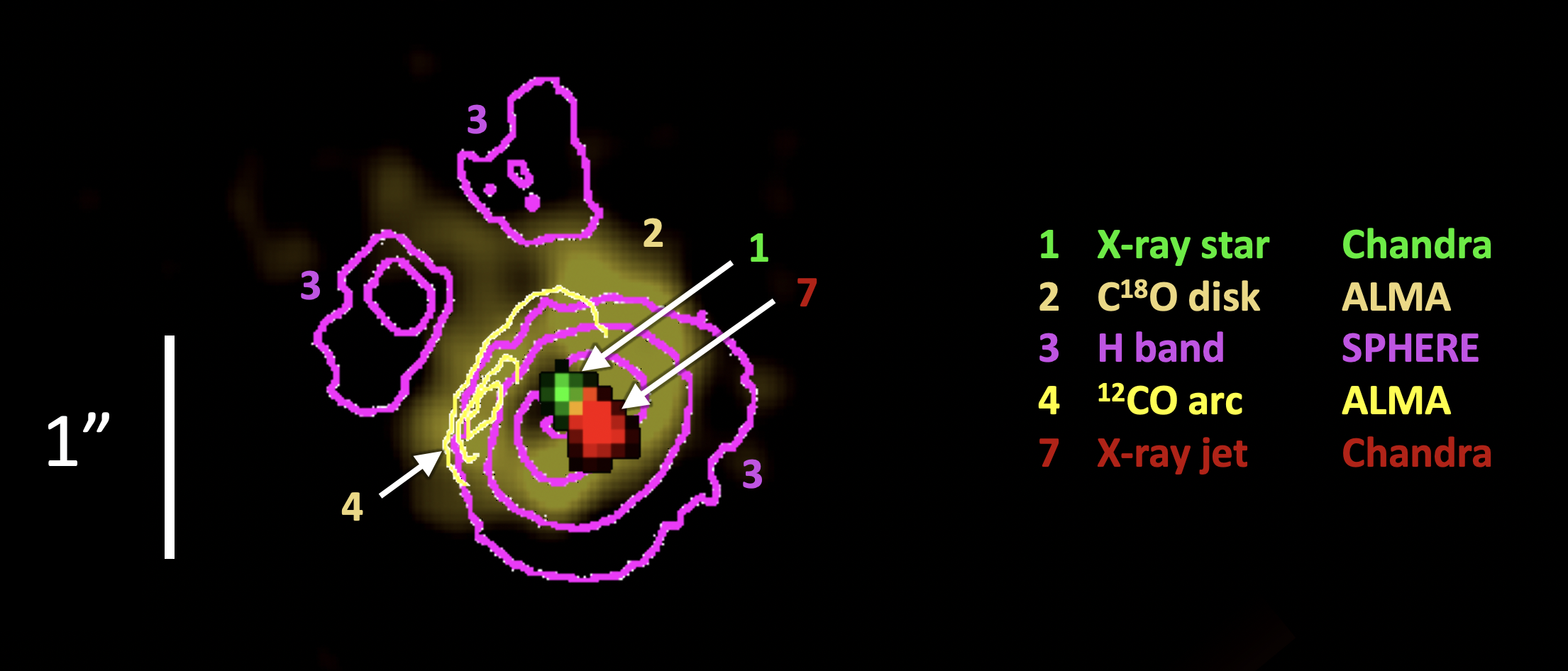}
    \caption[short]{  
    \textit{Top:} Overview of the large-scale structure of the DG~Tau system. The numbering of the features is the same as in Figs.~\ref{fig:overview} and \ref{fig:sketch}. White dot in center: The star DG Tau. 
     \textbf{2. }Faint yellow ellipsoidal structure around star: Protoplanetary disk in the C$^{18}$O line at 219.560~GHz observed by ALMA, smoothed  \citep{guedel2018}.    
     \textbf{4. }Bright yellow contours at the NE edge of the disk: Wind in the $^{12}$CO line at 230.538~GHz observed by ALMA (from moment-1 map; \citealt{guedel2018}). 
    \textbf{5. }Faint green broad arc to the NE of disk: Arc in H$_2$ S(1)~17.03~$\mu$m observed by MIRI/MRS (smoothed), probably located in front of the disk, as discussed in this paper. 
     \textbf{6. }Brown arc to the NE of H$_2$ arc: Structure in the $^{12}$CO line at 230.538 GHz observed by ALMA \citep{guedel2018}. 
      \textbf{8. }Red elongated feature in the lower right: Outer blueshifted jet in 0.3--1.5~keV X-rays observed by Chandra (X-ray data from ObsID~11009/11010/11011); \citealt{guedel2011, Frank2014}); the image has been smoothed and cut off toward the star because of strong overlap with the PSF wings from the central X-ray source.  \textbf{9. }Blue, elongated structure to the SW of the star and blob in the NE: The jets in the [Ne\,{\sc ii}]~12.81$~\mu$m line observed by MIRI/MRS discussed further in this paper. 
     \textbf{10. }Dashed gray arc overlaid on $^{12}${\rm CO} arc: Part of a projected circle at an inclination of 38\degr\ to guide the eye.  
    }\label{fig:overview_alt}
    \end{figure}
\begin{figure}[h!]
    \ContinuedFloat
    \caption[] {\textit{Bottom:} Inner structures.  
    \textbf{1. }Bright green dot in center: The star DG Tau as observed in 2--8~keV X-rays by Chandra \citep{guedel2011, Frank2014}.      \textbf{2. }Faint yellow ellipsoidal structure around star: Protoplanetary disk in the C$^{18}$O line at 219.560~GHz observed by ALMA, smoothed  \citep{guedel2018}.  \textbf{3. }Purple contours above the disk: Scattered-light image in $H$-band from SPHERE, showing a steep cavity-like structure pointing along the forward jet axis, and filamentary extensions to the NE \citep{garufi2024}.  A  similar structure is seen in H$_2$ \citep{agraamboage2014}.
    \textbf{4. }Bright yellow contours at the NE edge of the disk: wind in the $^{12}$CO line at 230.538~GHz observed by ALMA (from moment-1 map; \citealt{guedel2018}).     
     \textbf{7. }Red elongated feature near star: Inner jet in 0.3--1.5~keV X-rays observed by Chandra \citep{guedel2011, Frank2014}; the  X-ray stellar and jet sources have been deconvolved using sub-pixel event repositioning; the region between the inner and the outer jet cannot be reliably mapped. 
     All observations have been corrected for proper motion of DG Tau to a common date. However, the SPHERE image required manual adjustment by a fraction of an arcsecond. 
    }
    \end{figure}

\end{appendix}

\end{document}